%% file: JBES_submission.tex
\documentclass[12pt,letterpaper]{article}

\newif\ifanonymous
\anonymousfalse

\usepackage[margin=1in]{geometry} 

\usepackage{mathpazo}
\usepackage{float}
\usepackage{placeins}
\usepackage{booktabs}
\usepackage{amsmath}
\usepackage{mathtools}
\usepackage{amssymb}
\usepackage{graphicx}
\usepackage{setspace}
\usepackage{natbib,epstopdf}
\usepackage{multibib} 
\newcites{app}{References}
\usepackage{xcolor}
\usepackage{hyperref}
\usepackage{appendix}
\usepackage{longtable}
\usepackage{bm}
\usepackage{subfig}
\usepackage{titlesec} 

\usepackage[utf8]{inputenc} 
\usepackage[T1]{fontenc}    
\usepackage{xspace}

\usepackage{pifont} 

\definecolor{myr}{rgb}{0.6,0,0}
\definecolor{myb}{rgb}{0,0,0.6}
\definecolor{myg}{rgb}{0,0.4,0}
\hypersetup{colorlinks,
    citecolor=myb,
    filecolor=myb,
    linkcolor=myr,
     urlcolor=myg}
\emergencystretch=\maxdimen
\bibpunct{(}{)}{;}{a}{,}{,}

\input{dfn}

\renewcommand{\footnotesize}{\fontsize{9}{10.5}\selectfont}

\makeatletter
\def\@maketitle{%
  \newpage
  \null
  \vskip 0.3em%
  \begin{center}%
  \let \footnote \thanks
    {\LARGE \@title \par}%
    \vskip 0.3em%
    {\large
      \lineskip .2em%
      \begin{tabular}[t]{c}%
        \@author
      \end{tabular}\par}%
    \vskip 0.2em%
    {\large \@date}%
  \end{center}%
  \par
  \vskip 0.2em}
\makeatother

\titleformat{\section}{\normalfont\large\bfseries}{\thesection}{1em}{}
\titleformat{\subsection}{\normalfont\normalsize\bfseries}{\thesubsection}{1em}{}
\titleformat{\subsubsection}{\normalfont\normalsize\bfseries}{\thesubsubsection}{1em}{}
\titlespacing*{\section}{0pt}{1.5ex plus .8ex minus .2ex}{0.6ex plus .2ex}
\titlespacing*{\subsection}{0pt}{1.2ex plus .6ex minus .2ex}{0.5ex plus .2ex}
\titlespacing*{\subsubsection}{0pt}{1ex plus .5ex minus .2ex}{0.3ex plus .2ex}

\begin{document}
\setstretch{1.72} %

\ifanonymous
\title{Macroeconomic Forecasting with Large Language Models}
\author{}
\else
\title{Macroeconomic Forecasting with Large Language Models\thanks{We thank Todd Clark, Frank Diebold, Domenico Giannone, Frank Schorfheide, and seminar participants at the European Central Bank, the 2024 NBER-NSF Time Series Conference, the 2025 IAAE Conference, the 2025 NBER Summer Institute, and the 46th International Symposium on Forecasting  for their helpful comments and suggestions. The views and opinions expressed here are those of the authors and do not necessarily reflect the views or positions of any entities they are affiliated with.}}

\author{ Andrea Carriero\thanks{
School of Economics and Finance, Queen Mary University of London. Email:
\href{mailto:a.carriero@qmul.ac.uk}{a.carriero@qmul.ac.uk}} ,
\ Davide Pettenuzzo\thanks{
School of Business and Economics, Brandeis University. Email:
\href{mailto:dpettenu@brandeis.edu}{dpettenu@brandeis.edu}} ,
\ and Shubhranshu Shekhar\thanks{
School of Business and Economics, Brandeis University. Email:
\href{mailto:sshekhar@brandeis.edu}{sshekhar@brandeis.edu}} }
\fi
\date{August 2026}
\maketitle
\begin{abstract}
\singlespacing
\fontsize{11}{13.2}\selectfont

\noindent This paper evaluates the accuracy of Time Series Foundational Models (\tsfm{}s) against traditional macroeconomic forecasting approaches. \tsfm{}s have gained popularity for capturing intricate patterns in data and adapting across domains, but their effectiveness relative to conventional macro-forecasting methods remains an open question. We conduct a rigorous evaluation of \tsfm{}s against traditional macro-prediction methods using the FRED-MD database, documenting substantial heterogeneity in performance across economic domains and persistence regimes. We show that forecast errors from econometric models and \tsfm{}s are nearly orthogonal at short horizons, providing complementary information. Building on this, we propose a hybrid approach combining the structure of econometric models with the flexibility of \tsfm{}s, yielding the most robust and accurate forecasts.

\smallskip


\smallskip


\end{abstract}

\section{Introduction}

The recent emergence of Large Language Models (\llm{}s) has reshaped the landscape of natural language processing, ushering in a new era of computational linguistics. Bolstered by advancements in machine learning and deep neural networks, \llm{}s have garnered widespread attention for their remarkable ability to understand and generate human-like text. This transformative technology has revolutionized various applications, ranging from machine translation and sentiment analysis to chatbots and content generation. By leveraging vast amounts of text data and sophisticated algorithms, \llm{}s have demonstrated unparalleled proficiency in capturing linguistic nuances, contextual dependencies, and semantic meanings.\footnote{For example, the first draft of the abstract of this paper was generated with a single prompt by chatGPT 3.5.} As a result, they are increasingly becoming invaluable tools for researchers, developers, and businesses. 

In this paper we focus on an even more recent development, which is the use of \llm{}s to forecast time series data. While \llm{}s have predominantly been associated with natural language processing tasks, their versatility and adaptability have sparked interest in exploring their capabilities beyond linguistic domains. By leveraging the computational power and flexibility inherent in machine learning algorithms, \llm{}s promise to uncover intricate nonlinear relationships, capturing latent dynamics, and adapting to evolving data patterns. 

This latest iteration of \llm{}s, trained specifically with the goal of forecasting time series data, is referred to as either Time Series Language Models (\tslm{}s) or Time Series Foundational Models (\tsfm{}s) in the Machine Learning and Artificial Intelligence literature.\footnote{The \tslm acronym makes it explicit that these models are derived from textual \llm{}s and as such share many similarities with them both in terms of architecture and model design.} Several \tsfm{}s have already been produced and are publicly available, including IBM's Tiny Time Mixers \citep{ekambaram2024ttms}, \timellm \citep{jin2023time}, \lagllama \citep{rasul2023lag}, Google's \timesfm \citep{das2024decoder}, Salesforce's \moirai \citep{woo2024unified}, Amazon's \chronos \citep{ansari2024chronos}, and Nixtla's \timegpt \citep{garza2023timegpt1}. All these contributions show that \tsfm{}s can produce gains in forecast accuracy, but in their empirical evaluations they tend to focus on a variety of datasets from very different domains and do not offer any specific information on how these models would fair in forecasting macroeconomic variables specifically. 

Our first contribution in this paper is a thorough investigation on how \tsfm{}s perform in predicting macroeconomic time series. A second contribution is a detailed comparison of their performance versus state-of-the-art time series methods such as Bayesian Vector Autoregressions (BVARs) and Factor Models, as well as Neural Networks and Bayesian Additive Regression Trees (BART), which have both become very popular nonlinear models for macroeconomic forecasting. A third contribution is the development of a novel hybrid approach that combines the structural rigor of traditional econometric models with the adaptability of \tsfm{}s, leading to consistent improvements in forecast accuracy across variables and horizons. In these endeavors, we focus on forecasting the variables contained in the FRED-MD dataset, a comprehensive repository of more than one hundred monthly macroeconomic variables curated by the Federal Reserve Economic Data (FRED) system.\footnote{See \href{https://research.stlouisfed.org/econ/mccracken/fred-databases/}{https://research.stlouisfed.org/econ/mccracken/fred-databases/}.} This dataset encompasses a diverse array of economic indicators, and it is considered a benchmark in macroeconomic forecasting.\footnote{A non-exhaustive lists of contributions using this data set to study the performance of alternative models in macroeconomic forecasting includes  \citet{Stock:Watson:2006}, \citet{Banburaetal2010}, \citet{Carriero:etal:JBES:2016}, \citet{CARRIERO2019JOE}, \citet{KOROBILIS2019241}, \citet{KOOP2019135},  \citet{chan2022asymmetric}.} 

As noted in \citet{FL2024}, there are some characteristics of LLM models that makes it problematic to use them in the pseudo out-of-sample forecasting exercise that is typically performed in the empirical macroeconomics literature. One major challenge is that LLMs are pretrained by developers on datasets of their choice, giving the researcher little to no control over the training data. For example, of the LLMs considered in this work, three out of five of them (including one of the best performing ones, Salesforce's \moirai) list in their training data a large subset of the series we set out to forecast in this paper. To further complicate matters, since the training data lacks timestamps it's hard to retrain the model up to a certain date, making it difficult to reproduce the type of real-time analysis that an economic forecaster is mostly concerned with. Solving these issues by retraining LLMs on subsets of data and adding data to the training set is usually not possible due to high hardware requirements. But having said that, it is important to recognize that even including the dataset of interest in real time would not remove the issue of a contaminated training set, i.e. including information that the forecaster would not have access to in real-time. 

The picture emerging from our analysis is one in which only three of the six \tsfm{}s we consider are competitive against a simple AR benchmark (Salesforce's \moirai, Google's \timesfm, and Amazon's \chronos). Moreover, when these more competitive \tsfm{}s are stacked against Bayesian VARs and Factor Models, their forecasting performance is broadly comparable, if not slightly inferior. In particular, we find that the forecasting gains achieved by the econometric models tend to be more stable while \tsfm{}s can perform very well for a handful of series but also show less reliability at times, as they seem to be more prone to generating the occasional unreasonable forecast.\footnote{This result may not appear to some as totally surprising. When working with textual LLMs, we are sometimes exposed to ``Model hallucinations'', i.e. the tendency of the model to  generate content that is irrelevant, made-up, or inconsistent with the input data.} On the other hand, \tsfm{}s seem to work relatively better in the post-Covid-19 era (with the important caveat that the training set of these models does include information from the pandemic and post-pandemic period). While these results are based on using the \tsfm models in a zero-shot forecasting manner, we also find that fine-tuning the \tsfm{} models does not yield to significant improvements in their forecast accuracy.\footnote{By ``zero-shot'', we mean that the model is first trained on a large amount of data, and is then directly applied to solve a new task without re-training, or fine-tuning, its parameters on the new dataset.}  

A growing number of papers apply textual LLMs to macroeconomics and finance, and more are likely to follow: \citet{bybee2023surveying} prompt an LLM with historical WSJ news to construct time series of macroeconomic expectations; \citet{Chen_Kelly_Xiu_2022} use LLM-based sentiment scores from global news text to predict firm-level returns; \citet{Kim_et_al_WP_2024} find that GPT-4 can outperform financial analysts at predicting earnings direction from financial statements; and \citet{Zhou_et_al_WP_2025} show that a news-based sentiment ratio extracted by prompting ChatGPT predicts stock returns. These approaches leverage LLMs' natural language processing abilities, whereas this paper focuses on models trained specifically for time-series forecasting. The closest work we are aware of is \citet{FL2024}, which looks at time series LLMs but focuses on a single LLM and a single target variable (inflation). In contrast, we evaluate a variety of \tsfm{}s against state-of-the-art macroeconomic forecasting approaches across a large set of macroeconomic indicators. 

The paper is organized as follows. Section~\ref{sec:tsfm} introduces \tsfm{}s. Section~\ref{sec:models} describes our econometric benchmarks. Section~\ref{OOS} describes the data and the design of our pseudo out-of-sample forecasting exercise. Section~\ref{Results} presents our main results, and Section~\ref{TSLM-BVAR} introduces a hybrid approach combining econometric models with \tsfm{}s. Section~\ref{closerlook} examines the robustness, heterogeneity, and evolution of \tsfm{} performance in more detail, and Section~\ref{Conclusions} concludes. An Online Appendix provides further technical details, additional results, and data transformations.

\noindent 

\FloatBarrier
\section{Foundational Models and LLMs}\label{sec:tsfm}
In recent years, Foundational Models (FMs), i.e. very large neural network models with billions of parameters have received tremendous attention in computer science fields due to their ability to capture complex relationships within massive datasets. One of the most astonishing features of FMs is their flexibility, i.e. their applicability to various tasks across many different domains in a zero-shot manner, effectively eliminating the need to train specialized models for each separate task. Within this context, textual LLMs have emerged as a new way of understanding language, generating text and images, and conversing in natural language, and have completely revolutionized the way we interact with technology \citep{achiam2023gpt}. State-of-the-art textual LLMs and Chatbots, such as Open AI's ChatGPT and Meta's Llama, are now being used extensively across various natural language processing tasks, such as Information Retrieval and Scientific Research, but also Code Generation and Debugging.

\FloatBarrier
\subsection{Time Series Foundational Models (\tsfm{}s)}

Inspired by these successes, researchers are now exploring the usefulness of LLMs in other settings. Among these extensions, we are witnessing the emergence of Time Series Foundational Models (\tsfm), bridging the gap between LLMs' original text data training and the numerical nature of time series data. TSFMs are now being used to accomplish a variety of time series related tasks, ranging from prediction and classification to anomaly detection and data imputation. Notable examples of this new wave of model development include \citet{rasul2023lag}, \citet{goswami2024moment}, \citet{ekambaram2024ttms}, \citet{garza2023timegpt1}, \citet{das2024decoder}, and \citet{ansari2024chronos}. While there are in principle significant differences between a textual LLM, whose main task is essentially to predict the next word, and a time series forecasting model, whose main objective is to predict the next value of one or multiple time series jointly, inherently these two tasks are deeply intertwined: both aim at learning the sequential structure of the data from the past and present history of some features and with that attempt to predict a sequence of future outcomes.

In the case of a \tsfm{}, the pretraining is done on a possibly very large set of $N$ time series observed over the time periods $t=1,\dots,T$ which we collect in $\bm{X}_{1:T}=\left(x_{1,1:T},...,x_{N,1:T}\right)$. Pretraining yields a very large set of coefficients $\widehat{\theta} \in \Theta \subseteq \mathbb{R}^d$ which describe a forecasting function $f_{\widehat{\theta}}$. Then, given $f_{\widehat{\theta}}$ and the current and past values of a time series of interest $y_{1:T}$ (not necessarily in the training set), the \tsfm{} attempts to predict its future $h$ values $y_{T+1:T+h}$ using\footnote{To keep the notation simple, we have assumed here a balanced set of features, i.e. all available over the same time period. If that is not the case, then we could instead write this as $\bm{X}_{1:T}=\left(x_{1,\underline{t}_1:T_1},...,x_{N,\underline{t}_N:T_N}\right)$, where $\underline{t}_i$ and $T_i$ would denote, respectively, the first and last observation of feature $i$, $i \in \left\{1,...,N\right\}$.} 
    \begin{equation}
        \widehat{y}_{T+1:T+H|T}=f_{\widehat{\theta}} ( \left. y_{1:T}\right)=f ( \left. y_{1:T}|\bm{X}_{1:T} ,\widehat{\theta}  \right.).
    \end{equation}
The main difference with LLMs is that while natural language consists of words from a finite vocabulary, time series are real-valued.

\FloatBarrier
\subsection{Components of a \tsfm{}}
\label{subsec:tslm}

\begin{figure}[!htb]
 \centering
\includegraphics[width=0.9\textwidth]{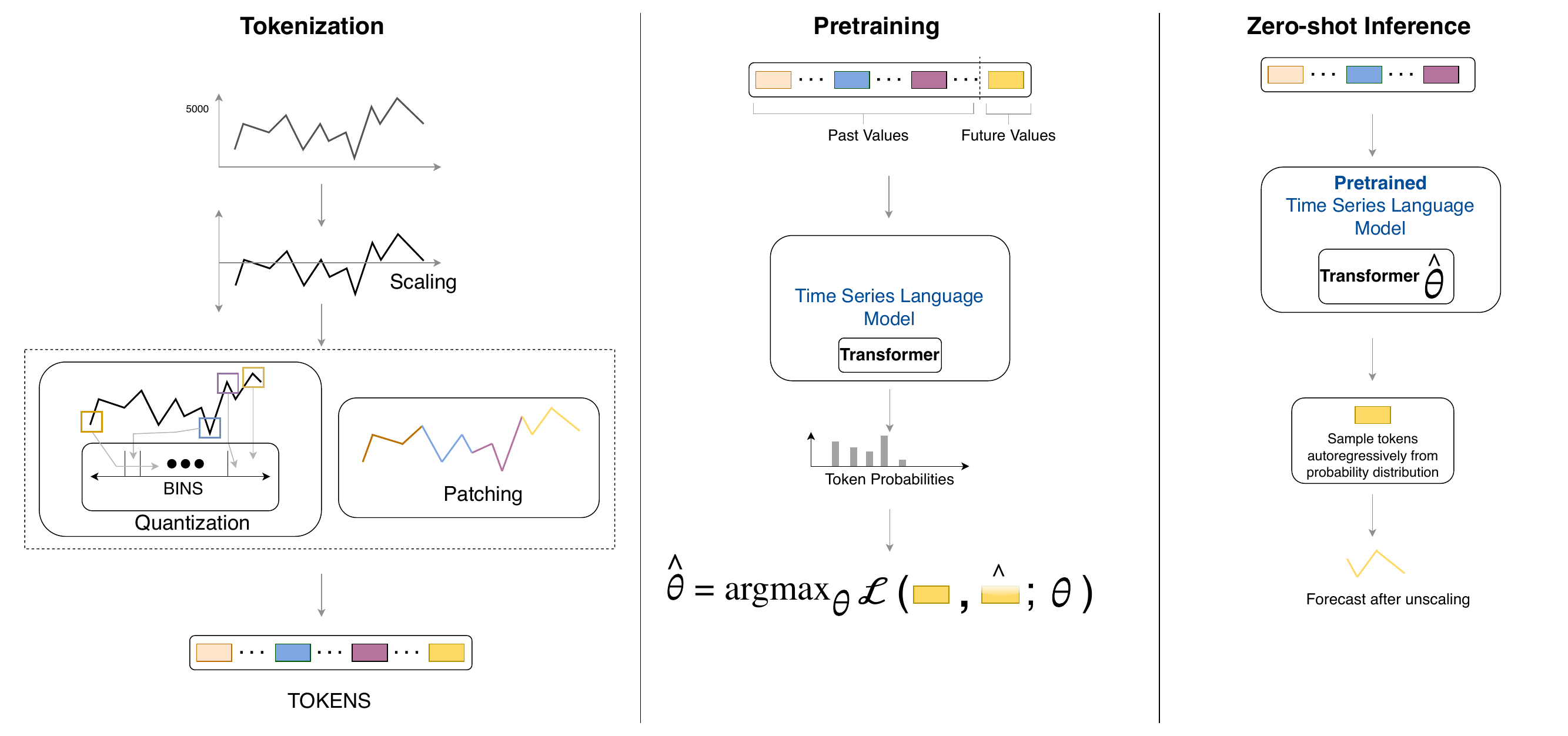}
\caption{Components of a \tsfm : The key components of a \tsfm include tokenization, pretraining, and zero-shot forecasting. First, raw time series  is tokenized into discrete tokens using techniques like scaling, quantization, and patching. Tokens are then used to pretrain a Transformer-based model that learns temporal dependencies and patterns. Finally, through zero-shot forecasting, the pretrained model can forecast in new and unseen time series domains.\label{fig:tslm}}
\end{figure}

The earliest applications of LLMs to time series forecasting simply prompted existing textual LLMs with time series data converted into text prompts; this approach proved sub-optimal, mainly due to the lack of real-world time series data in the training of these LLMs in the first place. The most recent versions of \tsfm{}s instead build on quantization-based methods, i.e. a process through which the numerical data is converted into discrete representations first, before becoming an input to the training of the model. \autoref{fig:tslm} provides a schematic overview of the three key components of a \tsfm{}: tokenization, pretraining, and zero-shot forecasting. Different choices and configurations at each of these stages define the unique characteristics of the various \tsfm{}s; \ref{App_TSFM} provides a detailed, tutorial-style description of tokenization, model architecture, data augmentation, and pretraining/fine-tuning.

We consider six \tsfm{}s in our empirical application -- \lagllama \citep{rasul2023lag}, \moirai \citep{woo2024unified}, \chronos \citep{ansari2024chronos}, \timesfm \citep{das2024decoder,timesfm25}, \ttm \citep{ekambaram2024ttms}, and \timegpt \citep{garza2023timegpt1} -- ranging in size from roughly 2.5M ($\sim$1M for \ttm) to 200M parameters, and pretrained on large collections of publicly available time series spanning energy, transport, web, sales, and other domains, among others. \ref{App_TSFM} describes each of these models in detail, along with their training data and size (\autoref{tab:TSLM_training_details}). The same Appendix also describes the newest generation of \tsfm{}s -- \chronos-2, \moirai 2.0, and \timesfm 2.5 -- released in the past several months, which we compare empirically to their predecessors in Section~\ref{newgen}.

\FloatBarrier
\section{Econometric Models} \label{sec:models}

Previous work focusing on macroeconomic forecasting has evidenced that models which perform particularly well are Bayesian VARs (BVARs) and Factor Models. See e.g. \citet{Stock:Watson:2006}, \citet{Banburaetal2010}, \citet{Carriero:etal:JBES:2016}, \citet{CARRIERO2019JOE}, \citet{KOROBILIS2019241}, \citet{KOOP2019135},  \citet{chan2022asymmetric}. We therefore start from these models. We also include neural networks and regression trees in the comparison, as recent work by \citet{hyndman2018forecasting}, \citet{HUBER202352}, and \citet{Clark_etal_2023} show that in some instances allowing for non-linearities can help improve the accuracy of macroeconomic forecasts. Below we describe the specific choices we made in terms of the model specifications. 

\FloatBarrier
\subsection{Bayesian VAR}

We consider the popular BVARs with natural conjugate Normal-inverted Wishart (N-IW)
priors. This prior dates back to \citet{Zellner1971} and was later studied by
\citet{KK1993,Kadiyala:Karlsson:1997}. In a seminal paper, \citet{Banburaetal2010} show that it can be successfully applied to a very large cross
section of macroeconomic data. Several contributions followed using this
model and prior to handle large macroeconomic datasets. Collect $N$ different variables in the vector $y_{t}=(y_{1t}\ y_{2t}\
...\ y_{Nt})^{\prime }$, and write the Vector Autoregression or order $p$, $VAR(p)$, as:%
\begin{equation} 
y_{t}=\Phi _{c}+\Phi _{1}y_{t-1}+\Phi _{2}y_{t-2}+...+\Phi
_{p}y_{t-p}+\varepsilon _{t};\ \varepsilon _{t}\sim i.i.d.N(0,\Sigma ),
\label{TSnotation}
\end{equation}%
where $t=1,\dots,T$ denotes time. Note that with $p$ lags and $N$ variables, each equation has $K=Np+1$ regressors. By grouping the
coefficient matrices in the $N\times $ $K$ matrix $\Phi ^{^{\prime }}=[\Phi
_{c}\ \Phi _{1}\ ...\ \Phi _{p}]$ and defining $x_{t}=(1\ y_{t-1}^{\prime }\
...\ y_{t-p}^{\prime })^{\prime }\ $as a vector containing an intercept and $%
p$ lags of $y_{t}$,\ the $VAR$ can be written as: $y_{t}=\Phi ^{\prime }x_{t}+\varepsilon _{t}.  $
Next, after stacking the $t=1,\dots,T$ equations by column and transposing we can rewrite the $VAR(p)$ using more compact matrix notation: %
\begin{equation}
Y=X\Phi +E,  \label{MATRIXnotation}
\end{equation}%
where $Y=$ $[y_{1},..,y_{T}]^{\prime }$, $X=$ $[x_{1},..,x_{T}]^{\prime }$,
and $E=[\varepsilon _{1},..,\varepsilon _{T}]^{\prime }$ are $%
T\times N$, $\ T\times K\ $and $T\times N$ matrices.\ 

The conjugate N-IW prior elicits the prior on the coefficients conditionally
on the error variance $\Sigma $:%
\begin{equation}
\Phi |\Sigma \sim N(\Phi _{0},\Sigma \otimes \Omega _{0}),\ \Sigma \sim
IW(S_{0},v_{0}).  
\label{NIWprio}
\end{equation}%
The conditional posterior distribution of this model is also N-IW:
\begin{equation}
\Phi |\Sigma ,Y\sim N(\bar{\Phi},\Sigma \otimes \bar{\Omega}),\ \Sigma
|Y\sim IW(\bar{S},\bar{v}),  \label{NIWpost}
\end{equation}%
where $%
\bar{\Phi}=\bar{\Omega}(\Omega _{0}^{-1}\Phi
_{0}+X^{\prime }Y)$, $\bar{\Omega}=(\Omega _{0}^{-1}+X^{\prime }X)^{-1}$,  $\bar{S}=S_{0}+Y^{\prime }Y+\Phi _{0}^{\prime }\Omega _{0}^{-1}\Phi
_{0}-\bar{\Phi}^{\prime }\bar{\Omega}^{-1}\bar{\Phi}$, and $\ 
\bar{v}=v_{0}+T$. 
We elicit $\Phi _{0},\Omega_{0},S_{0},v_{0}$ as in the Minnesota tradition. To these priors, we add the "sum of coefficients" and "single unit root" priors of \citet{Doanetal1984} and \citet{Sims:1993:nber:chapter}, which are both discussed also in \citet{Sims:Zha:IER:1998}. These priors were motivated by the need to avoid having an unreasonably large share of the sample period variation in the data accounted for by deterministic components. All the priors described above ultimately depend on a small set of hyperparemeters, which we estimate by maximizing the marginal likelihood of the model, which is available in closed form. More details can be found in Appendix \ref{Prior moments}.\footnote{This model can be used for very large cross sections because the Kronecker product in (\ref{NIWprio}) implies that the error variance $\Sigma$ enters
the prior variance of the coefficients in a way that is symmetric and
proportional to each equation. This fact, together with the fact that in a VAR every equation has the same regressors, implies that the posterior
distribution of $p(\Phi, \Sigma)$ is a matricvariate-t. See \citet{Carriero_jae.2315} for a discussion.}$^,$\footnote{It is worth pointing out that one important limitation of the natural conjugate prior described in this section is that the Kronecker structure of
the prior variance of the coefficients implies that the prior must be symmetric across equations. This in
turn rules out cross-variable shrinkage, i.e. the possibility to shrink coefficients on lags of other variables
more aggressively than those on own lags. \citet{chan2022asymmetric} developed an alternative
specification that allows for asymmetry in the prior, while maintaining conjugacy. We experimented with this alternative and found the results to be broadly similar; therefore, we did not include them in the paper. These results are available from the authors on request.}

\FloatBarrier
\subsection{Factor Model}

Factor models are another class of models that has repeatedly shown to be
particularly well suited for macroeocnomic forecasting. In this paper we use
the implementation of \citet{McCracken:Ng:2016}, which we now proceed to
summarize. First, a set of static factors
is estimated from the entire cross section of available data using Principal
Components Analysis and the EM of \citet{Stock:Watson:JBES:2002} to balance the panel by filling in any missing values. Second,
we use the extracted first factor to augment an autoregression of the $i$-th series with its lag values, i.e.: 
\begin{equation}\label{factor_reg}
y_{i,t}=\alpha _{h}+\beta _{h}(L)\hat{f}_{1t-h}+\gamma
_{h}(L)y_{i,t-h}+\varepsilon _{i,t}
\end{equation}%
We estimate this model via OLS, and rely on BIC to select the number of lags of $%
\hat{f}_{1t}$ and $y_{it}$. The $h$-step ahead forecast is then computed as\footnote{Note that this approach (known as the "direct approach") is directly
	minimizing the $h$-step ahead forecast error, and it requires to perform a
	different regression for each forecast horizon.} 
\begin{equation}
\hat{y}_{i,t+h}=\hat{\alpha}_{h}+\hat{\beta}_{h}(L)\hat{f}_{1t}+\hat{\gamma}%
_{h}(L)y_{i,t}
\end{equation}%

\FloatBarrier
\subsection{Neural Network Autoregression (NNAR)}

VARs and factor models assume linearity between endogenous variables and their lags or lags of their linear combinations. Although these models have proven quite successful at forecasting macroeconomic time series, the linearity assumption could be overly restrictive and this could have a negative impact on forecast accuracy.  To test this hypothesis, we also include in the comparison the Neural Network Autoregression (NNAR) model of \citet{hyndman2018forecasting} and popularized through the \texttt{forecast} package published by the same authors. The NNAR model combines the structure of classical autoregressive models with the flexibility of feed-forward neural networks. In particular, 
the model is specified as:
\[
y_{i,t} = f(y_{i,t-1}, y_{i,t-2}, \dots, y_{i,t-p}) + \varepsilon_{i,t},
\]
where \( f(\cdot) \) is a nonlinear function represented by a single-hidden-layer feed-forward neural network with \( k \) hidden nodes, and \( \varepsilon_{i,t} \) is a white noise error term. The model is denoted as \texttt{NNAR(p,k)}, analogous to an \texttt{AR(p)} model in traditional time series analysis. The NNAR model is based on a single hidden layer neural network trained using backpropagation, where we set the number of lagged inputs to \( p=12 \). Furthermore, to avoid overfitting, the model uses weight decay (L2 penalty) during training. Forecasts from the NNAR model are generated iteratively, and averaging over multiple networks trained with different initial weights, thereby capturing model uncertainty. Despite its simplicity compared to modern deep learning models, the NNAR model often performs competitively on short-term forecasting tasks, especially for time series with nonlinear dynamics and limited data.\footnote{We have also experimented with \texttt{DeepAR} \citep{salinas2020deepar}, a sigificantly more sophisticated approach leveraging recurrent neural networks that trains the network using all available time series at once, but found it overall to not be more accurate than this simpler alternative.}

\FloatBarrier
\subsection{Bayesian Additive Regression Trees}

 Lastly, we include in the pool of candidate models Bayesian Additive Regression Trees (BART). BART \citep{Chipman_etal_2010} models the conditional mean of the regression model by summing over a large number of trees which are, by themselves, constrained through a regularization prior. The main intuition behind the approach is that while each of these simple trees may only be able to explain a small fraction of the variation in the response variable, summing over a large number of them can describe the data extremely well. \citet{Pruser_2019}, \citet{Huber_Rossini_2022}, \citet{HUBER202352} and \citet{Clark_etal_2023} are recent examples of successful applications of BART with macroeconomic data.  For each series $i$, we implement BART by estimating the following model\footnote{We separate $y_{i,t-h}$ from the other predictors to avoid the model not recognizing that the lagged depedent variable is likely to have a non trivial role in predicting $y_{i,t}$. Our modeling choice is inspired by the approach of \citet{Lopez_et_al_WP_2025}, who modified the standard BART to introduce a Minnesota-type shrinkage specification into the node splitting selection. We also consider the alternative model $y_{i,t}=f\left(X_{t-h}\right)+\varepsilon _{i,t}$ but we found it to perform significantly below our preferred specification.}
\begin{equation}\label{bart_reg}
y_{i,t}=\alpha _{h}+\gamma
_{h}(L)y_{i,t-h}+f\left(X_{-i,t-h}\right)+\varepsilon _{i,t}
\end{equation}%
where $X_{-i,t-h}$ denotes all series at time $t-h$ other than $y_{i,t-h}$ and $g\left(X_{-i,t-h}\right)$ is a non-linear function obtained by summing over a larger number of regression trees,
\begin{align}
    f\left(X_{-i,t-h}\right) = \sum_{j=1}^M g\left(X_{-i,t-h}\vert \mathcal{T}_j,m_j\right)
\end{align}
where $g\left(X_{-i,t-h}\vert \mathcal{T}_j,m_j\right)$ identifies a single tree model with $\mathcal{T}_j$ denoting the tree structure and $m_j$ is the vector of terminal node parameters associated with $\mathcal{T}_j$. Following \citet{Chipman_etal_2010}, we set $M = 250$ in our application and specify a prior on the tree structure along the lines they suggested. 
The $h$-step ahead forecast is then computed as 
\begin{equation}
\widehat{y}_{i,t+h}=\widehat{\alpha}_{h}+\widehat{\gamma}%
_{h}(L)y_{i,t} + \widehat{f}\left(X_{-i,t}\right)
\end{equation}%
As with factor models, this approach requires a different regression for each forecast horizon.

\FloatBarrier
\section{Empirical Application}\label{OOS}

\FloatBarrier
\subsection{Data} \label{sec:data}

We collect 120 monthly variables for the US spanning the period January 1959 to December 2022. The data, which are obtained from the Federal Reserve Economic Data (FRED) and are available at \href{https://fred.stlouisfed.org}{https://fred.stlouisfed.org}, cover a wide range of key macroeconomic variables that applied economists monitor regularly, such as different measures of output, prices, interest and exchange rates, and stock market performance. We provide a full list of the data and their transformations in \ref{App_data}.

\FloatBarrier
\subsection{Forecasting Exercise}
We use the first twenty-five year of data, January 1960--December 1984, to obtain
initial parameter estimates for all the econometric models, which are then used to predict
outcomes from January 1985 ($h=1$) to December 1985 ($h=12$). The next period, we
include data for January 1985 in the estimation sample, and use the resulting
estimates to predict the outcomes from February 1985 to January 1986. We proceed
recursively in this fashion until December 2022, thus generating a time
series of forecasts for each forecast horizon $h$, with $h=1,...,12$. 

For the BVAR, point forecasts are obtained by taking the mean of the predictive densities. When $h>1$, the predictive densities are produced by iterating the model forward via simulation.\footnote{We set the number of lags in the BVARs to $p=6$.} For the factor models we first extract the first common factor, $\hat{f}_{1t}$. Next, for each variable and each forecast horizon, we run the factor-augment autoregression in \eqref{factor_reg} and forecast up to $12$ months ahead. Similarly, for BART we re-estimate the trees for each variable set and sample, as the number of series in $X_{-i,t-h}$ changes with each variable set.

We proceed in a similar fashion with all \tsfm{}s. More specifically, for each \tsfm{} considered, starting with the first forecast date (December 1984), we feed the historical data to the pretrained model (when working with the univariate \tsfm{}s we supplied one time series at a time, while with the multivariate \tsfm{}s we supplied all time series within the specific model size at once). Next, we query the model to generate a zero-shot forecast for each of the variables provided, up to 12 steps ahead.\footnote{For all our experiments, we downloaded the  open-sourced pretrained models and run them locally on our computer, using an Nvidia A6000 GPU with 48GB of memory.  The only exception was \timegpt, where we instead queried directly their API, with a separate query for each forecast date/time series pair.} We also include a benchmark approach which uses OLS forecasts from univariate AR(1)\ models.

We evaluate the predictive accuracy of the various models, for each of the variables considered and each forecast horizon. In particular, we measure the precision of the $h$-step-ahead point forecasts for model $i$ and variable $j$, relative to that from the univariate AR(1), by means of the ratio of Root MSFEs: 
\begin{equation}
RMSFE_{ijh}=\sqrt{\frac{{\sum_{\tau =\underline{t}}^{\overline{t}-h}e_{i,j,\tau
+h}^{2}}}{{\sum_{\tau =\underline{t}}^{\overline{t}-h}e_{bcmk,j,\tau +h}^{2}}}%
},  \label{R2_OOS}
\end{equation}%
where $\underline{t}$ and $\overline{t}$ denote the start and end of the
out-of-sample period, and where $e_{i,j,\tau +h}^{2}$ and $e_{bcmk,j,\tau
+h}^{2}$ are the squared forecast errors of variable $j$ at time $\tau $ and
forecast horizon $h$ associated with model $i$ ($i\in \left\{
\text{BVAR},\text{Factor model},\text{NNAR},\text{BART}\right\} \cup \text{TSFMs}$) and the AR(1) model, respectively.
The point forecasts used to compute the forecast errors are obtained by
averaging over the draws from the various models' $h$-step-ahead predictive
densities. Values of $RMSFE_{ijh}$ below one suggest that model $i$ produces
more accurate point forecasts than the AR(1) benchmark for variable $j$ and
forecast horizon $h$.

In closing this Section it is worth mentioning that in this paper we are only focusing on point forecast accuracy. This choice is intentional. While there is a rich literature on density forecasts for macroeconomic variables, using \tsfm{}s to construct density forecasts presents additional challenges, and we leave that to further research.

\FloatBarrier
\section{Results}\label{Results}

\FloatBarrier
\subsection{Do Some \tsfm{}s Perform Better than Others?}

\begin{figure}[t!]
    \centering
    \includegraphics[width=0.78\textwidth]{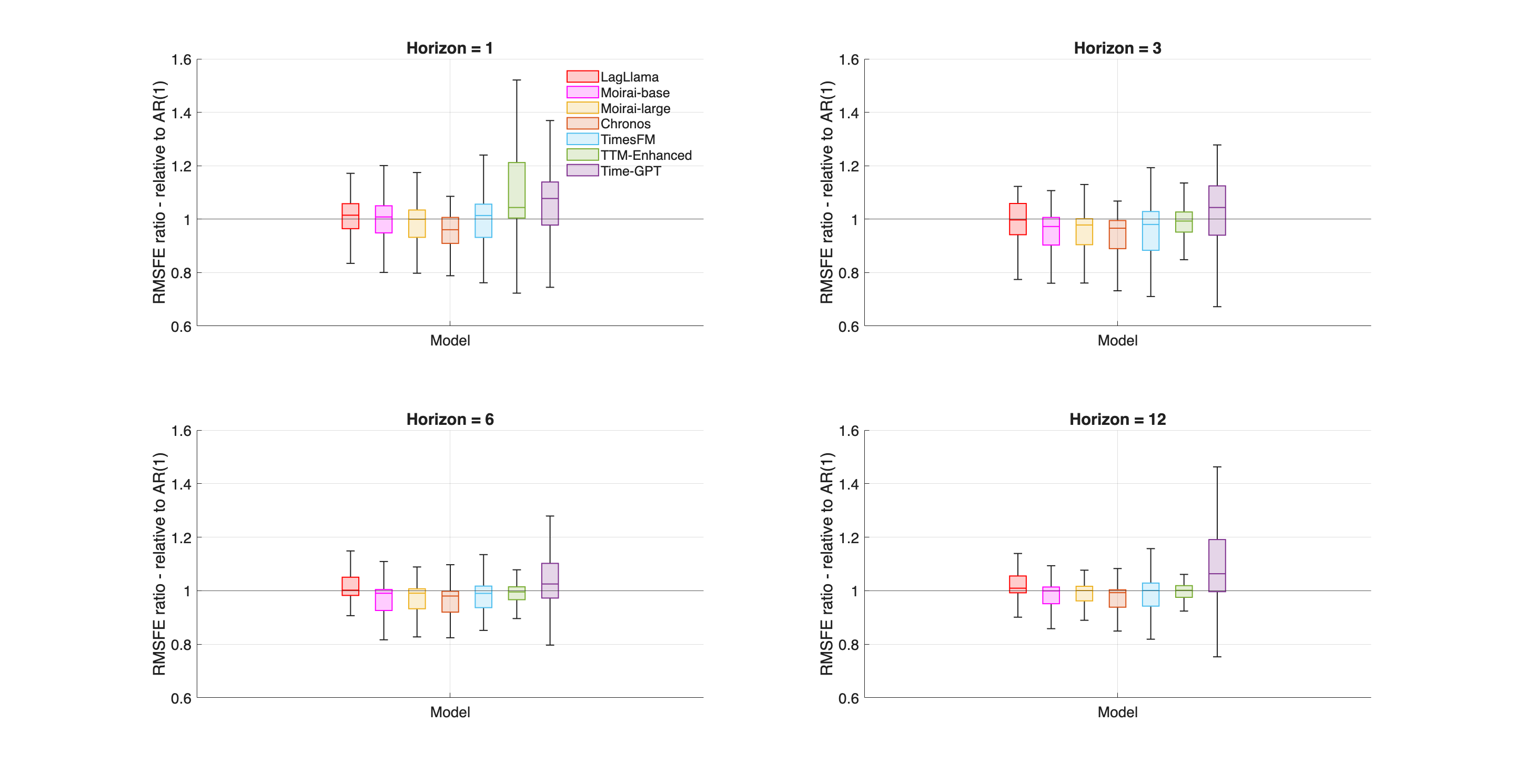}
    \caption{Distribution of RMSFEs (relative to an AR benchmark) for the \tsfm{}s. The evaluation sample is January 1985 to December 2019.}
    \label{fig:boxplot_llm_only}
\end{figure}

To begin, we focus on the overall forecasting performance of the various \tsfm{}s under examination. Given that the evaluation sample we are considering include the Covid-19 period, to avoid conflating the overall picture with the large idiosyncrasies brought about by the pandemic, we focus first on the accuracy of the various models by stopping before the onset of the pandemic. That is, our evaluation sample stops in December 2019. Later on, we analyze how stable these results are over time, and also look at the accuracy of the various models over the most recent period, January 2020 to December 2022. \autoref{fig:boxplot_llm_only} provides a synoptic view of the forecasting performance across all 120 series we focused on, as measured by RMSFE ratios relative to the AR benchmark. Each box-plot shows the interquartile range of the distribution of relative RMSFEs for each of the \tsfm{}s considered. The solid line within each box represents the median, while the whiskers represent the maximum and minimum relative RMSFEs that are not defined as outliers.\footnote{Outliers are defined as values that are more than 1.5 times the interquantile range away from the top or bottom of the box (i.e. from the lower and upper quartiles). } When the mass is concentrated above the line corresponding to $RMSFE=1$, this is an indication that the models do generally worse than the benchmark.

\begin{table}[h!]
  \centering
    {\footnotesize\input{tables/Table2_TSFM_stats.tex}}
    \caption{Median, Std. deviation, Min and Max RMSFE by \tsfm{} model type and forecast horizon. The evaluation sample is January 1985 to December 2019.}
  \label{tab:RMSFE_stats_LLMs}%
\end{table}%

Outliers are not depicted for scaling reasons, and instead we separately report the maximum and minimum RMSFEs attained by each model in \autoref{tab:RMSFE_stats_LLMs} (together with their median and standard deviation).  Indeed, it does happen on several occasions that for some series the \tsfm{}s produce forecasts that are patently unreasonable. These forecasts would be automatically disregarded by the user and would call for a fine-tuning of the models. This result highlights that while zero-shot \tsfm{}s may produce on average reasonable forecasts, they should not be used mechanically and do require careful monitoring.

Having said that, the pattern emerging from a careful inspection of \autoref{fig:boxplot_llm_only} is one of heterogeneity among the various \tsfm{}s. Specifically, some \tsfm{}s show a performance that is consistently worse than others. These are \ttm-enhanced and \timegpt, for which the box plots are invariably positioned higher than those of the remaining \tsfm{}s, signaling higher RMSFEs. These models also perform poorly compared to the benchmark, for example \timegpt has large part of the mass of the box-plot above 1 for all horizons, and \ttm for the one-step-ahead horizon. \lagllama performs a bit better than these two, however it is also systematically outperformed by the remaining models.

The \tsfm{}s showing the best forecasting performance are \moirai, \chronos, and \timesfm. As we discussed, \moirai comes in three sizes, and the evidence shows that the largest size performs the best, with a mass of RMSFEs being consistently below 1 at all horizons. Because of these considerations, in the remainder of this Section, we will only focus on these best performing \tsfm{}s (\moirai-\largemodel, \chronos and \timesfm), and compare their performance against the econometric models.

\FloatBarrier
\subsection{Differences in Point Forecast between \tsfm{}s and Econometric Models}

\begin{figure}[t!]
    \centering
    \includegraphics[width=0.78\textwidth]{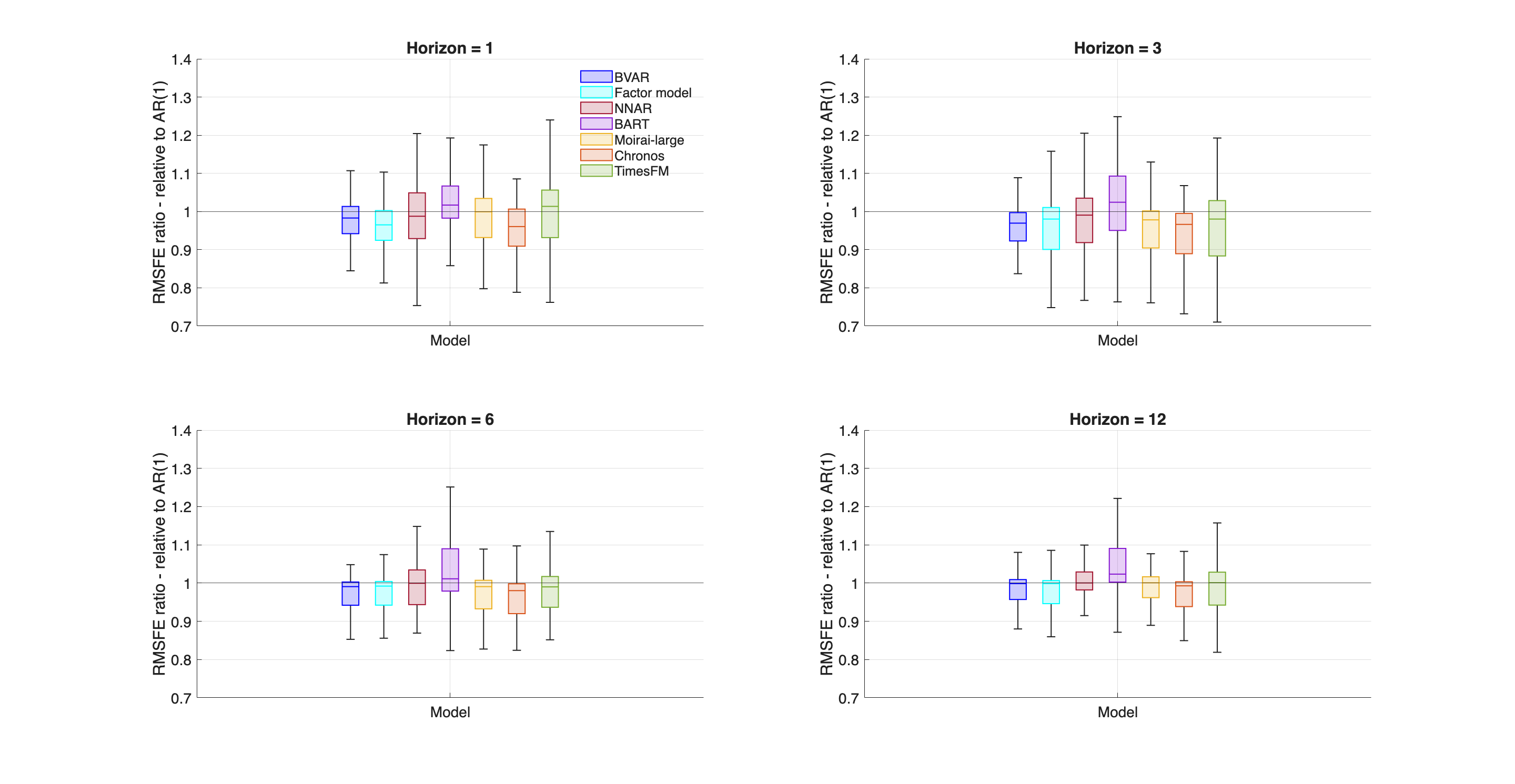}
    \caption{Distribution of RMSFEs (relative to an AR benchmark) for the econometric models and a selection of \tsfm{}s. The evaluation sample is January 1985 to December 2019.}
    \label{fig:boxplot_econometric_and_best_llm}
\end{figure}

\autoref{fig:boxplot_econometric_and_best_llm} provides a summary of the forecasting performance of the three best \tsfm{}s contrasted with the performance of the econometric models under consideration. As before, the box-plots represent the interquartile range of the distribution of RMSFEs (relative to the AR(1) benchmark) and the whiskers represent the maximum and minimum data points that are not flagged as outliers.

Focusing first on the econometric models, both the BVAR and the factor model are consistently showing a good performance relative to the AR model. This is a well known result, which has been obtained by several different studies using this dataset. 

As already shown in \autoref{fig:boxplot_llm_only}, \moirai-\largemodel, \chronos and \timesfm also perform consistently better than the AR benchmark, and \autoref{fig:boxplot_econometric_and_best_llm} now shows that the RMSFE distribution is similar to those obtained with the econometric models. Interestingly, while all models at almost all horizons have a median that is below the threshold of one, there is a tendency for the distribution of the RMSFEs from \tsfm{}s to have a more pronounced right tail, signaling an increased likelihood of observing large forecast errors. Instead, econometric methods produce RMSFE distributions that are more left-skewed, signaling a consistent better performance compared to the AR(1).  

Turning to the two non-linear econometric models, the evidence is mixed. The performance of BART is consistently worse than that of the benchmark. The simple NNAR model produces better forecast than BART, and is overall slightly better than the AR(1) benchmark. At 1-step ahead, the performance of NNAR is similar to that of the \tsfm{}s. However, at longer horizons this model is outperformed by both \tsfm{}s and the linear econometric models. This suggests that while non-linearities do play an important role, \tsfm{}s seem to have an edge at capturing them, over the simpler non-linear models we considered.

As before, additional descriptive statistics including the maximum and minimum relative RMSFEs (including outliers) are reported in \autoref{tab:RMSFE_stats_ecnmtr}. The linear econometric models (BVARs and factor models) do not present any extreme cases in which forecast errors are a multiple of those obtained by the AR benchmark, which it is instead the case for the \tsfm{}s results shown in \autoref{tab:RMSFE_stats_LLMs}, and - to a lesser extent - for the nonlinear models in the last two rows of \autoref{tab:RMSFE_stats_ecnmtr}. Moreover, it is evident that the RMFSEs coming out of the econometric models are less dispersed. To sum up, the picture is one in which BVARs and factor models seem to offer a more stable and reliable performance, when compared to zero-shot \tsfm{}s and other simpler nonlinear models. 

Note that so far we have not considered the statistical significance of the various forecasts, nor how these results hold up in the immediate aftermath of the Covid-19 pandemic. Appendix~\ref{DMtest} presents results on the statistical significance of the various forecasts (relative to the AR(1) benchmark), focusing for ease of exposition on a subset of key macroenomic variables, and shows that for several series the improvements of both the econometric models and the \tsfm{}s over the AR benchmark are statistically significant. Separately, Appendix~\ref{App_covid} reports the corresponding results for the post-Covid-19 period (January 2020 to December 2022), where \tsfm{}s exhibit relatively better performance, albeit with the important caveat that their training sample includes the pandemic period itself.

\begin{table}[t!]
  \centering
    {\footnotesize\input{tables/Table3_econometric_stats.tex}}
    \caption{Median, Std. deviation, Min and Max RMSFE by model type and forecast horizon. The evaluation sample is January 1985 to December 2019.}
  \label{tab:RMSFE_stats_ecnmtr}%
\end{table}%

\section{A TSFM-Augmented BVAR (TSFM-BVAR)}\label{TSLM-BVAR}

The results so far show that although the econometric model consistently outperforms the AR benchmark, in some cases \tsfm{}s can be very accurate. At the same time, in a number of cases the performance of \tsfm{}s can be significantly sub-par.  One way to rationalize these results is that while \tsfm{}s are capable of exploiting vast datasets, they still lack the distilled domain knowledge built from decades of macroeconomic analysis. Still,  in some cases, their ability to capture nonlinearities in the data can be quite advantageous. 

In this section, we consider an approach that can help leverage the best of both worlds: that is, exploiting the domain knowledge and simplicity of the econometric models while relying on \tsfm{}s to capture nonlinear behavior missed by the traditional approaches. Our proposal is to build a simple hybrid approach. First, an econometric model (we use the BVAR) is fit to the data and the in-sample residuals are computed. Second, the residuals from the first step are modeled using a \tsfm{}. Finally, forecasts are computed from both models and combined. Hence, the final forecast adjusts the econometric prediction by adding the predicted residual component, effectively correcting systematic biases and capturing nonlinearities.

Formally, let $y_{i,t}$ denote the target variable of interest. The procedure comprises the following steps:
\begin{enumerate}
    \item \textbf{Econometric Model Forecast.} Estimate a BVAR using historical data on all series and generate forecasts:
    \begin{equation}
        \hat{y}_{i,t+h}^{(E)} = \mathbb{E}\bigl[ y_{i,t+h} \,\big|\, \mathcal{I}_{t} \bigr],
    \end{equation}
    where $\mathcal{I}_{t}$ denotes the information set up to time $t$. Compute the in-sample residuals:
    \begin{equation}
        \hat{e}_{i,\tau} = y_{i,\tau} - \hat{y}_{i,\tau}^{(E)}, \quad \tau=1,...,t.
    \end{equation}
    
    \item \textbf{Foundational Model Training and forecasting.} Train a foundational time series model on the residuals $\{\widehat{e}_{i,\tau}\}_{\tau=1}^t$ to learn remaining patterns and then generate a zero-shot forecast:
    \begin{equation}
        \hat{e}_{i,t+h}^{(F)} = \mathbb{E}\bigl[ \hat{e}_{i,t+h} \,\big|\, \mathcal{H}_t \bigr],
    \end{equation}
    where $\mathcal{H}_t$ denotes the residual history up to time $t$.
    
    \item \textbf{Hybrid Forecast Construction.} The final hybrid forecast combines the econometric prediction and the TSFM forecast:
    \begin{equation}
        \hat{y}_{i,t+h}^{(\text{H})} = \hat{y}_{i,t+h}^{(E)} + \hat{e}_{i,t+h}^{(F)}.
    \end{equation}
\end{enumerate}

This approach retains the interpretability of the BVAR while flexibly capturing nonlinear residual dynamics through the foundational architecture. 

\begin{figure}[t!]
    \centering
    \includegraphics[width=0.78\textwidth]{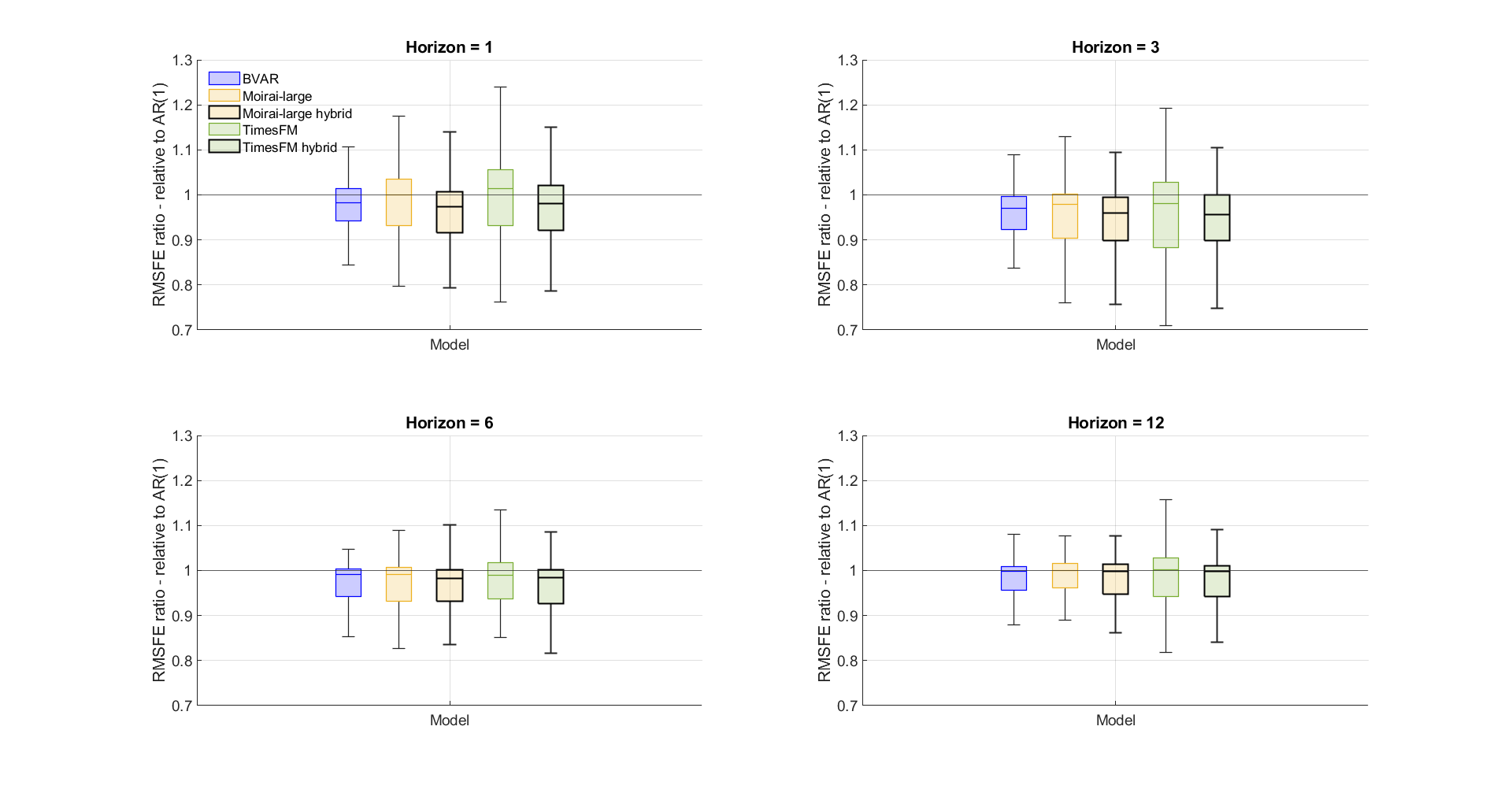}
    \caption{Distribution of RMSFEs (relative to an AR benchmark) for the original BVAR and the new procedure combining the BVAR with an error correction based on two of the best-performing \tsfm{}s, \timesfm and \moirai-large. The evaluation sample is January 1985 to December 2019.}
    \label{hybrid}
\end{figure}

\autoref{hybrid} and \autoref{tab:RMSFE_ratios_hybrid} present the results of this analysis. In particular, \autoref{hybrid} shows results for the original BVAR plus two of the best-performing \tsfm{}s, \moirai and \timesfm, and finally two hybrid versions combining the BVAR with both \tsfm{}s. The hybrid approach shifts the TSFM error distribution downward, placing most of its mass below one—the threshold for outperforming the AR benchmark.

\begin{table}[t!]
  \centering
  \footnotesize
    \begin{tabular}{lcccc|cccc}
    \toprule
     & \multicolumn{4}{c|}{\textbf{h=1}} & \multicolumn{4}{c}{\textbf{h=3}} \\
\cmidrule{2-9}          & \textbf{Median} & \textbf{Std} & \textbf{Min} & \textbf{Max} & \textbf{Median} & \textbf{Std} & \textbf{Min} & \textbf{Max} \\
    \textbf{BVAR} & 0.983 & 0.060 & 0.827 & 1.158 & 0.970 & 0.060 & 0.767 & 1.089 \\
    \textbf{\moirai-\largemodel} & 0.999 & 0.102 & 0.703 & 1.436 & 0.978 & 0.099 & 0.637 & 1.158 \\
    \textbf{\moirai-\largemodel hybrid} & 0.973 & 0.075 & 0.734 & 1.139 & 0.960 & 0.079 & 0.674 & 1.095 \\
    \textbf{\timesfm} & 1.014 & 0.129 & 0.706 & 1.482 & 0.980 & 0.127 & 0.635 & 1.318 \\
    \textbf{\timesfm hybrid} & 0.980 & 0.081 & 0.720 & 1.150 & 0.956 & 0.089 & 0.670 & 1.105 \\
    \midrule
    & \multicolumn{4}{c|}{\textbf{h=6}} & \multicolumn{4}{c}{\textbf{h=12}} \\
\cmidrule{2-9}          & \textbf{Median} & \textbf{Std} & \textbf{Min} & \textbf{Max} & \textbf{Median} & \textbf{Std} & \textbf{Min} & \textbf{Max} \\
    \textbf{BVAR} & 0.991 & 0.068 & 0.725 & 1.106 & 0.999 & 0.080 & 0.671 & 1.162 \\
    \textbf{\moirai-\largemodel} & 0.991 & 0.096 & 0.600 & 1.159 & 1.001 & 0.113 & 0.619 & 1.324 \\
    \textbf{\moirai-\largemodel hybrid} & 0.983 & 0.079 & 0.648 & 1.102 & 0.998 & 0.087 & 0.671 & 1.164 \\
    \textbf{\timesfm} & 0.990 & 0.142 & 0.593 & 1.629 & 1.001 & 0.158 & 0.482 & 1.440 \\
    \textbf{\timesfm hybrid} & 0.984 & 0.080 & 0.636 & 1.119 & 0.999 & 0.088 & 0.677 & 1.174 \\
    \bottomrule
    \end{tabular}%
    \caption{Summary statistics of RMSFE ratios by model and forecast horizon, including median, standard deviation, minimum, and maximum values.}
  \label{tab:RMSFE_ratios_hybrid}
\end{table}

 Moreover, \autoref{tab:RMSFE_ratios_hybrid} shows that the maximum relative RMSFEs from the \tsfm{}s are greatly reduced by the hybrid approach: at one step ahead, the error of \moirai decreases from 1.436 to 1.139, and that of \timesfm from 1.482 to 1.150. These figures are very close to the maximum error of the BVAR, which is 1.158. Thus, the use of econometric models in the first step substantially reduces the risk of hallucinations and other aberrant forecast behaviors. Incorporating the domain knowledge of econometric models helps \tsfm{}s across the board by lowering forecast errors for all series.  

Lastly, it is worth pointing out that combining the BVAR with the \tsfm{}s has also a positive effect on the left tail: for those forecasts that were already improving over the AR benchmark, the magnitude of the improvement increases. For example, at one-step ahead the best relative RMSFE drops from 0.827 to 0.734 or 0.720 when the BVAR is used in conjunction with \moirai and \timesfm, respectively. 

Overall, the hybrid approach improves upon both the econometric model and the \tsfm{}s used in isolation. Using the econometric model in the first step prevents \tsfm{}s from producing subpar forecasts, while using the \tsfm{}s on the residuals enhances the performance of the econometric model. This suggests that the domain knowledge of econometric models is robust and reliable, but can be further improved using \tsfm{}s to capture nonlinearities.\footnote{While these results are encouraging, one must not forget the caveat that the performance improvement form \tsfm{}s we observed here might be due to the fact it has a spuriously superior information set, as discussed in Section \ref{INFOset}.}

\FloatBarrier
\subsection{Interpreting the Performance Improvements}\label{correlations}

\begin{figure}[t!]
	\centering
	\subfloat[$h=1$]{\includegraphics[width=0.48\textwidth]{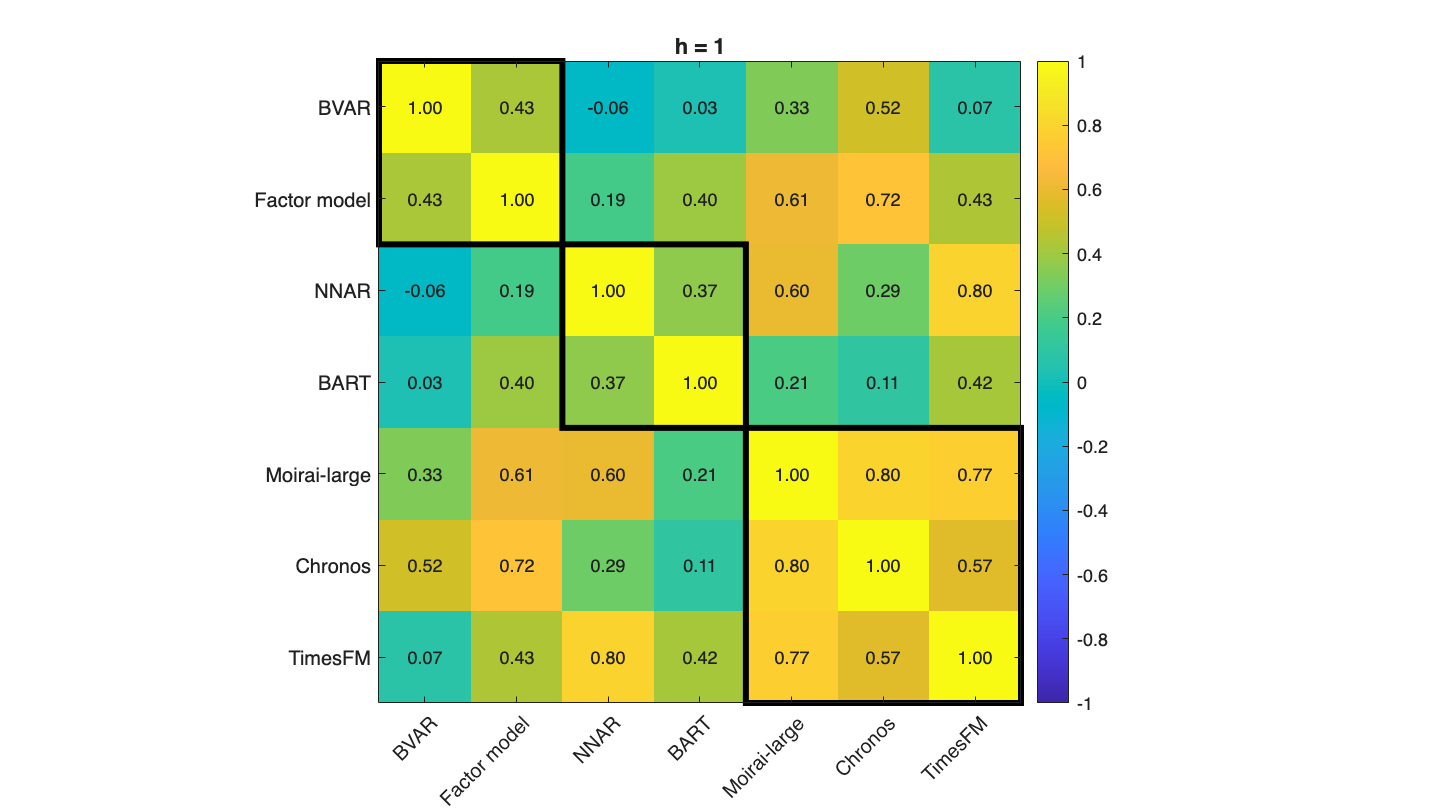}}
	\hfill
	\subfloat[$h=3$]{\includegraphics[width=0.48\textwidth]{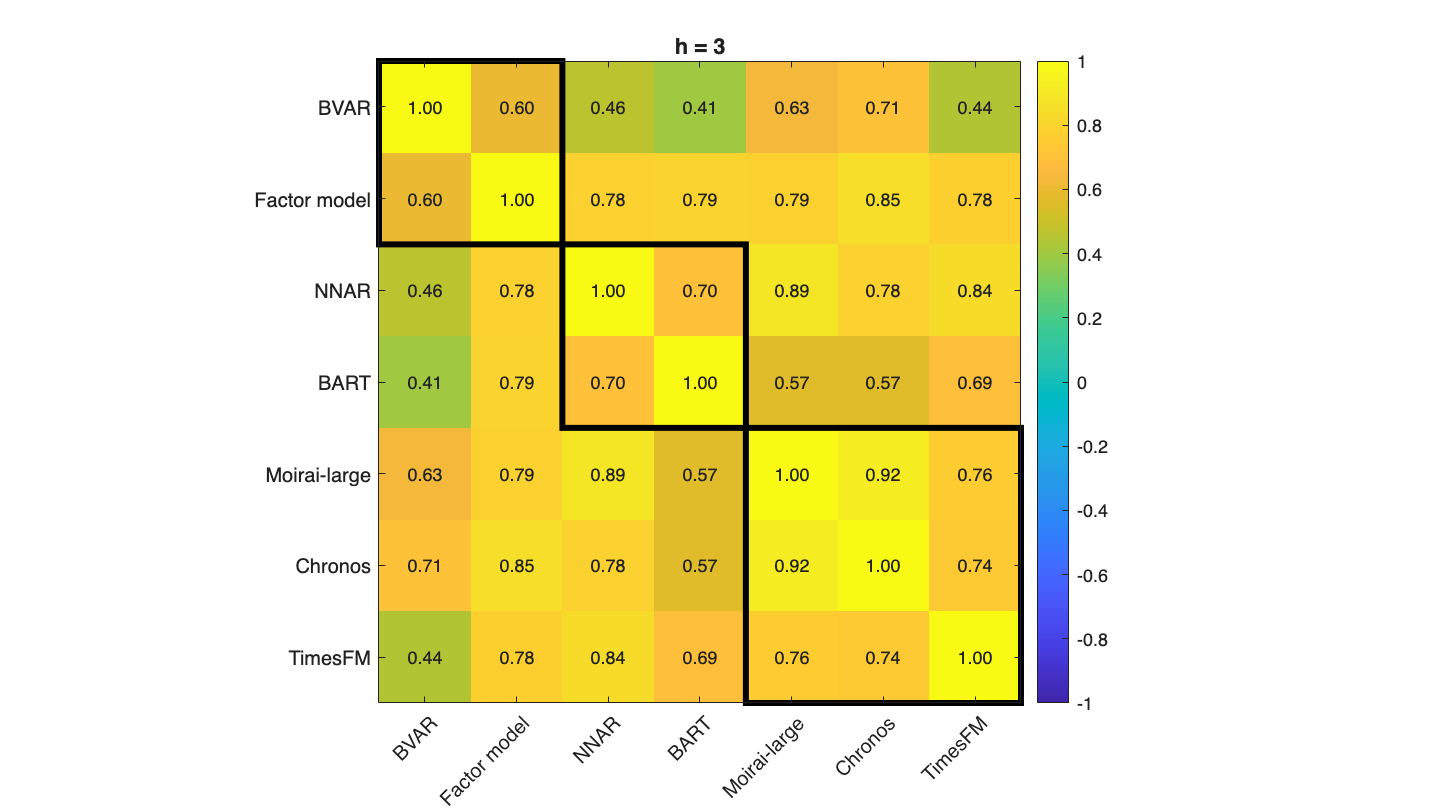}}\\
	\subfloat[$h=6$]{\includegraphics[width=0.48\textwidth]{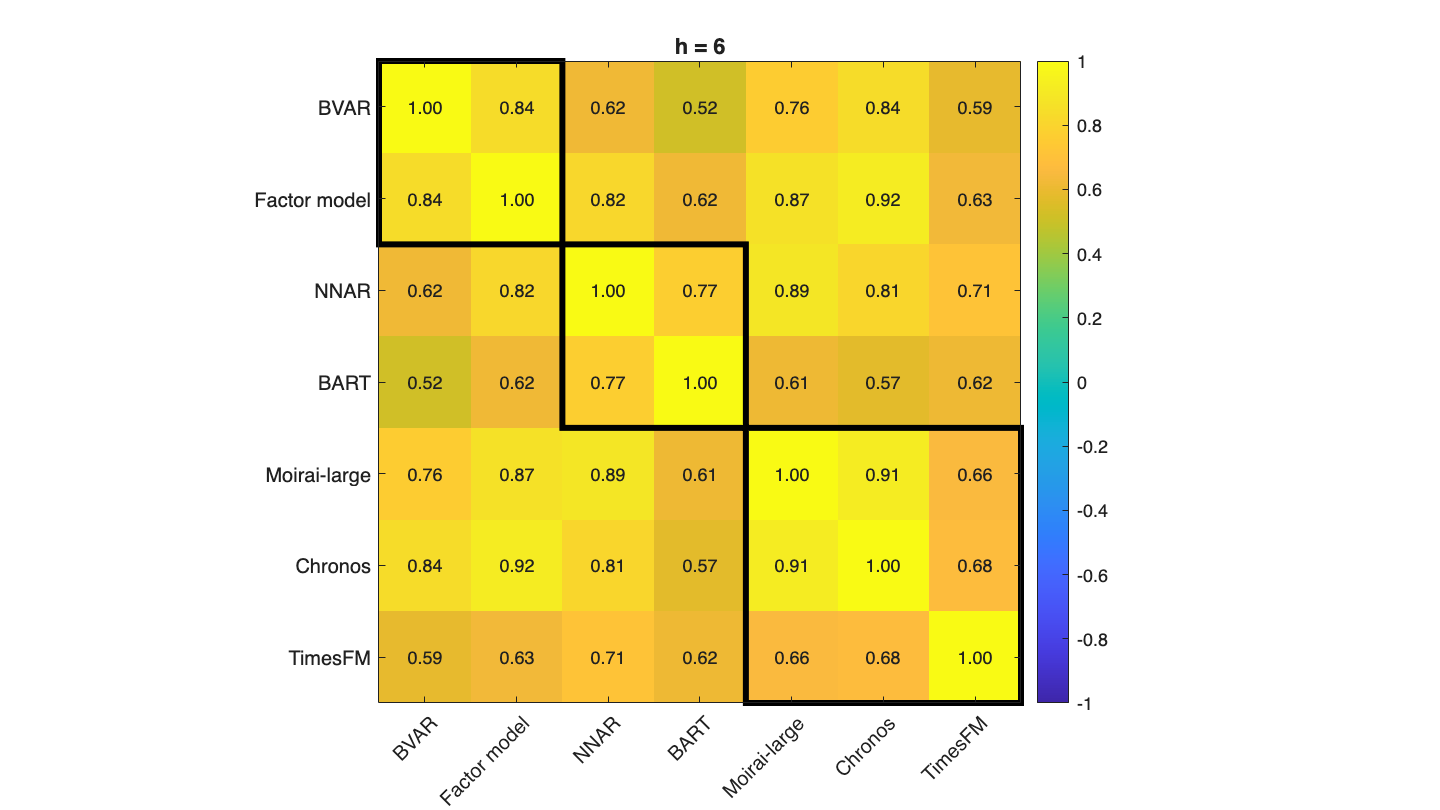}}
	\hfill
	\subfloat[$h=12$]{\includegraphics[width=0.48\textwidth]{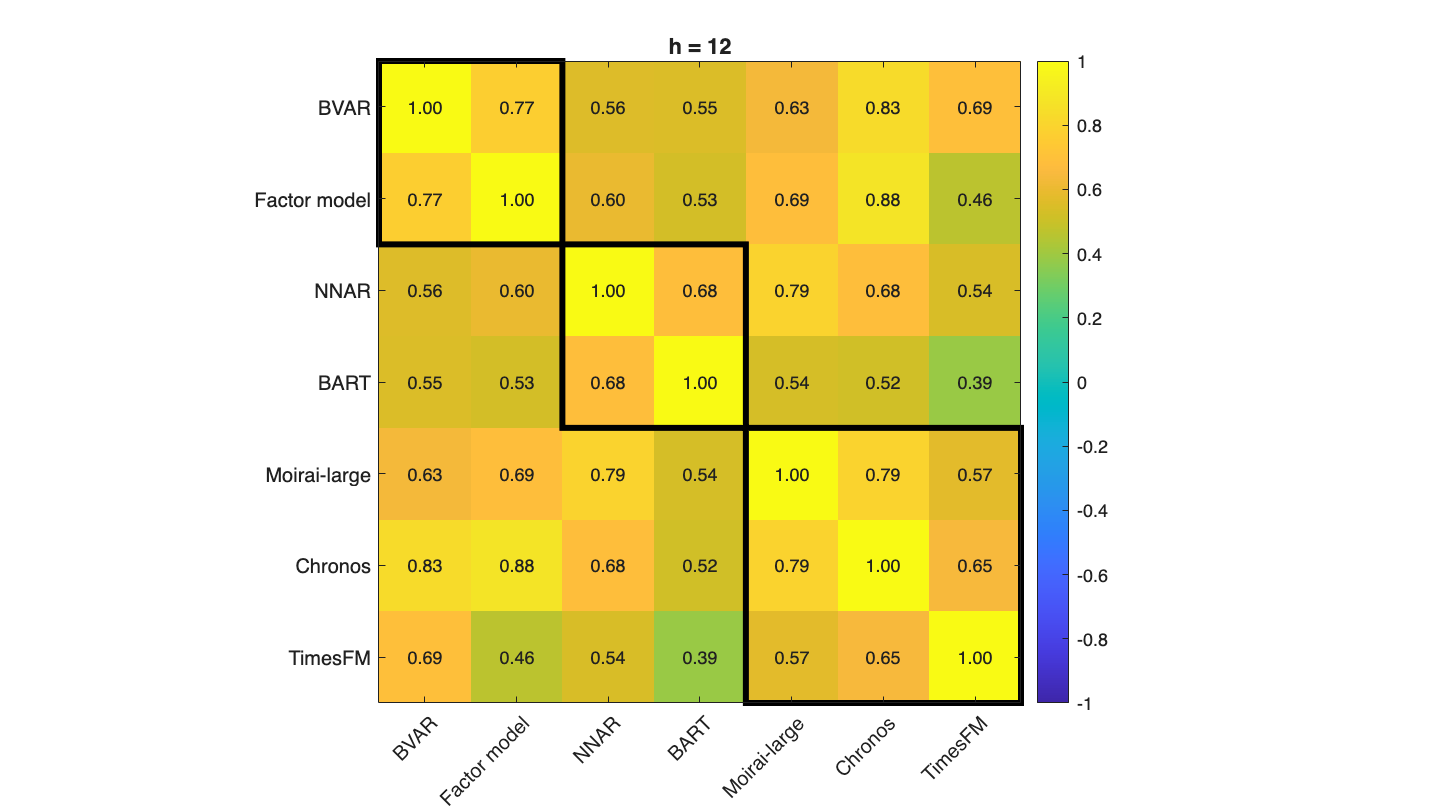}}
	\caption{Correlation of per-series RMSFE ratios across models, by forecast horizon. The evaluation sample is January 1985 to December 2019.}
	\label{fig:corr_heatmap}
\end{figure}

The hybrid approach (BVAR forecast plus a TSFM-modeled residual correction) only helps if the TSFM is picking up something the BVAR is missing. If a candidate TSFM tends to do well/poorly on the \emph{same} series as the BVAR, its residual correction is largely redundant; if it tends to do well/poorly on a \emph{different} set of series, the two are complementary and combining them should help. The results from the previous section indicate that this may be the case. To check this conjecture directly, we look at the correlation, across the 120 series, between each pair of models' per-series RMSFE ratios (Pearson correlation, pairwise-complete). At $h=1$ (\autoref{fig:corr_heatmap}, panel a), BVAR's errors are nearly orthogonal to every TSFM's: correlation with TimesFM is $0.07$, and even with the two best TSFMs it is modest (Moirai-large $0.33$, Chronos $0.52$). This is exactly the pattern that would make a BVAR+TSFM hybrid effective at short horizons -- the TSFM residual correction is targeting series the BVAR is not already getting right. By contrast, the TSFMs are highly correlated \emph{with each other}: Moirai-large and Chronos correlate at $0.80$, meaning the nonlinear/TSFM family tends to struggle on the same series as one another (largely the more persistent ones). 
This suggests limited additional diversification benefit from combining multiple TSFMs together, as opposed to combining any one of them with BVAR. Moving across the four panels of \autoref{fig:corr_heatmap}, this complementarity does not hold for long: BVAR's correlation with Chronos rises from $0.52$ ($h=1$) to $0.71$ ($h=3$) to $0.84$ ($h=6$), and BVAR-TimesFM rises from $0.07$ to $0.44$ to $0.59$ over the same horizons -- essentially all of the erosion happens by $h=6$, well before $h=12$, where correlations plateau or even retreat slightly for some pairs (BVAR-Moirai-large falls back from $0.76$ at $h=6$ to $0.63$ at $h=12$). By $h=6$, essentially every model is struggling on the same subset of series.  This gives a direct, data-driven explanation for a pattern already visible in the hybrid results reported in Section~\ref{TSLM-BVAR}: the improvement from hybridizing BVAR with a TSFM is largest at $h=1$, where the two error series are close to orthogonal, and shrinks at longer horizons, where they have become highly correlated and there is correspondingly less complementary information left for the TSFM residual stage to add.

\FloatBarrier
\section{Dissecting TSFM Performance: Robustness, Heterogeneity, and Model Evolution}\label{closerlook}

To better understand the results presented so far, this Section examines four aspects of the \tsfm{}s and how they shape the models' forecast performance. In Section \ref{INFOset} we consider the effects of possible "data leakage", which as we mentioned before can spuriously tilt the forecast accuracy scale in favor of the \tsfm{}s. In Section \ref{nonstationary} we show that highly persistent series are particularly problematic for \tsfm{}s to forecast. Section \ref{heterogeneity} shows that persistence is not the whole story: relative performance also varies systematically across broad categories of macroeconomic series -- real activity, prices, financial variables, and so on -- so that pooling all series together, as in Section \ref{Results}, can mask large offsetting gains and losses within specific domains. Finally, in Section \ref{newgen} we compare the original Moirai and Chronos models against their newly released second-generation counterparts, to gauge how much of the TSFM performance documented in this paper is attributable to genuine architectural progress rather than being an inherent feature of the modeling approach. Appendix~\ref{App_finetuning} additionally reports whether fine-tuning two of the best-performing pretrained \tsfm{}s on \fred data improves upon their zero-shot forecasts.

\FloatBarrier
\subsection{Investigating the Impact of Different Information Sets on Forecast Accuracy}\label{INFOset}

\begin{figure}[t!]
    \centering
    \includegraphics[width=0.78\textwidth]{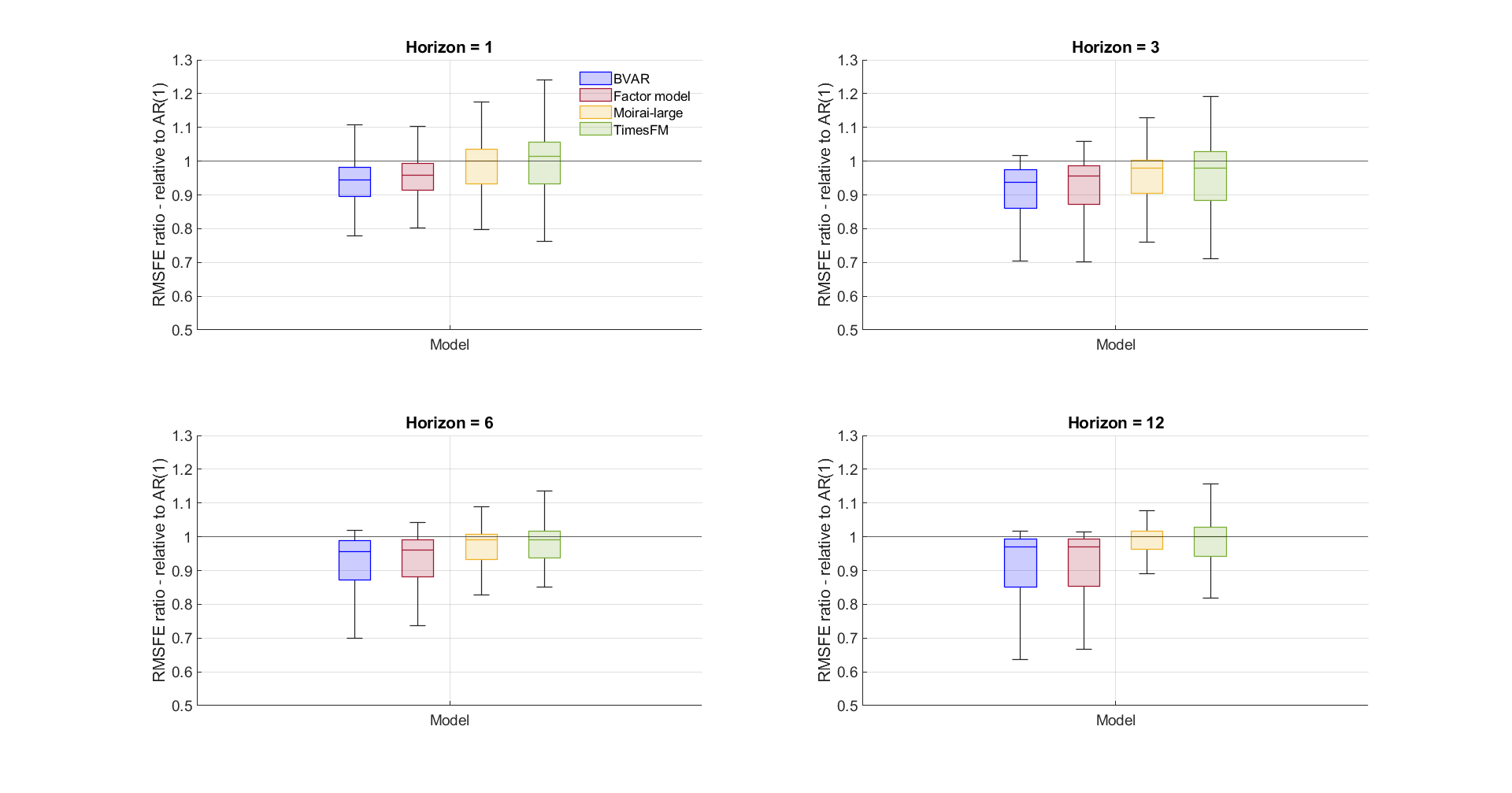}
    \caption{Distribution of RMSFEs (relative to an AR benchmark) for the econometric models and a selection of \tsfm{}s. BVARs and Factor models are trained using data up to December 2022. The evaluation sample is January 1985 to December 2019.}
    \label{fig:boxplot_econometric_and_best_llm_fullsample}
\end{figure}

In the Introduction, we emphasized that it is not straightforward to assess the real-time forecast accuracy of \tsfm{}s, mainly because these models are trained on data lacking timestamps. Thus, previous results may overstate \tsfm{} forecasting ability, introducing a potential look-ahead bias due to some of the training data postdating the forecast date. To further complicate things, \autoref{tab:TSLM_training_details} inthe Appendix shows that some of the \tsfm{} models do include in the training set the FRED-MD database, which means they potentially have access to the perfect forecast. Although retraining the \tsfm{} models in real time is difficult and costly, we can perform the opposite experiment of giving the same unfair advantage to the econometric models. This is the experiment we describe in this subsection. 

Specifically, we re-estimated the econometric models using the entire sample available, up to December 2022. This date precedes or coincides with the dates in which the \tsfm{}s were trained (based on the year of publication of the corresponding papers, during 2024). While this comparison is not perfect, it can provide a first order approximation of the potential impact of using different information sets in the model training phase. \autoref{fig:boxplot_econometric_and_best_llm_fullsample} shows the results of this experiment, where the BVAR and Factor model RMSFEs are replaced by their full-sample counterparts. \moirai and \timesfm results, taken from \autoref{fig:boxplot_llm_only}, are included to make the comparison easier.

The results are revealing. The forecasting performance of BVAR and Factor models is dramatically enhanced by the use of the extra information, with their respective box plots lying  entirely below one. In other words, these models consistently beat the AR benchmark across most series and horizons. They also perform significantly better than the \tsfm{}s, which appeared to perform comparably to the econometric models only when the latter were estimated *without* the benefit of hindsight.

\FloatBarrier
\subsection{How Do \tsfm{}s Work with Persistent Series?}\label{nonstationary}

As described above, many of the macroeconomic series are transformed before entering the estimation stage of the models. We now analyze how model performance varies with the persistence of the series, measured by the one-lag partial autocorrelation of the transformed series. Persistence ranges continuously from $-0.46$ to $1.00$ across the 120 series, with a median of $0.52$. Although 43 series (36\%) exceed a conventional threshold of 0.7, these are not evenly spread above it -- 34 series (28\% of the sample) are in fact clustered very close to unit-root behavior ($>0.9$), forming a distinct spike separate from the roughly uniform mass of series between $-0.5$ and $0.7$. 

\begin{figure}[t!]
    \centering
    \includegraphics[width=0.78\textwidth]{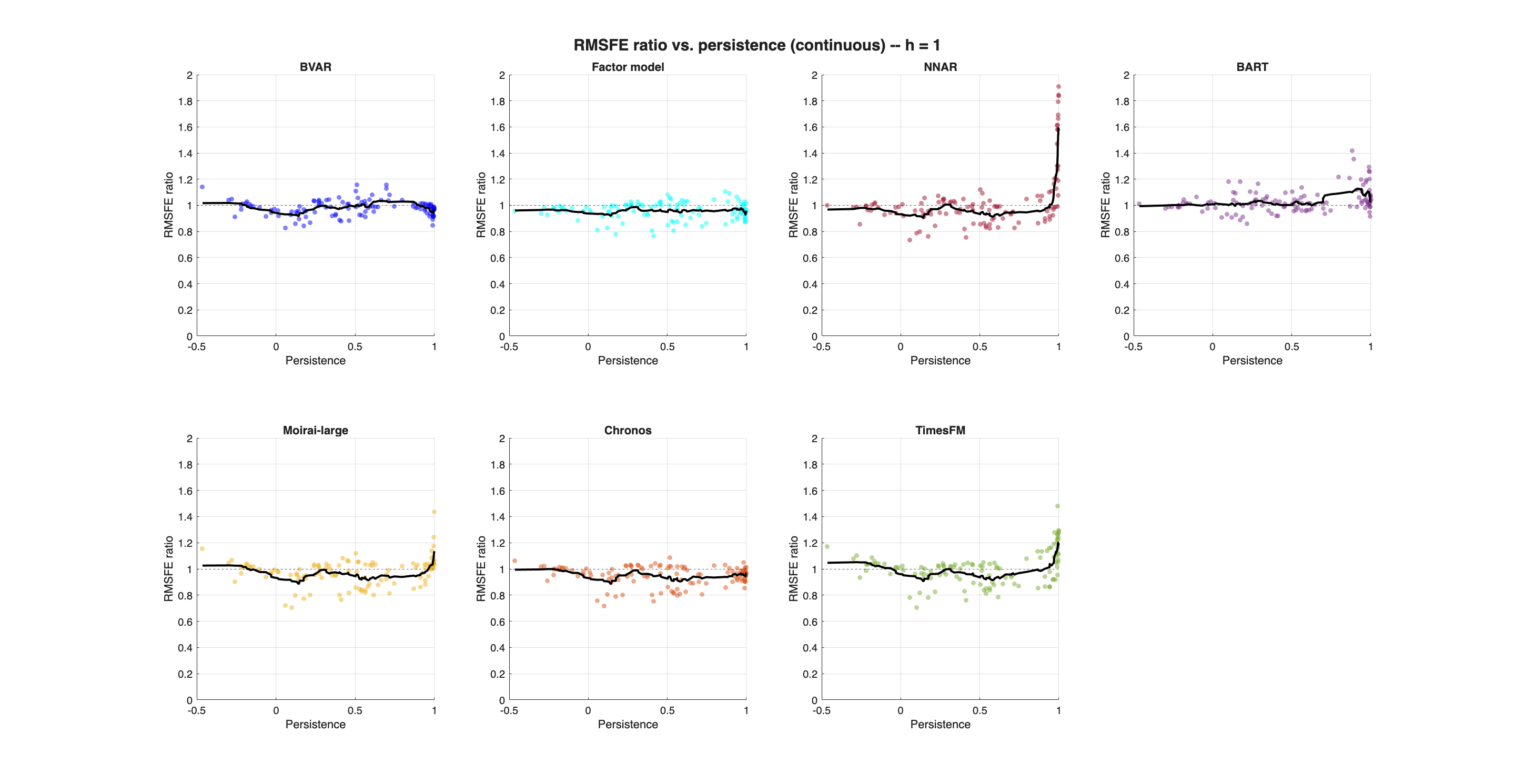}
    \caption{RMSFE ratio against continuous persistence, by model, one-step-ahead horizon. Each point is one series; the solid line is a 15-series moving average, sorted by persistence.}
    \label{fig:persistence_continuous_h1}
\end{figure}

\begin{figure}[t!]
    \centering
    \includegraphics[width=0.78\textwidth]{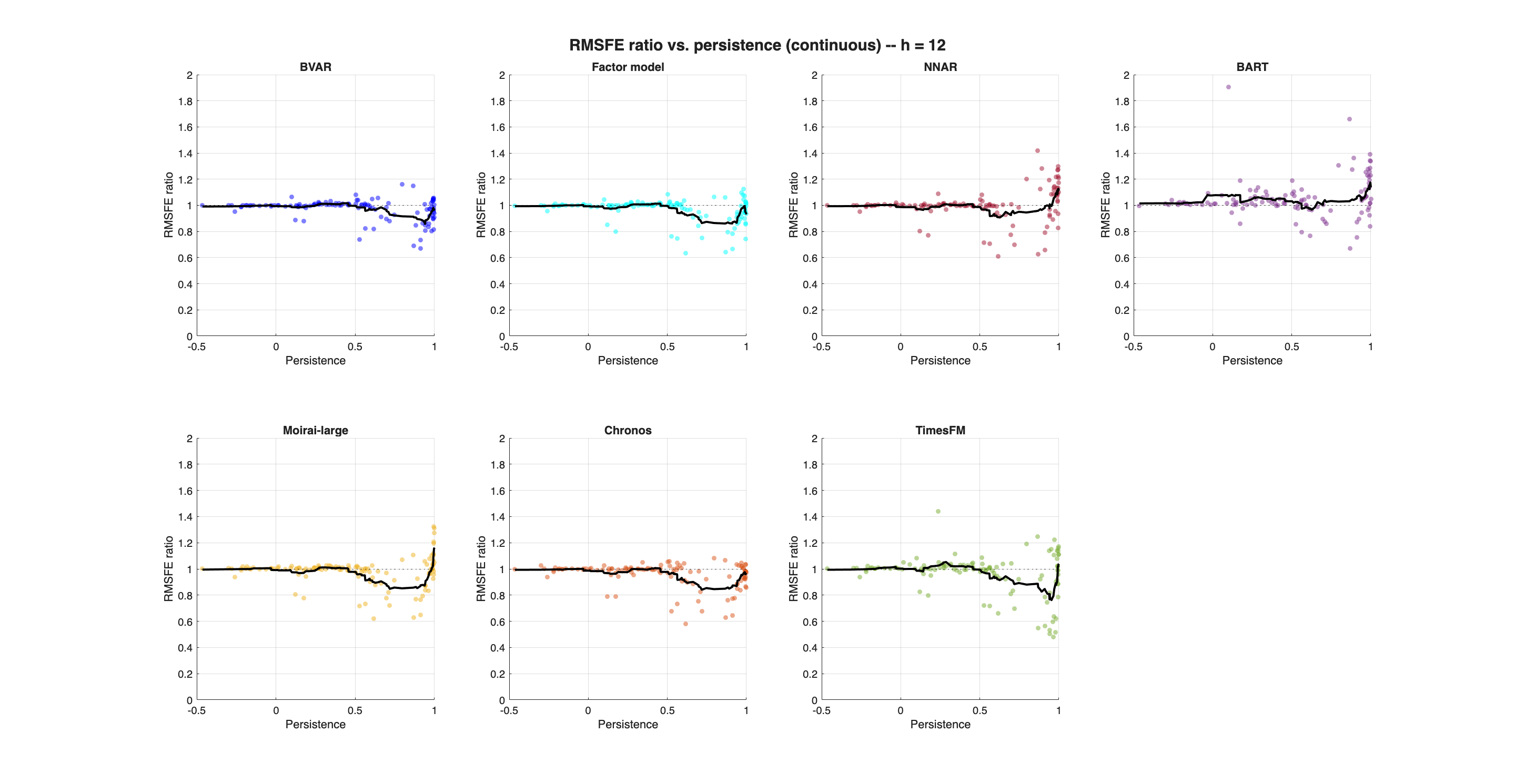}
    \caption{RMSFE ratio against continuous persistence, by model, twelve-step-ahead horizon. Each point is one series; the solid line is a 15-series moving average, sorted by persistence.}
    \label{fig:persistence_continuous_h12}
\end{figure}

\autoref{fig:persistence_continuous_h1} shows that at the one-step-ahead horizon, the relationship between persistence and relative accuracy differs sharply by model type. The two linear econometric models (BVAR, Factor model) and Chronos are essentially flat across the full range of persistence, meaning these three models are as reliable on a near-random-walk series as on a mean-reverting one. NNAR, in contrast, deteriorates sharply and visibly as persistence approaches one, with several near-unit-root series producing RMSFE ratios above 1.5; BART and TimesFM show the same qualitative pattern to a lesser degree, and Moirai-large is only mildly affected. At the twelve-step-ahead horizon (\autoref{fig:persistence_continuous_h12}) the picture is more nuanced than a simple correlation coefficient would suggest. For BVAR, Factor model, Moirai-large, Chronos, and TimesFM, the moving average \emph{improves} as persistence rises from moderate values up to roughly 0.8--0.9, but accuracy then \emph{reverses} sharply for the small subset of series with persistence above roughly 0.95, climbing back toward or above one. Moirai-large and TimesFM are the most affected by this reversal, while BVAR, Chronos and the Factor model remain closest to their best-case accuracy even at the extreme. NNAR and BART do not show this dip-then-reverse pattern at all: both degrade monotonically as persistence increases, at both horizons, indicating they are the two models least suited to highly persistent series regardless of forecast horizon. 

To give concrete examples of the extent of variation in performance across different types of series, Appendix~\ref{App_series_comparisons} examines the time series of out-of-sample forecasts produced by the factor model, the BVAR, and \timesfm for a few selected variables, illustrating cases where \timesfm tracks turning points (such as the 2007--09 housing collapse) that the econometric models miss, as well as cases where it underperforms.

\FloatBarrier
\subsection{Heterogeneity in Forecast Performance}\label{heterogeneity}

The persistence-based analysis considered so far implicitly assumes that forecastability is governed by a single scalar property of a series. In practice, series belonging to different segments of the macroeconomy -- real activity, prices, financial variables -- are generated by different underlying dynamics, and a model's relative strength may depend on the economic domain rather than (or in addition to) persistence alone. To investigate this, we partition the 120 series into the eight variable groups defined by \citet{McCracken:Ng:2016} (Output and Income, Labor Market, Housing, Consumption/Orders/Inventories, Money and Credit, Interest and Exchange Rates, Prices, and Stock Market), and recompute the RMSFE ratio distributions separately within each group, for the same seven models considered in \autoref{fig:boxplot_econometric_and_best_llm}.

\begin{figure}[t!]
    \centering
    \includegraphics[width=0.78\textwidth]{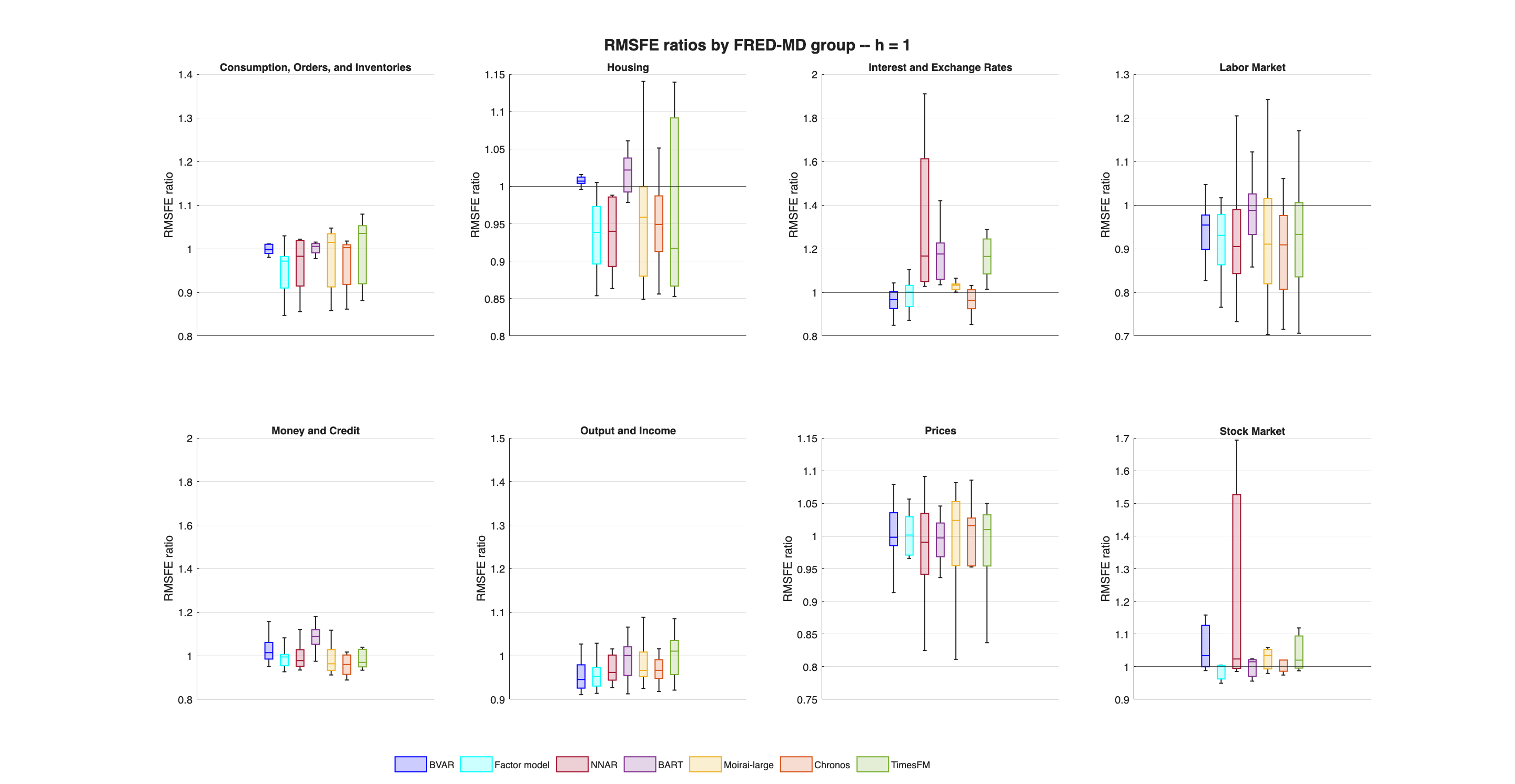}
    \caption{Distribution of RMSFEs (relative to an AR benchmark) by FRED-MD variable group, one-step-ahead horizon. The evaluation sample is January 1985 to December 2019.}
    \label{fig:boxplot_by_group_h1}
\end{figure}

\begin{figure}[t!]
    \centering
    \includegraphics[width=0.78\textwidth]{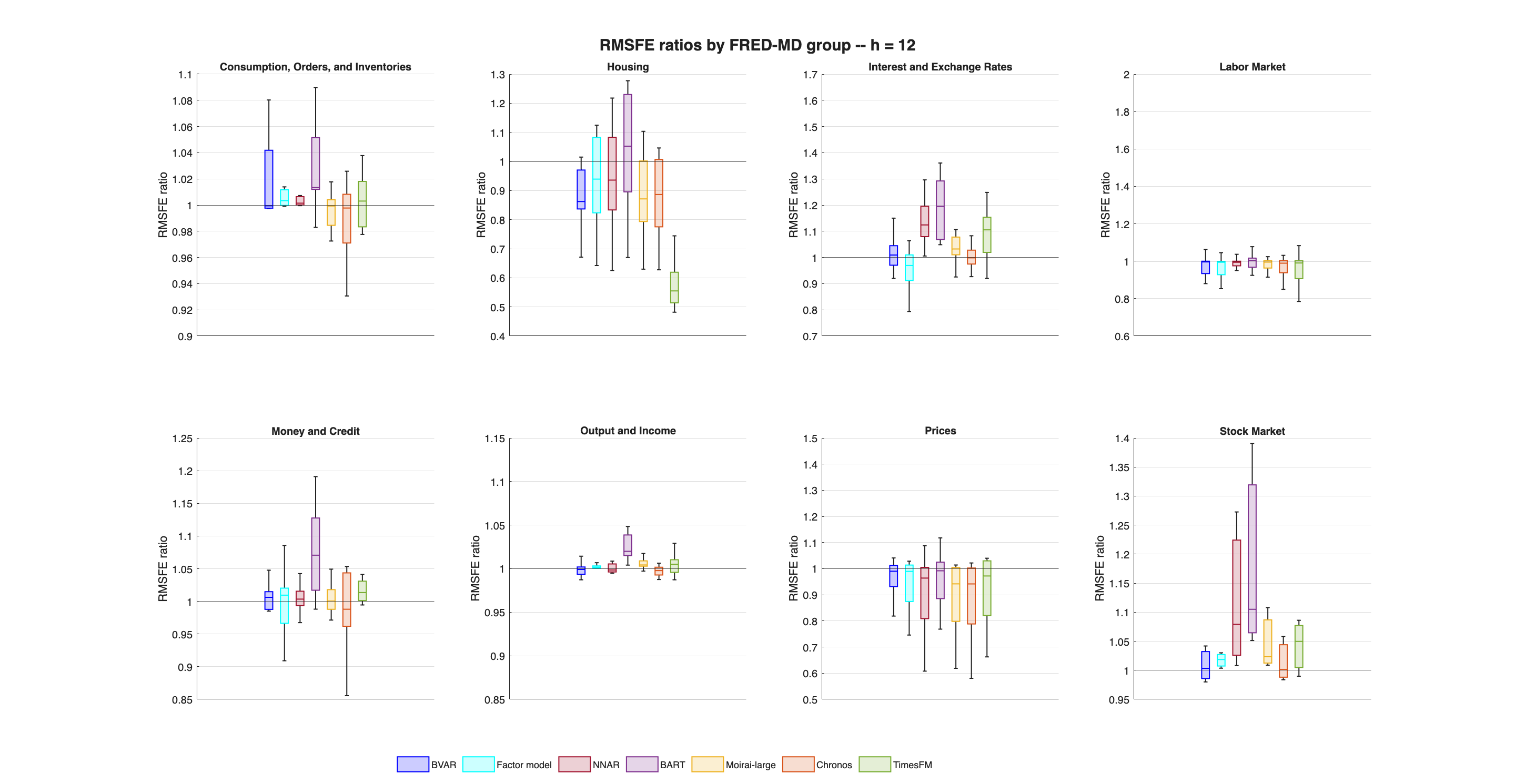}
    \caption{Distribution of RMSFEs (relative to an AR benchmark) by FRED-MD variable group, twelve-step-ahead horizon. The evaluation sample is January 1985 to December 2019.}
    \label{fig:boxplot_by_group_h12}
\end{figure}

\autoref{fig:boxplot_by_group_h1}, \autoref{fig:boxplot_by_group_h12}, and \autoref{tab:RMSFE_by_group_model} in the Appendix reveal substantial heterogeneity across economic domains that is masked by the aggregate results in Section~\ref{Results}. \textbf{Labor Market}, the largest group with 31 series, is where all seven models are both most accurate and most consistent with one another: each model's own group-level median at $h=1$ falls between 0.905 (NNAR) and 0.988 (BART), all at or below one, suggesting this is the domain where the econometric and TSFM approaches are most substitutable. At the opposite extreme, \textbf{Interest and Exchange Rates} is the one group where most models' medians exceed one at every horizon we consider, and where the nonlinear models struggle most: NNAR and BART's medians range from 1.09 to 1.22 across horizons, while TimesFM and Moirai-large also lag (1.01--1.16). Chronos is a conspicuous exception here, tracking the AR benchmark closely at every horizon (0.96--1.00) even as its TSFM peers deteriorate -- suggesting its relative strength is concentrated in exactly the domain where the other TSFMs are weakest. \textbf{Stock Market}, with only three series, should be read with the usual small-sample caveat, but is also uniformly difficult, consistent with the standard near-random-walk view of asset prices; BART is a notable outlier there, with a twelve-step-ahead median of 1.105 (rising monotonically from 1.015 at $h=1$).

\textbf{Housing} stands out for the size of the gains available at long horizons, and for how unevenly they are distributed across models. At $h=12$, TimesFM's median ratio for Housing falls to 0.555, driven by the same ability to anticipate the 2007--09 housing collapse already documented for HOUST in \autoref{fig:comparisons}, while BART's median for the same group and horizon \emph{rises} to 1.052. This is a useful, concrete illustration of why pooling across all 120 series can average away offsetting model-specific effects that are large in absolute terms. Detailed statistics for all groups, models, and horizons are reported in \autoref{tab:RMSFE_by_group_model} in \ref{App_heterogeneity}. Given this heterogeneity, a natural question is whether a \tsfm{} can be made more accurate, or at least more consistent, by fine-tuning it more closely to the data at hand; Appendix~\ref{App_finetuning} reports this exercise and finds that fine-tuning \timesfm and \moirai-\largemodel{} on \fred data yields only marginal gains over their zero-shot performance.

\FloatBarrier
\subsection{Performance Evolution: Old vs. New Generation TSFMs}\label{newgen}

Fine-tuning is one way of tailoring a pretrained \tsfm{} to \fred; another is to rely on the improvements that model developers themselves make from one model generation to the next. This subsection compares the forecasting performance of old and new generations of Moirai and Chronos (Moirai 1.0 large vs.\ 2.0 small; Chronos 1.0 T5-large vs.\ 2.0), to see how much of the TSFM improvement documented elsewhere in this paper is attributable to the newer model generations specifically.

\begin{figure}[t!]
	\centering
	\includegraphics[width=0.78\textwidth]{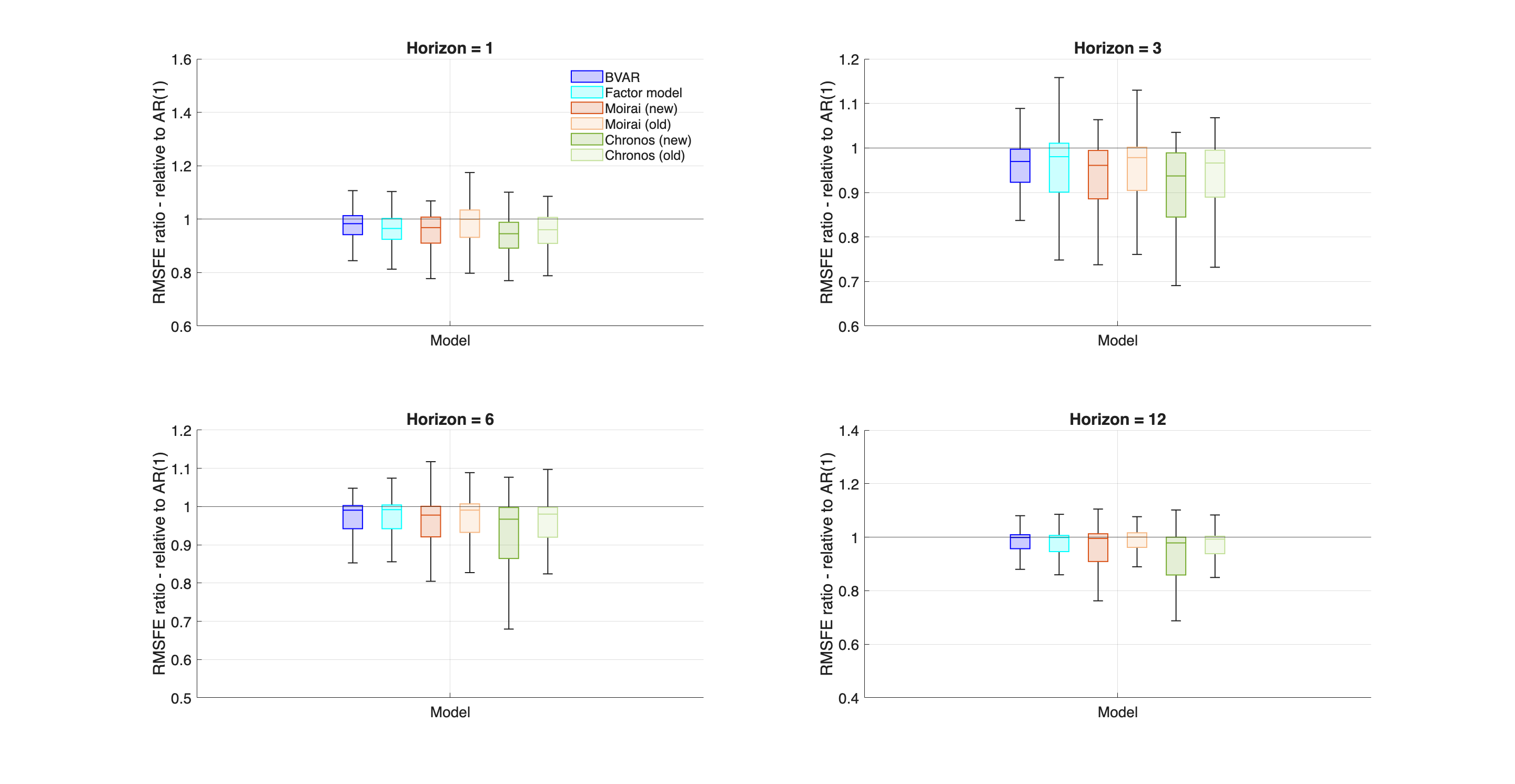}
	\caption{Distribution of RMSFEs (relative to an AR benchmark) for BVAR, the Factor model, and old vs.\ new generation Moirai/Chronos. The evaluation sample is January 1985 to December 2019.}
	\label{fig:boxplot_newgen}
\end{figure}

\begin{table}[h!]
	\centering
	{\footnotesize\input{tables/Table5_newgen_stats.tex}}
	\caption{Median, Std. deviation, Min and Max RMSFE by model type and forecast horizon: BVAR, Factor model, and old vs.\ new generation Moirai/Chronos. The evaluation sample is January 1985 to December 2019.}
	\label{tab:RMSFE_stats_newgen}%
\end{table}

\autoref{tab:RMSFE_stats_newgen} shows that the new generation improves on the old for both model families and at every horizon: Chronos-2's median RMSFE ratio is lower than Chronos-1's at $h=1,3,6,12$ (0.945 vs.\ 0.961, 0.937 vs.\ 0.966, 0.967 vs.\ 0.980, and 0.979 vs.\ 0.993, respectively), and Moirai 2.0 small similarly improves on Moirai 1.0 large at every horizon (e.g.\ 0.968 vs.\ 0.999 at $h=1$). The gain is not confined to the median: for Moirai, the new generation also curbs the worst-case outcomes, with the maximum RMSFE ratio falling from 1.436 to 1.068 at $h=1$ and from 1.324 to 1.189 at $h=12$, indicating a more reliable model rather than merely a better-centered one. Chronos-2 is in fact the single best-performing model in the table at every horizon, edging out both the BVAR and the Factor model, whereas Moirai 2.0 small is competitive with but does not clearly surpass the two econometric benchmarks, particularly at longer horizons ($h=6,12$). This suggests that at least part of the TSFM gains documented earlier in the paper reflect genuine progress across model generations, rather than being an artifact specific to the particular \tsfm{} vintages used in our main analysis.

\FloatBarrier
\section{Conclusions}\label{Conclusions}

In this paper, we have provided a comprehensive evaluation of the forecasting performance of Time Series Language Models (\tsfm{}s) in predicting macroeconomic variables. Our results indicate that their performance is mixed: only three of the six models we evaluated, Salesforce's \moirai, Google's \timesfm{}, and Amazon's \chronos{}, consistently outperform a simple autoregressive benchmark, and even these struggle to deliver consistently superior forecasts compared to established methods such as Bayesian Vector Autoregressions (BVARs) and Factor Models. \tsfm{}s can therefore offer valuable insights, but are not yet a clear replacement for state-of-the-art econometric models. A key complicating factor is the lack of control over their training data: many of these models are pretrained on the \fred dataset itself, introducing potential biases and making a clean pseudo out-of-sample exercise difficult, while practical and computational constraints make retraining them in real time non-trivial. Despite these limitations, \tsfm{}s exhibit some promising features, particularly in capturing nonlinearities and adapting to evolving economic conditions -- their relatively stronger performance in the post-Covid-19 period suggests an advantage in handling structural breaks. 

Crucially, econometric models tend to be strongest in exactly the regimes where \tsfm{}s are weakest, and vice versa: this complementarity, rather than a uniform ranking of one class of models over the other, is what motivates combining the two. This is precisely the logic behind our TSFM-augmented BVAR (TSFM-BVAR) hybrid, one of the central contributions of this paper. The TSFM-BVAR hybrid generates forecasts with a BVAR and then models the residuals using TSFMs, and consistently reduces forecast errors relative to either component in isolation, particularly by mitigating the extreme errors that TSFMs occasionally produce. This confirms that integrating structured econometric information with flexible machine learning models can substantially improve forecasting performance. Finally, our comparison of old versus new generations of TSFMs (Moirai 1.0 vs.\ 2.0, Chronos 1.0 vs.\ 2.0) shows newer models improving on their predecessors, suggesting continued gains ahead as the field evolves.

\bibliographystyle{chicago}
{\footnotesize
\singlespacing
\setlength{\bibsep}{4pt}
\bibliography{All_references_LLMs}
}

\clearpage
\begin{appendices}
	\renewcommand{\thesection}{Appendix \Alph{section}}
	\renewcommand{\thesubsection}{\Alph{section}.\arabic{subsection}}
	\renewcommand{\theequation}{\Alph{section}.\arabic{equation}}
	\renewcommand\thetable{\Alph{section}.\arabic{table}}
	\renewcommand\thefigure{\Alph{section}.\arabic{figure}}
	\setcounter{equation}{0}
	\setcounter{section}{0}
	\setcounter{table}{0}
	\setcounter{figure}{0}

\begin{center}
\ifanonymous
{\large\bfseries Online Appendix for  \\``Macroeconomic Forecasting with Large Language Models''}
\else
{\large\bfseries Online Appendix for  \\``Macroeconomic Forecasting with Large Language Models''\\by \\ Andrea Carriero, Davide Pettenuzzo, and Shubhranshu Shekhar}
\fi
\end{center}

\FloatBarrier
\section{Details on Time Series Foundational Models}\label{App_TSFM}

\setcounter{table}{0}
	\setcounter{figure}{0}

This appendix complements the overview of Time Series Foundational Models (\tsfm{}s) given in Section~\ref{sec:tsfm}. Appendix~\ref{App_TSFM_components} describes in detail the building blocks introduced in \autoref{fig:tslm} -- tokenization, model architecture, data augmentation, and pretraining/fine-tuning. Appendix~\ref{App_TSFM_dl} discusses the relationship between \tsfm{}s and standard Deep Learning models. Appendix~\ref{App_TSFM_models} describes each of the six \tsfm{}s used in this paper. Finally, Appendix~\ref{App_TSFM_newgen} describes the newest generation of \tsfm{}s, which we compare empirically to their predecessors in Section~\ref{newgen}.

\FloatBarrier
\subsection{Components of a \tsfm{} in Detail}\label{App_TSFM_components}

\subsubsection{Tokenization}
In textual LLMs e.g. ChatGPT, the process of text understanding begins with tokenization, which breaks down sentences into their fundamental semantic units -- tokens. Tokens can be individual words, characters, or subwords. Tokens allow LLMs to identify and learn the relationships between semantic units, leading to a better text understanding. As we discussed above, time series data lack inherent well-defined semantic units such as words. To circumvent this issue, \tsfm{}s rely on scaling, patching, and quantization techniques to transform numerical sequence data to a format "token" suitable for learning patterns and trends within data.\footnote{It is worth noting here that a lack of fundamental semantic units presents also an opportunity to apply domain knowledge to derive better tokens, and thus potentially shepherding field of \tsfm{}s for macroeconomic forecasting.} The common schemes utilized by current \tsfm{}s are listed below. Note that not all \tsfm{}s implement each scheme; they choose what works for their training process.

\noindent \textbf{Patching}

Patching is a process used to divide time series into shorter, fixed-length segments or windows called patches. These patches can be treated as tokens, capturing local patterns. For example, \moirai, \timesfm, \ttm use patching to create tokens, while \lagllama augments each patch with lag features and time-based features to construct tokens. Importantly, the patch size is itself a hyper-parameter that can be adjusted based on the specific characteristics of the time series data. For example,  larger patch sizes are more suitable for high-frequency time series, while smaller patch sizes tend to be more suitable for low frequency time series. Consider as an example a time series $x_{1:T}=\{4.7, 4.76, 6.8, 7.2, 6.1\}$. Tokenization with patching size $=3$ and overlap $=2$, will yield following tokens $\{4.7, 4.76, 6.8\}$, $\{4.76, 6.8, 7.2\}$, $\{6.8, 7.2, 6.1\}$. Without overlap the tokens will be $\{4.7, 4.76, 6.8\}$, $\{7.2, 6.1\}$. 

\noindent\textbf{Quantization}

Quantization is used to convert numerical time series values into a set of discrete tokens, similar to finite words in textual LLMs. This is achieved by dividing the value range of each time series, or a patch or set of patches (depending on the \tsfm) into a predefined number of bins $\mathbb{B}$.  Each data point is then assigned a token, a number between $\{1, \cdots \mathbb{B}\}$ based on the bin it falls into. There are different approaches to binning  - uniform quantization divides the value range into bins of equal size, and data-dependent quantization adjusts the bin sizes based on the data distribution. \chronos \citep{ansari2024chronos}, for example, uses uniform quantization to encode time series into tokens. To continue with the previous example, let's apply uniform quantization with  $\mathbb{B} = 4$ to the time series $x_{1:T}$. The mapping from $x_{1:T}$ to the four bins is carried out as follows:
\begin{align*}
    q_t = \left\{ 
    \begin{matrix}
        1 & \text{if } 4 \le x_t < 5 \\
        2 & \text{if } 5 \le x_t < 6 \\
        3 & \text{if } 6 \le x_t < 7 \\
        4 & \text{if } 7 \le x_t < 8 
    \end{matrix}
    \right.
\end{align*}
Then the tokenized sequence is given by $q_{1:T}=\{1, 1, 3, 4, 3\}$.

\noindent \textbf{Scaling}\footnote{Scaling can also be referred to as instance normalization or standardization.} 

When working with multiple time series at once, even when those are within the same dataset, it is common to have variables expressed with different scales. For example, in the case of macroeconomic variables we often see interest rates expressed in percentages, GDP growth reported in annualized growth rates, and unemployment rates displayed in absolute numbers. To ensure consistent processing, data points are typically scaled. Scaling also helps in optimization/learning for deep learning models, as varying scales may distort gradient computation. Scaling a value $x_t$ can be done as follows,
    \begin{displaymath}
    	\tilde{x}_t = \frac{x_t - M}{S}
    \end{displaymath}
    where $\tilde{x}_t$ is the scaled value. Different choices for $M$ and $S$ lead to the types of scaling used by the various \tsfm{}s. For example, \lagllama uses $M=$ median of the context window and $S=$ inter-quartile range within the context window.\footnote{A context window is the length of a time series considered as history for forecasting.} It is also worth noting that scaling can be applied at the global level (using the value range of the entire time series), at a context window level, or at a patch level.

\noindent \textbf{Embeddings}

In \tsfm{}s, tokenization --- through scaling, patching, or quantization --- converts raw sequence data into discrete tokens. This process is similar to how textual LLMs tokenize sentences into words or subwords. The time series tokens are then transformed into embeddings, which are dense numerical vectors. Textual embeddings capture semantic relationships between words, while in \tsfm embeddings encode local temporal patterns such as trend or seasonality. This transformation occurs via an embedding layer, a learnable component of the transformer architecture (we discuss the architecture in the next subsection), which maps tokens to a vector space suitable for processing by attention mechanisms.  While architecturally part of the transformer, the embedding process creates inputs capturing the contextual information of the time series data for the transformers.

\subsubsection{Model Architecture}

Similar to textual LLMs, the transformer~\citepapp{vaswani2017attention} is the core model architecture used in recent \tsfm{}s.\footnote{Additionally, there are non-tranformer-based architectures such as multi-layer perceptrons (MLP) used in some \tsfm{}s. Of the various models we considered, all of them use a transformer architecture, except for TTM which uses a traditional MLP-based architecture. }
Transformer-based models leverage self-attention mechanisms to capture long-range dependencies within time series, offering significant advantages in handling sequential data. Self-attention refers to the ability of transformers to focus on relevant parts of the time series data. In a transformer model, each token is projected into a higher dimensional space, allowing the model to capture richer information within that segment. Then, the self-attention layer applies linear transformations and uses the dot product between transformed vectors to identify relevant parts of the time series for each token. This helps in identifying long-range dependencies and relationships, agnostic of the token positions in the input time series.  The original transformer architecture utilized two components, namely (i) an encoder, responsible for extracting the relevant information from the input time series and projecting/embedding it into a vector representation and (ii) a decoder, responsible for autoregressive prediction of next token conditioned on observed input sequence of tokens, and encoder output. Recently, several transformer architectures have evolved, notably encoder-decoder, decoder-only, and encoder-only.
In the encoder-decoder setup, the encoder maps an input sentence of some language to a continuous representation, and the decoder generates the translation token-by-token using the input representation and previously decoded tokens. In the decoder-only case \citep[e.g. \timesfm,][]{das2024decoder}, the model only attends to tokens up to the current token, while in the encoder-only case \citep[e.g. \moirai,][]{woo2024unified}, the model uses information from the entire sequence to generate tokens. Note that encoder-decoder and decoder-only generate output tokens in an autoregressive manner in contrast to encoder-only architecture.

\subsubsection{Time Series Augmentation}
Textual LLMs train on massive text data readily available from books, articles, code, and web crawls. This abundance allows them to learn underlying patterns and relationships effectively. 
Similarly, \tsfm models are intended for plug-and-play forecasting, which requires large-scale time series data with diverse patterns for training. However, time series data are inherently scarce. Augmentation helps mitigate the scarcity and variability of time series datasets by generating more diverse training data. There are several approaches to time series data augmentation (see \citetapp{iglesias2023data} for a detailed survey). These techniques typically involve convex combinations of existing sequences, combinations of ARMA processes, seasonal patterns, trends, and step functions, and combining the frequency spectrum of sequences and then converting it back to the time domain.

\subsubsection{Pretraining and Fine-Tuning}\label{subsec:TSFM}
\subsubsection*{Pretraining} Pretraining refers to the process of feeding to the \tsfm{}s a large amount of time series data so that the model can learn general temporal patterns and relationships and with that perform well across a wide spectrum of tasks. The time series data fed into the \tsfm can be from various domains and span different frequencies and, as previously discussed, each time series will first be scaled and tokenized before any training occurs. Once the pretraining phase is completed, the \tsfm{}s will have built a foundational knowledge base from a very general corpus of time series data, and with that can be used to perform zero-shot forecasting in a new domain. 

Pretraining relies on self-supervised techniques that create tasks for the model so that it can learn directly from the input data without requiring human labeled examples. 
The success rate in these tasks can be measured in various ways, such as minimizing MSE or accurately predicting the quantized bins or the quantiles of the training data. 
As a concrete example, \lagllama pretraining looks for the model parameters that maximize the likelihood of the observed future values, under the assumption that these follow a Student-t distribution. 

\subsubsection*{Fine-Tuning}
Fine-tuning is a training process that is used to adapt a pretrained \tsfm, containing general foundational knowledge, to a specific task such as forecasting, or classification. Fine-tuning updates existing knowledge (parameters) of \tsfm to better handle unique characteristics of the task-specific data. This can potentially improve the performance of a \tsfm compared to zero-shot predictions for the given task.
Fine-tuning can also be used to continually learn the model parameters and incorporate fresh data, improving \tsfm{}'s forecasting accuracy over time. For example, in macroeconomic forecasting, we can use fine-tuning to incorporate the data from the current quarter to update the parameters of the \tsfm.

\FloatBarrier
\subsection{Relationship with Deep Learning}\label{App_TSFM_dl}

Before moving on to describe the existing \tsfm{}s, it is worth briefly discussing the difference between Deep Learning (\dl) models (see for example \citetapp{torres2021deep}, \citetapp{lim2021time} and \citetapp{wen2022transformers}) and \tsfm{}s. \dl models share many similarities with traditional time-series methods, requiring problem specific training data to learn model parameters and make forecasts. However, they differ the traditional methods in a number of ways -- using non-linear architectures, large numbers of learnable parameters, and requiring much larger training data than traditional time series methods. \tsfm{}s, on the other hand, are pretrained on large number of time series datasets. As with FMs, this pretraining allows them to perform ``zero-shot'' forecasting. This plug-and-play approach makes \tsfm{}s particularly useful and powerful. Furthermore, pretraining allows \tsfm{}s to require much less domain-specific knowledge data compared to deep learning models, and if needed can be fine-tuned on specific datasets.

\FloatBarrier
\subsection{State-of-the-Art \tsfm{}s}\label{App_TSFM_models}
We now briefly describe the \tsfm models considered in this paper.\footnote{In addition to the models listed in this section, we have also experimented with \moment\citep{goswami2024moment},  a family of time series transformer models that is pretrained on masked time series prediction task using patching. However this model is still in its very early stages and we have found its performance to be lagging significantly below the other \tsfm{}s we have considered. We therefore did not include it in the comparison below.}

\paragraph{\bfseries \lagllama} \citep{rasul2023lag} \lagllama is a probabilistic foundation model that extends the Llama 2 architecture~\citepapp{touvron2023llama} to time series data. \lagllama follows a z-score normalization at a window level, and tokenizes the input time series by extracting lagged features from the past values of the time series at different time lags. This allows the model to learn how past observations influence future ones. \lagllama is pretrained on 27 publicly available time series datasets from various domains. The model has about $\sim 2.5$M parameters.

\paragraph{\bfseries \moirai} \citep{woo2024unified} Salesforce's \moirai is pretrained on $27$B observations from time series datasets across nine domains. \moirai uses multiple patch (window) size projection layers to capture temporal patterns across various frequencies. It comes in 3 different model sizes -- largest of which has about $311$M parameters, which is significant in size compared to earlier models.

\paragraph{\bfseries Chronos} \citep{ansari2024chronos} .  Chronos tokenizes time series into discrete bins through simple scaling and quantization of real values. The model is pretrained on a comprehensive collection of publicly available datasets, which are then combined with data augmentation strategies, including TSMix and KernelSynth. TSMix randomly samples a set of base time series from different training datasets, and generates new time series based on a convex combination of them; KernelSynth uses Gaussian processes to generate synthetic time series by randomly composing kernel functions.

\paragraph{\bfseries \timesfm}~\citep{das2024decoder,timesfm25} Google's \timesfm is a $200$M parameter \tsfm{}. It is trained on largest corpus of 100 billion time points that includes both real and synthetic time series. \timesfm breaks time series in patches and learns to predict subsequent patches, and uses a standard normalization at patch level.

\paragraph{\bfseries \ttm}~\citep{ekambaram2024ttms} IBM's \ttm (Tiny Time Mixer), as the name suggests, is a small \tsfm with $\leq 1$M parameters that is pretrained on the Monash time series repository~\citetapp{godahewa2021monash} and LibCity~\citetapp{libcitylong} with $\sim 1$B data points. \ttm is based on TSMixer architecture~\citepapp{ekambaram2023tsmixer} which uses MLP-based (multi-layer perception) mixing across time steps and features. This enables multivariate forecasting through efficient extraction of temporal and cross-variate patterns for time series forecasting. \ttm first normalizes each sequence to have zero mean and unit standard deviation. Then it considers patches (windows) of varying length and varying resolution, since each dataset may have different optimal context to consider. \ttm also applies downsampling on high resolution time series (e.g. seconds, minutes) to augment the training dataset.

\paragraph{\bfseries \timegpt} \citep{garza2023timegpt1}
\timegpt was trained on a collection of publicly available time series, collectively encompassing over 100 billion data points. This training set incorporates time series from a broad array of domains, including finance, economics, demographics, healthcare, weather, IoT sensor data, energy, web traffic, sales, transport, and banking. Due to this diverse set of domains, the training dataset contains time series with a wide range of characteristics, ranging from multiple number of seasonalities, cycles of different lengths, and various types of trends.

\autoref{tab:TSLM_training_details} lists a few additional details on the various \tsfm{}s described in this section. In particular, it provides information on when the model was released, i.e. when the pretraining sample ends, the list of all the data domains used in the pretraining phase of the model, and the size of training sample. It also flags whether each \tsfm{} is univariate or multivariate.

\begin{table}[t!]
    \centering
    \footnotesize
    \resizebox{\linewidth}{!}{
    \begin{tabular}{p{1.6cm} l p{7cm} p{1.5cm} c}
    \toprule
    \textbf{Model} &  \textbf{Release date} & \textbf{Training datasets (domains)} & \textbf{Size} & \textbf{Multivariate}\\
    \hline
       \lagllama  & Feb 2024 &  Traffic, Uber TLC, Electricity, London Smart Meters, Solar power, Wind farms, KDD Cup 2018, Sunspot, Beijing Air quality, Air Quality UC Irvine Repository, Huawei cloud, Econ/Fin$^*$ & 352M tokens & \xmark\\
       \cline{1-5}
       \moirai & Mar 2024 & Energy, Transport, Climate, CloudOps, Web, Sales, Nature, Econ/Fin$^*$, Healthcare & 27B obs. & \cmark\\
       \cline{1-5}
       \chronos & Mar 2024 & Energy, Finance and economics, Healthcare, Nature, Retail, Mobility and transport, Web, Synthetic Time Series Data & 84B tokens & \xmark\\
       \cline{1-5}
       \ttm  & April 2024 & Electricity, Web traffic, Solar power, Wind farms, Energy consumption, KDD Cup 2018, Sunspot, Australian weather, US births, Bitcoin, Econ/Fin$^*$ & 1B obs.& \cmark\\
       \cline{1-5}
       Time-GPT & May 2024 & Finance, economics, Demographics, Healthcare, Weather, IoT sensor data, Energy, Web traffic, Sales, Transport, and Banking & 100B obs. & \xmark\\
       \cline{1-5}
       \timesfm & May 2024 & Google Trends, Wiki Page views, M4 Competion, Electricity and the Traffic data, Weather data, Synthetic Time Series Data & 100B obs. & \xmark\\
     \bottomrule    
    \end{tabular}
    }
    \caption{Training datasets for the various \tsfm{}s considered in this paper. For each model, we provide a summary of when the model was pretrained, the list of all the datasets used in the pretraining phase of the model, the size of training sample, and whether it is multivariate. Additional details and references on the various datasets can be found in the individual \tsfm{}s' technical papers. $^*$ indicates that the training data include the Monash forecast repository (available at \href{https://forecastingdata.org}{https://forecastingdata.org}), and therefore includes a large part of the FRED-MD dataset.}
    \label{tab:TSLM_training_details}
\end{table}

\FloatBarrier
\subsection{The New Generation of \tsfm{}s: What Has Changed?}\label{App_TSFM_newgen}

The \tsfm{}s described above represent the first wave of pretrained time series models, most of which were released in early 2024. Since then, the leading model families have all been succeeded by substantially redesigned second-generation versions -- \chronos-2 \citepapp{ansari2025chronos2}, \moirai 2.0 \citepapp{liu2025moirai2}, and \timesfm 2.5 \citep{das2024decoder}, released in the past several months. We briefly describe here what is new in each family relative to its predecessor, and then we empirically compare the two generations directly in Section~\ref{newgen}. Two broad themes recur across all three families: (i) a move to \emph{quantile} (rather than distributional or cross-entropy) forecasting objectives, and (ii) a convergence toward architectures that better exploit long context and cross-series information.

\paragraph{\bfseries \moirai 2.0} \moirai 1.0 \citep{woo2024unified} is a masked-encoder model with multiple patch-size projection layers and a mixture-distribution output head, trained on the 27B-observation LOTSA corpus. \moirai 2.0 \citepapp{liu2025moirai2} is a ground-up redesign that replaces the masked encoder with a \emph{decoder-only} autoregressive transformer, drops the multi-patch inputs in favor of a single patch size, and swaps the mixture likelihood for a quantile loss with multi-token prediction. Trained on a larger and more curated corpus, the redesigned model is reported to be roughly twice as fast and an order of magnitude smaller than the best 1.0 configuration while improving accuracy.

\paragraph{\bfseries \chronos-2} The original \chronos \citep{ansari2024chronos} tokenizes a scaled time series into a fixed number of discrete bins and was trained, language-model style, with a cross-entropy loss on a strictly univariate corpus. \chronos-2 \citepapp{ansari2025chronos2} uses continuous, patch-based inputs and a direct multi-horizon quantile objective. It introduces a \emph{group attention} mechanism that lets the model share information in context across related series so that a single pretrained model handles univariate, multivariate, and covariate-informed forecasting in a zero-shot manner. The model (roughly 120M parameters, encoder-only) is trained largely on synthetic datasets constructed to impose diverse multivariate dependence structures on univariate series.

\paragraph{\bfseries \timesfm 2.5} It retains the decoder-only, patched architecture of \timesfm 1.0 \citep{das2024decoder} but scales and refines it. The context length is extended from $512$ to $16{,}000$ time points, allowing a single forward pass to capture multiple seasonalities, low-frequency components, and regime shifts. The required frequency indicator of the original model is removed, making the model frequency-agnostic, and native probabilistic forecasts are supported through an optional calibrated quantile head. The model size remains comparable at $\sim$200M.

\section{Details on Priors}\label{Priors}

\FloatBarrier
\subsection{Prior Moments of Natural Conjugate Prior}\label{Prior moments}

We elicit $\Phi _{0}$ and $\Omega _{0}$ in such a way that the resulting moments of the matrices $\Phi _{l},l=c,1,\dots,p$  are as follows:%
\begin{equation}
E[\Phi _{l}^{(ij)}]=\left\{ 
\begin{array}{c}
\Phi ^{\ast }\ \ \ \ \ \ \ \ \ \ \text{if}\ i=j\text{, }l=1 \\ 
0\ \ \ \ \ \ \ \ \ \ \ \ \ \ \ \ \ \text{otherwise}%
\end{array}%
\right. ,\ Var[\Phi _{l}^{(ij)}]=\left\{ 
\begin{array}{c}
\frac{\lambda _{1}^{2}}{l^{2}}\frac{\sigma _{i}^{2}}{\hat{\sigma}_{j}^{2}},\
l=1,...,p, \\ 
\lambda _{0}^{2}\sigma _{i}^{2},\ l=c%
\end{array}%
\right.   \label{MInn}
\end{equation}%
where $\Phi _{l}^{(ij)}$denotes the element in position $(i,j)$ in the
matrix $\Phi _{l}$. The prior mean $\Phi ^{\ast }$ is set to either $1$ or $0$ depending on the presumed order of integration of the series. For the intercept ($l=c$) we assume an uninformative prior with mean $%
0$ and standard deviation $\lambda _{0}\sigma _{i}$. The shrinkage parameter 
$\lambda _{1}$ measures the overall tightness of the prior: when $\lambda
_{1}\rightarrow 0$ the prior is imposed exactly and the data do not
influence the estimates, while as $\lambda _{1}\rightarrow \infty \ $the
prior becomes loose and the prior information does not influence the
estimates, which will approach the standard $OLS$\ estimates. To set each
scale parameter $\hat{\sigma}_{j}$ in (\ref{MInn})\ we follow common
practice (see e.g. \citetapp{Litterman1986}; \citet{Sims:Zha:IER:1998}) and set it equal to
the standard error of regression from a univariate autoregressive model. The
parameters $\sigma _{i}^{2}$ appearing in (\ref{MInn})\ are instead coming
directly from the error variance $\Sigma $ on which this prior is
conditioning upon, trough the Kronecker multiplication shown in (\ref{NIWprio}).\footnote{Incidentally, note that such multiplication also induces correlation across coefficients belonging to different equations, which is not apparent in (\ref{MInn}) because it focuses only on the variances of the individual coefficients.}

The prior specification is completed by choosing $v_{0}$ and $S_{0}$.
Following \citet{Kadiyala:Karlsson:1997} we set these in a way that they are
as uninformative as possible and such that the prior expectation of the
error variance$\ \Sigma $ is finite and diagonal with diagonal elements
given by the standard errors of regression of univariate autoregressive
models, i.e. $E[\Sigma ]=diag(\hat{\sigma}_{1}^{2},...,\hat{\sigma}_{N}^{2})$%
. To achieve this we set $v_{0}=N+2$ and $S_{0}\ $diagonal with diagonal
elements $(v_{0}-N-1)\hat{\sigma}_{j}^{2}$, $j=1,...N$.\ 

\subsubsection{Dummy Initial Observations}\label{Dummy initial observations}
\citet{Doanetal1984} and \citet{Sims:1993:nber:chapter} have proposed to complement the priors
described above with additional priors which favour unit roots and
cointegration. Both these priors were motivated by the
need to avoid having an unreasonably large share of the sample period
variation in the data accounted for by deterministic components. They are both calibrated using the average of the first $p$ observations in the sample and are typically implemented as dummy observations. 

The \textquotedblleft sum of coefficients" prior expresses a belief that
when the average of lagged values of a variable is at some level $\bar{y}%
_{0i}$, that same value $\bar{y}_{0i}$ is likely to be a good forecast of
future observations, and is implemented by augmenting the system in (\ref%
{MATRIXnotation}) with the dummy observations $Y_{d_{1}}$ and $X_{d_{1}}$
with generic elements:%
\begin{equation}
y_{d}(i,j)=\left\{ 
\begin{array}{c}
\bar{y}_{0i}/\mu _{1}\ \ \ \ \ \ \ \ \ \ \ \ \ \ \ \ \text{if}\ i=j \\ 
0\ \ \ \ \ \ \ \ \ \ \ \ \ \ \ \ \ \text{otherwise}%
\end{array}%
\right. ;\ \ x_{d}(i,s)=\left\{ 
\begin{array}{c}
\bar{y}_{0i}/\mu _{1}\ \ \ \ \ \ \ \ \ \ \ \text{if}\ i=j\text{, }s<M \\ 
0\ \ \ \ \ \ \ \ \ \ \ \ \ \ \ \ \ \ \ \ \ \ \ \text{otherwise},%
\end{array}%
\right.   \label{sumc}
\end{equation}%
where $i$ and $j$ go from $1$ to $N$ while $s$ goes from $1$ to $K.$ When $%
\mu _{1}$ $\rightarrow 0$\ the model tends to a form that can be expressed
entirely in terms of differenced data, there are as many unit roots as
variables and there is no cointegration.

The \textquotedblleft single unit root" prior introduces a single dummy
observation such that all values of all variables are set equal to the
corresponding averages of initial conditions up to a scaling factor ($%
1/\lambda _{4}$). It is implemented by adding to the system in (\ref%
{MATRIXnotation}) the dummy variables $Y_{d_{2}}$ and $X_{d_{2}}$ with
generic elements:%
\begin{equation}
y_{d}(j)=\bar{y}_{0j}/\mu _{2};\ \ x_{d}(s)=\left\{ 
\begin{array}{c}
\bar{y}_{0j}/\mu _{2}\ \ \ \ \ \ \ \ \ \text{for}\ s<M \\ 
1/\mu _{2}\ \ \ \ \ \ \ \ \ \text{for}\ \ s=M,%
\end{array}%
\right.   \label{Init}
\end{equation}%
where $j$ goes from $1$ to $N$ while $s$ goes from $1$ to $K.$ As $\mu
_{2}\rightarrow 0$ the model tends to a form in which either all variables
are stationary with means equal to the sample averages of the initial
conditions, or there are unit root components without drift terms, which is
consistent with cointegration.

\subsubsection{Hyperparameters } \label{Hyperparameters }

To make the prior operational, one needs to choose the value of the
hyperparameters which control
the overall tightness of the Minnesota, sum of coefficients, and single unit
root priors. As mentioned above, in the natural conjugate NIW framework the
joint priors and joint posteriors are all matricvariate t distributions,
which implies that the marginal likelihood $p(Y)$ can be computed in closed
form simply by inverting Bayes formula and consists in the ratio of the
integrating constants of prior, likelihood, and posterior:%
\begin{equation}
p(Y)=\pi ^{\frac{-TN}{2}}\times \frac{|\Omega _{1}|^{\frac{N}{2}}}{|\Omega
_{0}|^{\frac{N}{2}}}\times \frac{|\bar{S}|^{-\frac{v_{0}+T}{2}}}{|S_{0}|^{-%
\frac{v_{0}}{2}}}\times \frac{\Gamma _{N}(\frac{v_{0}+T}{2})}{\Gamma _{N}(%
\frac{v_{0}}{2})},  
\end{equation}%
where $\Gamma _{N}(\cdot )$ denotes the $N$-variate gamma function.\ A
derivation based on theorem A.19 in \citetapp{BLR} can
be found in \citetapp{Carriero2012}. 

We follow \citetapp{Carriero2012} and \citet{KOROBILIS2019241} and choose the
vector of hyperparameters $\theta $\ by maximizing
the marginal data density: 
$\theta ^{\ast }=\arg \max_{\theta }\ln p(Y).$\footnote{\citetapp{Giannone:etal:RESTAT:2015} have developed a
hierarchical Bayes approach in which a prior is elicited for $\theta $ and its posterior distribution is derived via a random walk Metropolis step. This approach is more sophisticated but it requires MCMC methods, a feature that is less desirable when handling very large datasets. In our experience with the FRED-MD dataset, this approach and the simple maximization of the marginal likelihood produce very similar results.} 
As detailed above, the vector $\theta $\ comprises the parameters $\lambda _{1}$, $\mu _{1}$, and $\mu _{2}$ that control the overall tightness of the Minnesota, sum of coefficients, and single unit root priors, respectively.

\newpage
\FloatBarrier
\section{Data and Transformations} \label{App_data}

\setcounter{table}{0}
	\setcounter{figure}{0}

\input{appendix_data}

\newpage
\FloatBarrier
\section{Heterogeneity Analysis: Detailed Statistics by Variable Group}\label{App_heterogeneity}

\setcounter{table}{0}
	\setcounter{figure}{0}

\autoref{tab:RMSFE_by_group_model} provides detailed statistics (median, standard deviation, minimum, and maximum RMSFE ratios) for each of the eight FRED-MD variable groups across all models and forecast horizons. These statistics complement the visual analysis in \autoref{fig:boxplot_by_group_h1} and \autoref{fig:boxplot_by_group_h12}, providing numerical precision on the heterogeneity in forecast performance across economic domains.

\clearpage
{\footnotesize
\setlength{\LTpre}{0pt}
\setlength{\LTpost}{0pt}
\setlength{\tabcolsep}{3pt}
\renewcommand{\arraystretch}{0.85}
\input{tables/Table4_group_model_stats.tex}
}
\clearpage

\FloatBarrier
\section{Additional Results}\label{App_additional}

\FloatBarrier
\subsection{Are the Differences Statistically Significant?}\label{DMtest}

\setcounter{table}{0}
	\setcounter{figure}{0}

In the main body of the paper, we have considered the distribution of RMSFEs relative to an AR benchmark where the distribution was computed across all of 120 series in the dataset. The fact that such distribution was to the left (right) of the threshold of 1 meant that overall a model was producing better (worse) forecasts than the AR model, but no assessment was made on whether each forecast individually was statistically different from the one produced by the benchmark. In this section we tackle this specific issue. To save space, we only consider a subset of 19 key macroeconomic variables. A full set of result, covering all 120 variables, is available upon request. 

To provide a rough gauge of whether the RMSE ratios are significantly different from 1, we use the \citetapp{Diebold:Mariano:1995} t-statistic (DM). Our use of the DM test with models that are in some case nested is a deliberate choice. Monte Carlo evidence in \citetapp{CLARK2012JBES,CLARK2015160} indicates that, with nested models, the Diebold-Mariano test compared against normal critical values can be viewed as a somewhat conservative (conservative in the sense of tending
to have size modestly below nominal size) test for equal accuracy in the finite sample. Differences that are
statistically significant are denoted by one, two, or three stars, corresponding to significance levels of 10\%, 5\%, and 1\%, respectively. The underlying p-values are based on t-statistics computed with a serial correlation-robust variance, using a rectangular kernel, $h-1$ lags, and the small-sample adjustment of \citetapp{Harvey:etal:1997}.

Results for forecast horizons of one and three steps ahead are displayed in \autoref{tab:DM1_dupBART}. At these short forecast horizons the BVAR and factor model outperform the AR benchmark in many cases (12 and 14 cases respectively for BVAR and Factor model). In these instances, the forecasting gains are both large and statistically significant. In the remaining few cases in which the econometric models do not outperform the benchmark, the forecasting losses are small and never statistically significant. The \tsfm{}s outperform the AR in fewer instances (5 for \moirai and 6 for \timesfm), also with gains that are both large and statistically significant. In the remaining cases, the forecasting losses are not statistically significant. As we have seen before, there is a tendency for \tsfm{}s to produce the occasional more extreme results, with larger losses than those observed for the econometrics models. 

\begin{table}[!h]
  \centering
  \scriptsize
  \begin{tabular}{rcccccccccccc}
    \toprule
    & \multicolumn{12}{c}{$h=1$} \\
    \cmidrule{2-13}
    & \multicolumn{2}{c}{BVAR} 
    & \multicolumn{2}{c}{Factor model}      
    & \multicolumn{2}{c}{NNAR}      
    & \multicolumn{2}{c}{BART}      
    & \multicolumn{2}{c}{Moirai Large}  
    & \multicolumn{2}{c}{TimesFM}   \\
    \multicolumn{1}{l}{PAYEMS} & 0.92 & **  & 0.87 & **  & 0.90 & *** & 0.97 &       & 0.83 & *** & 0.82 & *** \\
\multicolumn{1}{l}{INDPRO} & 0.92 & *** & 0.93 & **  & 0.92 & *** & 1.02 &       & 0.95 & **  & 0.94 & **  \\
\multicolumn{1}{l}{FEDFUNDS} & 0.85 & *** & 0.91 &     & 2.34 &     & 1.34 &       & 1.00 &     & 1.23 &     \\
\multicolumn{1}{l}{UNRATE} & 0.92 & *** & 0.91 & **  & 1.03 &     & 0.94 &       & 1.04 &     & 1.11 &     \\
\multicolumn{1}{l}{RPI} & 0.98 &     & 0.98 &     & 1.02 &     & 1.05 &       & 1.00 &     & 1.04 &     \\
\multicolumn{1}{l}{DPCERA3M086SBEA} & 1.01 &     & 0.97 & **  & 1.04 &     & 1.03 &       & 1.02 &     & 1.06 &     \\
\multicolumn{1}{l}{CMRMTSPLx} & 1.01 &     & 0.97 &     & 1.05 &     & 1.01 &       & 1.01 &     & 1.04 &     \\
\multicolumn{1}{l}{CUMFNS} & 0.92 & *** & 0.96 &     & 1.93 &     & 0.96 &       & 1.09 &     & 1.48 &     \\
\multicolumn{1}{l}{CES0600000007} & 0.97 & *   & 0.93 & *   & 1.54 &     & 1.05 &       & 0.93 & *   & 0.94 & **  \\
\multicolumn{1}{l}{HOUST} & 0.99 &     & 0.94 & *** & 1.19 &     & 1.01 &       & 1.00 &     & 0.92 & *** \\
\multicolumn{1}{l}{S\&P 500} & 1.03 &     & 1.00 &     & 1.05 &     & 1.03 &       & 1.03 &     & 1.02 &     \\
\multicolumn{1}{l}{T1YFFM} & 0.99 &     & 1.10 &     & 1.33 &     & 1.28 &       & 1.04 &     & 1.13 &     \\
\multicolumn{1}{l}{T10YFFM} & 0.94 & **  & 1.04 &     & 1.24 &     & 1.09 &       & 1.04 &     & 1.16 &     \\
\multicolumn{1}{l}{BAAFFM} & 0.93 & *   & 0.97 &     & 1.19 &     & 1.07 &       & 1.01 &     & 1.28 &     \\
\multicolumn{1}{l}{EXUSUKx} & 1.04 &     & 1.01 &     & 1.06 &     & 1.05 &       & 1.02 &     & 1.01 &     \\
\multicolumn{1}{l}{WPSFD49207} & 0.98 &     & 1.03 &     & 1.00 &     & 1.03 &       & 1.01 &     & 1.01 &     \\
\multicolumn{1}{l}{PPICMM} & 1.06 &     & 0.99 &     & 1.06 &     & 1.02 &       & 1.03 &     & 1.04 &     \\
\multicolumn{1}{l}{PCEPI} & 1.04 &     & 0.98 &     & 0.98 &     & 1.02 &       & 0.99 &     & 0.95 & *   \\
\multicolumn{1}{l}{CES0600000008} & 0.84 & *** & 0.78 & *** & 0.80 & *** & 0.91 & ***   & 0.77 & *** & 0.78 & *** \\
    \cmidrule{2-13}
    & \multicolumn{12}{c}{$h=3$} \\
    \cmidrule{2-13}
    & \multicolumn{2}{c}{BVAR} 
    & \multicolumn{2}{c}{Factor model}      
    & \multicolumn{2}{c}{NNAR}      
    & \multicolumn{2}{c}{BART}      
    & \multicolumn{2}{c}{Moirai Large}  
    & \multicolumn{2}{c}{TimesFM}   \\
   \multicolumn{1}{l}{PAYEMS} & 0.89 & *    & 0.75 & *** & 0.82 & *** & 0.99 &     & 0.77 & *** & 0.75 & ** \\
\multicolumn{1}{l}{INDPRO} & 0.97 &      & 0.96 &     & 0.97 &     & 0.98 &     & 0.97 &     & 0.94 &    \\
\multicolumn{1}{l}{FEDFUNDS} & 0.87 & *** & 1.14 &     & 2.11 &     & 1.09 &     & 0.91 &     & 1.16 &    \\
\multicolumn{1}{l}{UNRATE} & 0.85 & *** & 0.78 & *   & 0.99 &     & 0.98 &     & 0.91 &     & 0.78 & ** \\
\multicolumn{1}{l}{RPI} & 0.99 & *    & 0.99 & *   & 0.99 &     & 1.02 &     & 1.00 &     & 1.01 &    \\
\multicolumn{1}{l}{DPCERA3M086SBEA} & 0.99 & *** & 0.99 &     & 0.98 & **  & 1.01 &     & 0.99 &     & 1.00 &    \\
\multicolumn{1}{l}{CMRMTSPLx} & 0.98 & *    & 0.98 &     & 0.97 & **  & 1.01 &     & 1.00 &     & 1.01 &    \\
\multicolumn{1}{l}{CUMFNS} & 0.88 & *** & 0.91 &     & 1.89 &     & 1.07 &     & 0.93 &     & 1.32 &    \\
\multicolumn{1}{l}{CES0600000007} & 0.80 & *** & 0.78 & *** & 2.13 &     & 0.80 & *** & 0.79 & *** & 0.83 & *** \\
\multicolumn{1}{l}{HOUST} & 0.93 & *    & 0.94 & *** & 1.48 &     & 1.04 &     & 1.01 &     & 0.78 & *** \\
\multicolumn{1}{l}{S\&P 500} & 1.01 &      & 1.01 &     & 1.02 &     & 1.03 &     & 1.00 &     & 1.03 &    \\
\multicolumn{1}{l}{T1YFFM} & 1.00 &      & 1.18 &     & 1.14 &     & 1.32 &     & 1.03 &     & 1.12 &    \\
\multicolumn{1}{l}{T10YFFM} & 0.89 & *** & 1.08 &     & 1.09 &     & 1.05 &     & 0.96 &     & 1.13 &    \\
\multicolumn{1}{l}{BAAFFM} & 0.92 & *    & 1.05 &     & 1.08 &     & 1.05 &     & 0.93 &     & 1.17 &    \\
\multicolumn{1}{l}{EXUSUKx} & 1.02 &      & 1.01 &     & 1.02 &     & 1.06 &     & 1.02 &     & 1.02 &    \\
\multicolumn{1}{l}{WPSFD49207} & 1.00 &      & 1.05 &     & 0.99 &     & 1.06 &     & 0.99 &     & 1.03 &    \\
\multicolumn{1}{l}{PPICMM} & 1.01 &      & 1.00 &     & 1.00 &     & 1.05 &     & 1.00 &     & 1.03 &    \\
\multicolumn{1}{l}{PCEPI} & 1.07 &      & 0.98 &     & 0.91 & *** & 0.98 &     & 0.89 & *** & 0.91 & *** \\
\multicolumn{1}{l}{CES0600000008} & 0.84 & *** & 0.76 & *** & 0.75 & *** & 0.91 & *** & 0.74 & *** & 0.75 & *** \\

    \bottomrule
  \end{tabular}
  \caption{Forecast horizons 1 and 3 step-ahead. RMSFE relative to the AR(1). Differences in accuracy that are statistically significant at 10\%, 5\%, and 1\% levels are denoted by one, two, or three stars, respectively. The evaluation sample is January 1985 to December 2019.}
  \label{tab:DM1_dupBART}
\end{table}

\begin{table}[!ht]
  \centering
  \scriptsize
  \begin{tabular}{rcccccccccccc}
    \toprule
    & \multicolumn{12}{c}{$h=6$} \\
    \cmidrule{2-13}
    & \multicolumn{2}{c}{BVAR} 
    & \multicolumn{2}{c}{Factor model}      
    & \multicolumn{2}{c}{NNAR}      
    & \multicolumn{2}{c}{BART}      
    & \multicolumn{2}{c}{Moirai Large}  
    & \multicolumn{2}{c}{TimesFM}   \\
    \multicolumn{1}{l}{PAYEMS} & 1.00 &       & 0.99 & **    & 0.99 & **    & 0.81 & **    & 0.99 & **    & 0.99 & * \\
\multicolumn{1}{l}{INDPRO} & 1.00 &       & 1.00 &       & 0.99 &       & 1.01 &       & 0.99 &       & 0.98 &  \\
\multicolumn{1}{l}{FEDFUNDS} & 0.94 &       & 1.02 &       & 1.99 &       & 1.36 &       & 0.93 &       & 1.21 &  \\
\multicolumn{1}{l}{UNRATE} & 0.97 & **    & 0.91 & *     & 1.01 &       & 0.85 &       & 0.95 &       & 0.94 & ** \\
\multicolumn{1}{l}{RPI}   & 1.00 &       & 1.00 &       & 1.00 &       & 1.03 &       & 1.00 &       & 1.01 &  \\
\multicolumn{1}{l}{DPCERA3M086SBEA} & 1.00 &       & 1.00 &       & 1.00 &       & 1.01 &       & 1.00 &       & 1.00 &  \\
\multicolumn{1}{l}{CMRMTSPLx} & 1.00 &       & 1.00 &       & 1.01 &       & 0.98 &       & 1.01 &       & 1.02 &  \\
\multicolumn{1}{l}{CUMFNS} & 0.91 & **    & 0.97 &       & 1.69 &       & 1.01 &       & 0.92 &       & 1.14 &  \\
\multicolumn{1}{l}{CES0600000007} & 0.73 & ***   & 0.78 & ***   & 2.43 &   & 0.87 & ***   & 0.76 & ***   & 0.79 & *** \\
\multicolumn{1}{l}{HOUST} & 0.88 & *     & 0.94 & **    & 1.80 &       & 0.99 &       & 0.98 &       & 0.66 & ** \\
\multicolumn{1}{l}{S\&P 500} & 1.01 &       & 1.01 &       & 1.01 &       & 1.05 &       & 1.00 &       & 1.03 &  \\
\multicolumn{1}{l}{T1YFFM} & 1.05 &       & 1.03 &       & 1.25 &       & 1.33 &       & 1.07 &       & 1.20 &  \\
\multicolumn{1}{l}{T10YFFM} & 0.89 & **    & 0.92 & *     & 1.13 &       & 1.09 &       & 0.97 &       & 1.14 &  \\
\multicolumn{1}{l}{BAAFFM} & 0.94 &       & 1.00 &       & 1.08 &       & 1.07 &       & 0.93 &       & 1.15 &  \\
\multicolumn{1}{l}{EXUSUKx} & 1.01 &       & 1.00 &       & 1.03 &       & 1.09 &       & 1.00 &       & 1.02 &  \\
\multicolumn{1}{l}{WPSFD49207} & 1.01 &       & 1.03 &       & 0.98 &       & 1.07 &       & 0.99 &       & 1.02 &  \\
\multicolumn{1}{l}{PPICMM} & 1.02 &       & 1.01 &       & 1.01 &       & 1.04 &       & 1.01 &       & 1.06 &  \\
\multicolumn{1}{l}{PCEPI} & 1.03 &       & 0.93 &       & 0.89 & ***      & 1.00 &       & 0.85 & ***   & 0.85 & *** \\
\multicolumn{1}{l}{CES0600000008} & 0.85 & ***   & 0.77 & ***   & 0.76 & ***   & 0.83 & ***   & 0.76 & ***   & 0.76 & *** \\

    \cmidrule{2-13}
    & \multicolumn{12}{c}{$h=12$} \\
    \cmidrule{2-13}
    & \multicolumn{2}{c}{BVAR} 
    & \multicolumn{2}{c}{Factor model}      
    & \multicolumn{2}{c}{NNAR}      
    & \multicolumn{2}{c}{BART}      
    & \multicolumn{2}{c}{Moirai Large}  
    & \multicolumn{2}{c}{TimesFM}   \\
    \multicolumn{1}{l}{PAYEMS} & 1.00 &       & 1.00 & **    & 1.00 &       & 1.00 &       & 1.00 &       & 0.99 &  \\
\multicolumn{1}{l}{INDPRO} & 1.00 &       & 1.00 &       & 1.00 &       & 1.04 &       & 1.01 &       & 0.99 &  \\
\multicolumn{1}{l}{FEDFUNDS} & 1.03 &       & 0.95 &       & 1.78 &       & 0.86 &       & 1.00 &       & 1.08 &  \\
\multicolumn{1}{l}{UNRATE} & 0.96 & *     & 0.85 & **    & 1.03 &       & 0.99 &       & 0.96 &       & 0.89 & * \\
\multicolumn{1}{l}{RPI}   & 1.00 &       & 1.00 &       & 1.00 &       & 1.01 &       & 1.00 &       & 1.00 &  \\
\multicolumn{1}{l}{DPCERA3M086SBEA} & 1.00 &       & 1.00 &       & 1.00 &       & 1.01 &       & 1.00 &       & 1.00 &  \\
\multicolumn{1}{l}{CMRMTSPLx} & 1.00 &       & 1.01 &       & 1.01 &       & 1.03 &       & 1.02 &       & 1.02 &  \\
\multicolumn{1}{l}{CUMFNS} & 0.90 & *     & 0.98 &       & 1.80 &       & 1.08 &       & 0.92 &       & 1.02 &  \\
\multicolumn{1}{l}{CES0600000007} & 0.73 & ***   & 0.78 & ***   & 3.08 &       & 0.80 & ***   & 0.76 & ***   & 0.78 & *** \\
\multicolumn{1}{l}{HOUST} & 0.86 &       & 0.99 &       & 2.20 &       & 1.18 &       & 0.96 &       & 0.48 & ** \\
\multicolumn{1}{l}{S\&P 500} & 1.00 &       & 1.00 &       & 1.01 &       & 1.04 &       & 1.01 &       & 1.05 &  \\
\multicolumn{1}{l}{T1YFFM} & 1.15 &       & 1.06 &       & 1.32 &       & 1.33 &       & 1.11 &       & 1.25 &  \\
\multicolumn{1}{l}{T10YFFM} & 0.97 &       & 0.87 & ***   & 1.13 &       & 0.88 & **    & 1.03 &       & 1.15 &  \\
\multicolumn{1}{l}{BAAFFM} & 0.97 &       & 0.94 &       & 1.14 &       & 0.82 &       & 0.96 &       & 1.22 &  \\
\multicolumn{1}{l}{EXUSUKx} & 1.00 &       & 1.00 &       & 1.01 &       & 1.09 &       & 1.03 &       & 1.00 &  \\
\multicolumn{1}{l}{WPSFD49207} & 1.03 &       & 1.03 &       & 0.98 &       & 1.05 &       & 1.00 &       & 1.03 &  \\
\multicolumn{1}{l}{PPICMM} & 1.01 &       & 1.01 &       & 1.01 &       & 1.04 &       & 1.01 &       & 1.03 &  \\
\multicolumn{1}{l}{PCEPI} & 1.02 &       & 0.91 &       & 0.87 & ***   & 0.94 &       & 0.82 & ***   & 0.83 & *** \\
\multicolumn{1}{l}{CES0600000008} & 0.88 & ***   & 0.80 & ***   & 0.76 & ***   & 0.91 & **    & 0.78 & ***   & 0.80 & *** \\

    \bottomrule
  \end{tabular}
  \caption{Forecast horizons 6 and 12 step-ahead. RMSFE relative to the AR(1). Differences in accuracy that are statistically significant at 10\%, 5\%, and 1\% levels are denoted by one, two, or three stars, respectively. The evaluation sample is January 1985 to December 2019.}
  \label{tab:DM2_dupBART}
\end{table}

Results for forecast horizons of 6 and 12 step-ahead are displayed in \autoref{tab:DM2_dupBART}. At these horizons, for the 19 series in this medium dataset, improvements versus the benchmark are less clear-cut. In general, neither the econometric models nor the \tsfm{}s systematically outperform the AR benchmark, even though it is worth noting that there are no instances in which they significantly under perform. Overall, \moirai is the model that more often produced better forecasts (in 13 cases at the 6-step ahead and 10 cases at the 12-step ahead) followed closely by the BVAR and factor model. For some series, there are significant gains. In particular the \tsfm{}s produce good forecasts for Personal Consumption Expenditures (PCEPI) and Hourly Earnings (CES0600000008), with gains that are significant and larger than those obtained by the econometric models. The econometric models produce good forecasts of Unemployment rate (UNRATE) and Weekly hours (CES0600000007).

\FloatBarrier
\subsection{Forecast Accuracy During the Covid-19 Pandemic}\label{App_covid}

\setcounter{table}{0}
	\setcounter{figure}{0}

The results presented in the main text gave an overall picture of the accuracy of the various models across the different forecast horizons, focusing on the evaluation sample January 1985 to December 2019. In this appendix, we focus instead on the post-Covid-19 period sample, January 2020 to December 2022. We exclude from this analysis the forecasts made between March and June 2020, to reduce the impact on the results of the extreme outliers right around the onset of the pandemic.\footnote{The choice of excluding from the sample the most problematic Covid-19 related observation is based exclusively on simplicity. Still, there is evidence that simple approaches such as this might work well, see \citetapp{SchorfheideSong:2021:nber:var}. One could also consider more sophisticated ways to deal with the Covid-19 observations such as \citetapp{CCMM:RESTAT:svo} or \citetapp{LenzaPrimiceri:2022:jae:VARafter2020}.}

\begin{figure}[t!]
    \centering
    \includegraphics[width=0.78\textwidth]{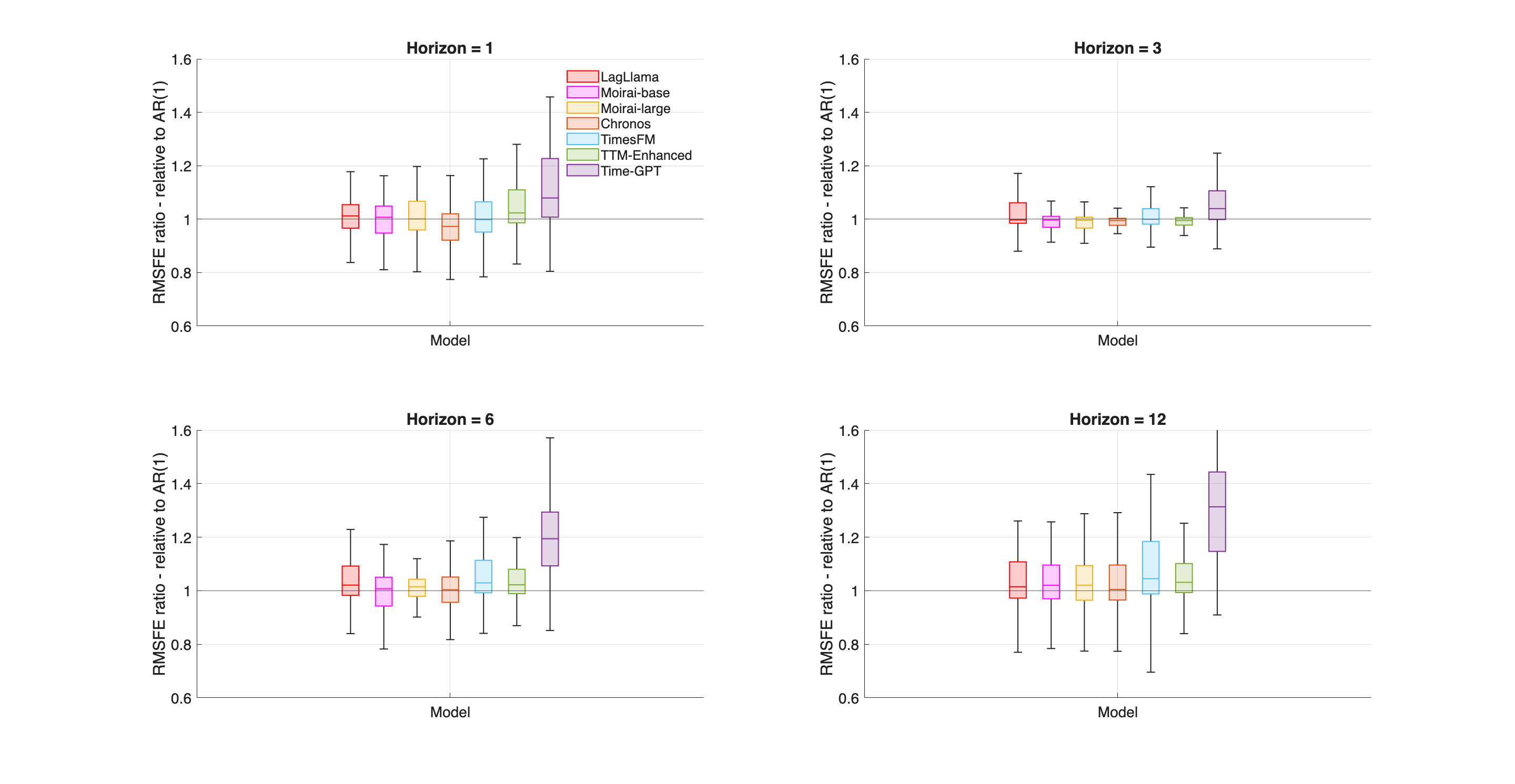}
    \caption{Distribution of RMSFEs (relative to an AR benckmark) for the \tsfm{}s. The evaluation sample is January 2020 to December 2022 (excluding the months from March to June 2020).}
    \label{fig:Covid-19_boxplot_llm_only}
\end{figure}

\begin{figure}[t!]
    \centering
    \includegraphics[width=0.78\textwidth]{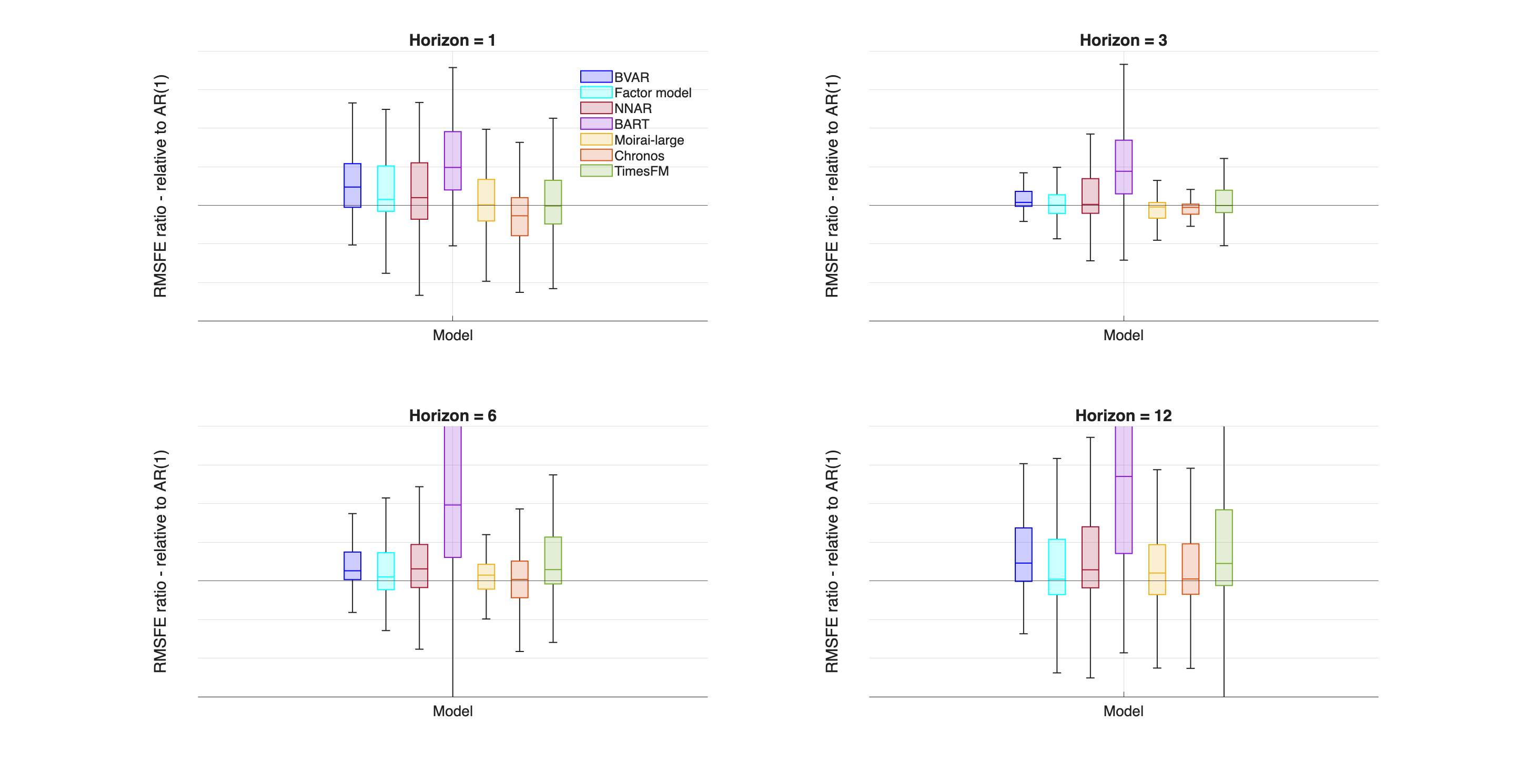}
    \caption{Distribution of RMSFEs (relative to an AR benckmark) for the econometric models and a selection of \tsfm{}s. The evaluation sample is January 2020 to December 2022 (excluding the months from March to June 2020).}
    \label{fig:Covid-19_boxplot_econometric_and_best_llm}
\end{figure}

\autoref{fig:Covid-19_boxplot_llm_only} compares the forecast accuracy of the various \tsfm{}s, and \autoref{fig:Covid-19_boxplot_econometric_and_best_llm} considers the best performing \tsfm{}s against the various econometric models.

Compared to what we shown in \autoref{fig:boxplot_llm_only} and \autoref{fig:boxplot_econometric_and_best_llm} there is an overall deterioration in forecast accuracy for all the models, which all struggle in overperforming the benchmark. Within the \tsfm{}s, only \moirai-\largemodel and \chronos stand out, with a performance comparable with that of the factor model and the BVAR. In this sample the econometric models tend to make at times large forecast errors, mainly during the second half of 2020. For the factor model it happens especially for interest rates and spreads.

The usual caveat applies that, as shown in \autoref{tab:TSLM_training_details}, namely that the \tsfm{}s include the pandemic period in their training sample (and a few also include many of the variables we set out to forecasts), and this makes it harder to pin down to which extent this forecasting performance could have been achieved in real time.

\FloatBarrier
\subsection{Does Fine-Tuning Improve on Zero-Shot Forecasts?}\label{App_finetuning}

\setcounter{table}{0}
	\setcounter{figure}{0}

Up until now, all \tsfm{}s results have been based on zero-shot predictions of the original pretrained models. As discussed in Appendix~\ref{App_TSFM_components}, alternatively one could consider the option of fine-tuning the \tsfm{}s, i.e. updating the pretrained model's parameters on a task-specific dataset, with the goal of improving forecasting performance. Broadly speaking, to implement fine-tuning one could follow one of two approaches, (i) Continued-pretraining or (ii) Parameter-efficient fine-tuning.\footnote{See \citeapp{han2024parameter} for a detailed review.} Continued-pretraining involves further training the model on a domain-specific dataset starting with pretrained parameters, and allowing all model parameters to adapt to the characteristics of the given data. This can lead to overfitting when the fine-tuning data is small. Parameter-efficient methods instead freeze some of the original model's parameters and introduce trainable low-rank matrices (e.g. low-rank adaptation (LoRA) methods as in \citepapp{hu2022lora}), significantly reducing the number of trainable parameters, addressing overfitting, and requiring comparatively less computational resources.

Here, we selected the two best models from our zero-shot experiments -- \timesfm and \moirai-\largemodel{} -- and fine-tuned them on \fred data. Note that some of the pretrained \tsfm{s}, including \moirai, are trained on datasets that overlap with the \fred. However, the influence of any single dataset in the pretraining phase gets diluted within the vast and diverse pretraining corpus. Consequently, the models learn general time series patterns, rather than specializing in the specific characteristics of \fred. To make a \tsfm more dataset-specific, fine-tuning is required.
The main challenge we face in this context is that FRED-MD is a relatively small dataset, which may lead to overfitting problems.\footnote{\citetapp{kumar2022fine} discusses how in these situations the model may also lose the knowledge acquired during pretraining, leading to the so-called catastrophic forgetting.} To address these challenges, we employ a simple parameter-efficient fine-tuning strategy as opposed to continued-pretraining. In particular, we freeze the transformer layers of the two \tsfm{}s, thereby reducing the number of trainable parameters.\footnote{\citetapp{kumar2022fine} refers to this as Linear Probing.} Note that freezing the pretrained model's weights (parameters) is a special case of LoRA, which additionally injects trainable rank decomposition matrices into each layer of \tsfm. After freezing the transformer layers, the number of trainable parameters reduced to approximately $\leq 10$M for both \timesfm (instead of $200$M) and \moirai-\largemodel (instead of $311$M), primarily part of the final layers.\footnote{In the case of \moirai, we also consider fine-tuning using the continued-pretraining strategy, as this was the recommended approach provided by the authors in the model's repository, but in this case the model ended up severely overfitting and producing poor out-of-sample forecasts.} Throughout, to keep computational costs manageable, we chose to fine-tune each model once every 12 months, and the resulting models were then used for zero-shot forecasting over the subsequent 12 months.\footnote{Within each fine-tuning round, the training \fred data was split into an 80:20 ratio, with 20\% reserved for validation. We trained these models for 200 epochs, utilizing early stopping with a patience of 5 to mitigate overfitting. To further counter overfitting and ensure stable training on our limited dataset, we reduced the learning rate, preventing large weight updates.} We run the fine-tuning experiment on a stock server equipped with a single NVIDIA A6000 GPU. In terms of wall-clock time, fine-tuning took about 10 hours for each of the two models. Hence, comparatively, fine-tuning remains significantly more resource-intensive than the BVARs and factor models we considered, which can be trained on standard CPUs in a fraction of the time.\footnote{We also considered the continued-pretraining option for the same models, and in that case the total computing time was approximately 90 hours.}

\begin{figure}[t!]
    \centering
    \includegraphics[width=0.78\textwidth]{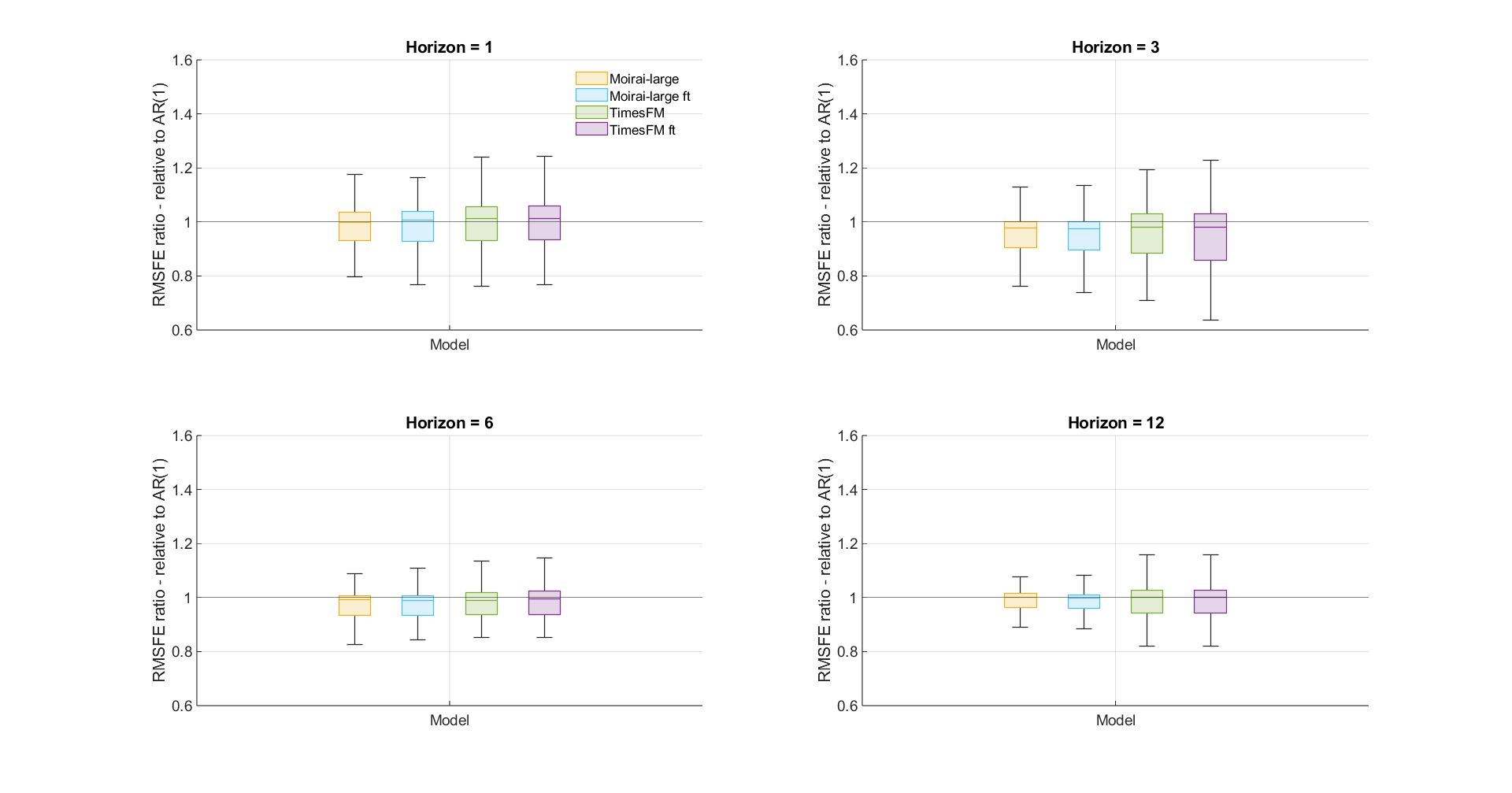}
    \caption{Distribution of RMSFEs (relative to an AR benckmark) for \timesfm and \moirai-large for zero shot and fine-tuning. The evaluation sample is January 1985 to December 2019.}
    \label{finetuning}
\end{figure}

\autoref{finetuning} shows the comparison of zero-shot and fine-tuned performance for \timesfm and \moirai-\largemodel. We observe that fine-tuning yields results that are similar or slightly better than the zero-shot performance for both models.
This result overall isn't particularly surprising for our setup: a large model and a small dataset.
While fine-tuning offers the potential for adapting models to a domain, it comes with its own set of challenges, such as hyperparameter optimization and increased computational resource (e.g. GPUs) requirements, thus needs careful consideration. Given the marginal performance gains achieved through fine-tuning in our context, the inherent "plug-and-play" simplicity of zero-shot forecasting presents itself as a potentially more practical approach.

\subsection{Detailed Forecast Comparisons for Persistent Series}\label{App_series_comparisons}

\setcounter{table}{0}
	\setcounter{figure}{0}

\begin{figure}[t!]
    \centering
    \includegraphics[width=0.78\textwidth]{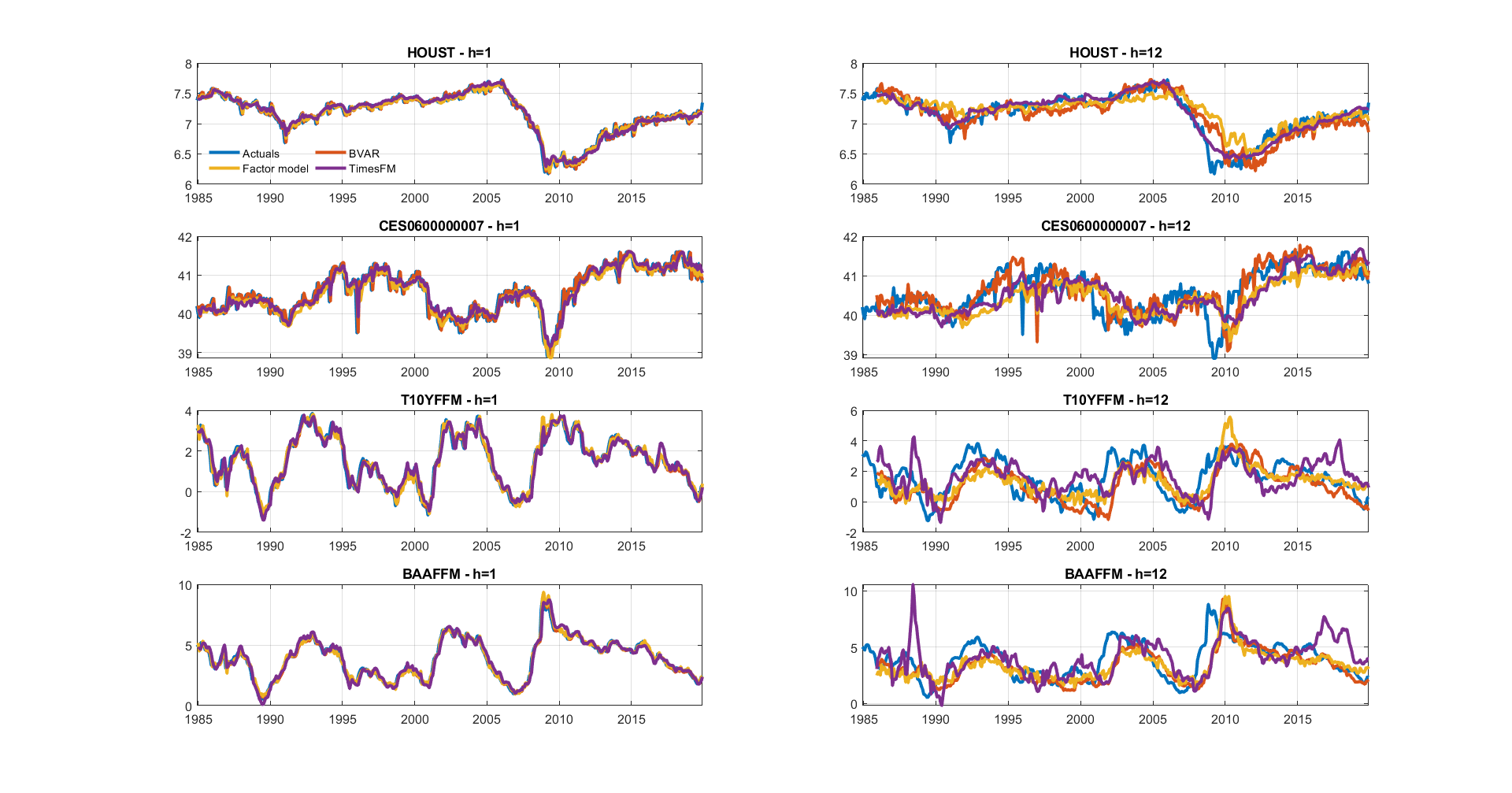}
    \caption{Time series of out-of-sample forecasts produced by the factor model, the BVAR, and the \timesfm model for selected variables. Left panels show 1-step ahead forecasts, and right panels show 12-step ahead forecasts. The sample is January 1985 to December 2019.}
    \label{fig:comparisons}
\end{figure}

To give concrete examples of the extent of variation in performance across different types of series, \autoref{fig:comparisons} shows the time series of out-of-sample forecasts produced by the factor model, the BVAR and \timesfm for a few selected variables. While for the 1-step ahead horizon the forecasts are quite similar (left side panels), large differences emerge at the 12-step ahead horizon (right side panels).

For Housing Starts (HOUST), the 12-step ahead forecasts produced by \timesfm are remarkably good, as this model is able to almost perfectly predict the pronounced slowdown that happened during the 2007 financial crisis and the ensuing recession. This results in an astonishing 50 percent improvement in RMSFE compared to the AR benchmark. Instead, the econometric models only pick up this slowdown with a delay, and the factor model even fails to recognize the depth of the trough. Of course, one must keep in mind that \timesfm includes this recessionary time period in its training sample, so it seems to be on some level aware that a change in time series behavior is to be expected within these dates. For weekly hours (CES0600000007) \timesfm outperforms the factor model but underperforms the BVAR. For variables such as the spread between the 10 year interest rate and the Federal Funds rate (T10YFFM) and the spread between corporate bonds and the Federal Funds rate (BAAFFM), the performance of \tsfm{}s is subpar, featuring entire periods of "hallucination" in the second half of the 80s and in 2017-2019, resulting in RMSFE losses on the order of 15-20 percent.

\bibliographystyleapp{chicago}
{\footnotesize
\singlespacing
\setlength{\bibsep}{4pt}
\bibliographyapp{All_references_LLMs}
}

\end{appendices}

\end{document}

%% file: dfn.tex
\newcommand{\tsfm}{{\scshape{TSFM}}\xspace}
\newcommand{\dl}{{\scshape{DL}}\xspace}

\newcommand{\llm}{{LLM}\xspace}
\newcommand{\tslm}{{TSLM}\xspace}

\newcommand{\lagllama}{{ LagLlama}\xspace} 
\newcommand{\moirai}{{Moirai}\xspace}
\newcommand{\timesfm}{{TimesFM}\xspace}
\newcommand{\timellm}{{Time-LLM}\xspace}
\newcommand{\ttm}{{TTM}\xspace}

\newcommand{\timegpt}{{Time-GPT}\xspace}
\newcommand{\moment}{{MOMENT}\xspace}
\newcommand{\chronos}{{Chronos}\xspace}

\newcommand{\largemodel}{{large}\xspace}

\newcommand{\fred}{{FRED-MD}\xspace}

\newcommand{\xmark}{\ding{55}}%
\newcommand{\cmark}{\ding{51}}

%% file: tables/Table2_TSFM_stats.tex
\begin{tabular}{lcccc|cccc}
\toprule
 & \multicolumn{4}{c|}{\textbf{h=1}} & \multicolumn{4}{c}{\textbf{h=3}} \\
\cmidrule{2-9} & \textbf{Median} & \textbf{Std} & \textbf{Min } & \textbf{Max} & \textbf{Median} & \textbf{Std} & \textbf{Min } & \textbf{Max} \\
\textbf{Lag-Llama} & 1.015 & 1.057 & 0.726 & 7.271 & 0.997 & 0.787 & 0.633 & 4.843 \\
\textbf{Moirai-base} & 1.008 & 0.097 & 0.704 & 1.204 & 0.973 & 0.100 & 0.634 & 1.107 \\
\textbf{Moirai-large} & 0.999 & 0.102 & 0.703 & 1.436 & 0.978 & 0.099 & 0.637 & 1.158 \\
\textbf{Chronos} & 0.961 & 0.075 & 0.715 & 1.086 & 0.966 & 0.092 & 0.632 & 1.068 \\
\textbf{TimesFM} & 1.014 & 0.129 & 0.706 & 1.482 & 0.980 & 0.127 & 0.635 & 1.318 \\
\textbf{TTM-Enhanced} & 1.044 & 0.352 & 0.723 & 2.959 & 0.993 & 0.108 & 0.718 & 1.448 \\
\textbf{TimeGPT} & 1.077 & 0.124 & 0.745 & 1.531 & 1.044 & 0.134 & 0.672 & 1.278 \\
\hline
 & \multicolumn{4}{c|}{\textbf{h=6}} & \multicolumn{4}{c}{\textbf{h=12}} \\
\cmidrule{2-9} & \textbf{Median} & \textbf{Std} & \textbf{Min } & \textbf{Max} & \textbf{Median} & \textbf{Std} & \textbf{Min } & \textbf{Max} \\
\textbf{Lag-Llama} & 1.002 & 0.461 & 0.568 & 3.577 & 1.009 & 0.260 & 0.597 & 2.431 \\
\textbf{Moirai-base} & 0.990 & 0.093 & 0.567 & 1.109 & 0.999 & 0.098 & 0.594 & 1.168 \\
\textbf{Moirai-large} & 0.991 & 0.096 & 0.600 & 1.159 & 1.001 & 0.113 & 0.619 & 1.324 \\
\textbf{Chronos} & 0.980 & 0.090 & 0.573 & 1.097 & 0.993 & 0.095 & 0.581 & 1.083 \\
\textbf{TimesFM} & 0.990 & 0.142 & 0.593 & 1.629 & 1.001 & 0.158 & 0.482 & 1.440 \\
\textbf{TTM-Enhanced} & 0.995 & 0.097 & 0.643 & 1.400 & 1.002 & 0.198 & 0.663 & 2.257 \\
\textbf{TimeGPT} & 1.025 & 0.140 & 0.600 & 1.363 & 1.063 & 0.169 & 0.611 & 1.515 \\
\bottomrule
\end{tabular}

%% file: tables/Table3_econometric_stats.tex
\begin{tabular}{lcccc|cccc}
\toprule
 & \multicolumn{4}{c|}{\textbf{h=1}} & \multicolumn{4}{c}{\textbf{h=3}} \\
\cmidrule{2-9} & \textbf{Median} & \textbf{Std} & \textbf{Min } & \textbf{Max} & \textbf{Median} & \textbf{Std} & \textbf{Min } & \textbf{Max} \\
\textbf{BVAR} & 0.983 & 0.060 & 0.827 & 1.158 & 0.970 & 0.060 & 0.767 & 1.089 \\
\textbf{Factor model} & 0.965 & 0.065 & 0.766 & 1.103 & 0.980 & 0.102 & 0.682 & 1.183 \\
\textbf{NNAR} & 0.988 & 0.234 & 0.733 & 1.910 & 0.991 & 0.132 & 0.653 & 1.368 \\
\textbf{BART} & 1.017 & 0.093 & 0.858 & 1.420 & 1.024 & 0.132 & 0.763 & 1.381 \\
\textbf{Moirai-large} & 0.999 & 0.102 & 0.703 & 1.436 & 0.978 & 0.099 & 0.637 & 1.158 \\
\textbf{Chronos} & 0.961 & 0.075 & 0.715 & 1.086 & 0.966 & 0.092 & 0.632 & 1.068 \\
\textbf{TimesFM} & 1.014 & 0.129 & 0.706 & 1.482 & 0.980 & 0.127 & 0.635 & 1.318 \\
\hline
 & \multicolumn{4}{c|}{\textbf{h=6}} & \multicolumn{4}{c}{\textbf{h=12}} \\
\cmidrule{2-9} & \textbf{Median} & \textbf{Std} & \textbf{Min } & \textbf{Max} & \textbf{Median} & \textbf{Std} & \textbf{Min } & \textbf{Max} \\
\textbf{BVAR} & 0.991 & 0.068 & 0.725 & 1.106 & 0.999 & 0.080 & 0.671 & 1.162 \\
\textbf{Factor model} & 0.992 & 0.087 & 0.602 & 1.105 & 0.999 & 0.089 & 0.633 & 1.125 \\
\textbf{NNAR} & 0.999 & 0.111 & 0.591 & 1.229 & 1.000 & 0.127 & 0.608 & 1.418 \\
\textbf{BART} & 1.011 & 0.131 & 0.757 & 1.457 & 1.023 & 0.154 & 0.670 & 1.905 \\
\textbf{Moirai-large} & 0.991 & 0.096 & 0.600 & 1.159 & 1.001 & 0.113 & 0.619 & 1.324 \\
\textbf{Chronos} & 0.980 & 0.090 & 0.573 & 1.097 & 0.993 & 0.095 & 0.581 & 1.083 \\
\textbf{TimesFM} & 0.990 & 0.142 & 0.593 & 1.629 & 1.001 & 0.158 & 0.482 & 1.440 \\
\bottomrule
\end{tabular}

%% file: tables/Table5_newgen_stats.tex
\begin{tabular}{lcccc|cccc}
\toprule
 & \multicolumn{4}{c|}{\textbf{h=1}} & \multicolumn{4}{c}{\textbf{h=3}} \\
\cmidrule{2-9} & \textbf{Median} & \textbf{Std} & \textbf{Min } & \textbf{Max} & \textbf{Median} & \textbf{Std} & \textbf{Min } & \textbf{Max} \\
\textbf{BVAR} & 0.983 & 0.060 & 0.827 & 1.158 & 0.970 & 0.060 & 0.767 & 1.089 \\
\textbf{Factor model} & 0.965 & 0.065 & 0.766 & 1.103 & 0.980 & 0.102 & 0.682 & 1.183 \\
\textbf{Moirai (new)} & 0.968 & 0.073 & 0.715 & 1.068 & 0.961 & 0.094 & 0.612 & 1.063 \\
\textbf{Moirai (old)} & 0.999 & 0.102 & 0.703 & 1.436 & 0.978 & 0.099 & 0.637 & 1.158 \\
\textbf{Chronos (new)} & 0.945 & 0.082 & 0.697 & 1.101 & 0.937 & 0.098 & 0.614 & 1.035 \\
\textbf{Chronos (old)} & 0.961 & 0.075 & 0.715 & 1.086 & 0.966 & 0.092 & 0.632 & 1.068 \\
\hline
 & \multicolumn{4}{c|}{\textbf{h=6}} & \multicolumn{4}{c}{\textbf{h=12}} \\
\cmidrule{2-9} & \textbf{Median} & \textbf{Std} & \textbf{Min } & \textbf{Max} & \textbf{Median} & \textbf{Std} & \textbf{Min } & \textbf{Max} \\
\textbf{BVAR} & 0.991 & 0.068 & 0.725 & 1.106 & 0.999 & 0.080 & 0.671 & 1.162 \\
\textbf{Factor model} & 0.992 & 0.087 & 0.602 & 1.105 & 0.999 & 0.089 & 0.633 & 1.125 \\
\textbf{Moirai (new)} & 0.978 & 0.098 & 0.558 & 1.117 & 0.995 & 0.108 & 0.569 & 1.189 \\
\textbf{Moirai (old)} & 0.991 & 0.096 & 0.600 & 1.159 & 1.001 & 0.113 & 0.619 & 1.324 \\
\textbf{Chronos (new)} & 0.967 & 0.101 & 0.558 & 1.077 & 0.979 & 0.107 & 0.575 & 1.102 \\
\textbf{Chronos (old)} & 0.980 & 0.090 & 0.573 & 1.097 & 0.993 & 0.095 & 0.581 & 1.083 \\
\bottomrule
\end{tabular}

%% file: appendix_data.tex
{\scriptsize
	\begin{longtable}{llc|cc}
		\toprule
		\textbf{Abbreviation} & \textbf{Description} & \textbf{Transf. code} & \textbf{Small set} & \textbf{Full set} \\
		\hline
		PAYEMS & All Employees: Total nonfarm & 5 & $\mathrm{x}$ & $\mathrm{x}$ \\ 
		INDPRO & IP Index & 5 & $\mathrm{x}$ & $\mathrm{x}$ \\
		FEDFUNDS & Effective Federal Funds Rate & 1 & $\mathrm{x}$ & $\mathrm{x}$ \\
		UNRATE & Civilian Unemployment Rate & 1 & $\mathrm{x}$ & $\mathrm{x}$ \\
		RPI & Real personal income & 5 & $\mathrm{x}$ & $\mathrm{x}$ \\
		DPCERA3M086SBEA & Real  PCE & 5 & $\mathrm{x}$ & $\mathrm{x}$ \\
		CMRMTSPLx & Real Manu. and TradeIndustries Sales & 5 & $\mathrm{x}$ & $\mathrm{x}$ \\
		CUMFNS & Capacity Utilization: Manufacturing & 1 & $\mathrm{x}$ & $\mathrm{x}$\\
		CES0600000007 & Avg Weekly Hours: Goods-Producing & 1 & $\mathrm{x}$ & $\mathrm{x}$\\
		HOUST & Housing Starts, Total & 4 & $\mathrm{x}$ & $\mathrm{x}$\\
		S\&P 500 & S\&P's Common Stock Price Index: Composite & 5 & $\mathrm{x}$ & $\mathrm{x}$ \\
		T1YFFM & 1-Year Treasury C Minus FEDFUNDS & 1 & $\mathrm{x}$ & $\mathrm{x}$\\
		T10YFFM & 10-Year Treasury C Minus FEDFUNDS & 1 & $\mathrm{x}$ & $\mathrm{x}$\\
		BAAFFM & Moodys Baa Corporate Bond Minus FEDFUNDS & 1 & $\mathrm{x}$ & $\mathrm{x}$\\
		EXUSUKx & U.S.-UK Foreign Exchange Rate & 5 & $\mathrm{x}$ & $\mathrm{x}$ \\
		WPSFD49207 & PPI: Final Demand: Finished Goods & 5 & $\mathrm{x}$ & $\mathrm{x}$\\
		PPICMM & PPI: Metals and metal products & 5 & $\mathrm{x}$ & $\mathrm{x}$\\
		PCEPI & Personal Consumption Expenditures & 5 & $\mathrm{x}$ & $\mathrm{x}$\\
		CES0600000008 & Avg Hourly Earnings: Goods-Producing & 6 & $\mathrm{x}$ & $\mathrm{x}$\\
		\hline
		HWI & Help-Wanted Index for U & 1 & & $\mathrm{x}$\\
		HWIURATIO & Help Wanted to Unemployed ratio & 1 & & $\mathrm{x}$\\
		CLF16OV & Civilian Labor Force & 5 & & $\mathrm{x}$\\
		M1SL & M1 Money Stock & 5 & & $\mathrm{x}$\\
		M2SL & M2 Money Stock & 5 & & $\mathrm{x}$\\
		M2REAL & Real M2 Money Stock & 5 & & $\mathrm{x}$\\
		S\&P div yield & S\&P's Composite Common Stock: Dividend Yield & 1 & & $\mathrm{x}$\\
		S\&P PE ratio & S\&P's Common Stock: Price/Earnings ratio & 5 & & $\mathrm{x}$\\
		TB6MS & 6-Month Treasury Bill & 1 & & $\mathrm{x}$\\
		GS1 & 1-Year Treasury Rate & 1 & & $\mathrm{x}$\\
		GS5 & 5-Year Treasury Rate & 1 & & $\mathrm{x}$\\
		AAA & Moody's Seasoned Aaa Corporate Bond Yield & 1 & & $\mathrm{x}$\\
		BAA & Moody's Seasoned Baa Corporate Bond Yield & 1 & & $\mathrm{x}$\\
		OILPRICEx & Crude Oil, spliced WTI and Cushing & 5 & & $\mathrm{x}$\\
		INVEST & Securities in Bank Credit at All Commercial Banks & 5 & & $\mathrm{x}$\\
		\hline
		W875RX1 & Real personal income ex transfer receipts & 5 & & $\mathrm{x}$\\
		RETAILx & Retail and Food Services Sales & 5 & & $\mathrm{x}$\\
		IPFPNSS & IP: Final Products & 5 & & $\mathrm{x}$\\
		IPFINAL & IP: Final Products (Market Group) & 5 & & $\mathrm{x}$\\
		IPCONGD & IP: Consumer Goods & 5 & & $\mathrm{x}$\\
		IPDCONGD & IP: Durable Consumer Goods & 5 & & $\mathrm{x}$\\
		IPNCONGD & IP: Non-Durable Consumer Goods & 5 & & $\mathrm{x}$\\
		IPBUSEC & IP: Business \& Equipment & 5 & & $\mathrm{x}$\\
		IPMAT & IP: Materials & 5 & & $\mathrm{x}$\\
		IPDMAT & IP: Durable Goods Materials & 5 & & $\mathrm{x}$\\
		IPNMAT & IP: Non-Durable Goods Materials & 5 & & $\mathrm{x}$\\
		IPMANSICS & IP: Manufacturing (SIC) & 5 & & $\mathrm{x}$\\
		IPB51222S & IP: Residential Utilities & 5 & & $\mathrm{x}$\\
		IPFUELS & IP: Fuels & 5 & & $\mathrm{x}$\\
		CE16OV & Civilian Employment & 5 & & $\mathrm{x}$\\
		UEMPMEAN & Average Duration of Unemployment (Weeks) & 1 & & $\mathrm{x}$\\
		UEMPLT5 & Civilians Unemployed: Less Than 5 Weeks & 5 & & $\mathrm{x}$\\
		UEMP5TO14 & Civilians Unemployed for 5-14 Weeks & 5 & & $\mathrm{x}$\\
		UEMP15OV & Civilians Unemployed: 15 Weeks \& Over & 5 & & $\mathrm{x}$\\
		UEMP15T26 & Civilians Unemployed for 15-26 Weeks & 5 & & $\mathrm{x}$\\
		UEMP27OV & Civilians Unemployed for 27 Weeks and Over & 5 & & $\mathrm{x}$\\
		CLAIMSx & Initial Claims & 5 & & $\mathrm{x}$\\
		USGOOD & All Employees: Goods-Producing Industries & 5 & & $\mathrm{x}$\\
		CES1021000001 & All Employees: Mining and Logging: Mining & 5 & & $\mathrm{x}$\\
		USCONS & All Employees: Construction & 5 & & $\mathrm{x}$\\
		MANEMP & All Employees: Manufacturing & 5 & & $\mathrm{x}$\\
		DMANEMP & All Employees: Durable goods & 5 & & $\mathrm{x}$\\
		NDMANEMP & All Employees: Nondurable goods & 5 & & $\mathrm{x}$\\
		SRVPRD & All Employees: Service-Providing Industries & 5 & & $\mathrm{x}$\\
		USTPU & All Employees: TT\&U & 5 & & $\mathrm{x}$\\
		USWTRADE & All Employees: Wholesale Trade & 5 & & $\mathrm{x}$\\
		USTRADE & All Employees: Retail Trade & 5 & & $\mathrm{x}$\\
		USFIRE & All Employees: Financial Activities & 5 & & $\mathrm{x}$\\
		USGOVT & All Employees: Government & 5 & & $\mathrm{x}$\\
		AWOTMAN & Avg Weekly Overtime Hourse: Manufacturing & 1 & & $\mathrm{x}$\\
		AWHMAN & Avg Weekly Hours: Manufacturing & 1 & & $\mathrm{x}$\\
		HOUSTNE & Housing Starts, Northeast & 4 & & $\mathrm{x}$\\
		HOUSTMW & Housing Starts, Midwest & 4 & & $\mathrm{x}$\\
		HOUSTS & Housing Starts, South & 4 & & $\mathrm{x}$\\
		HOUSTW & Housing Starts, West & 4 & & $\mathrm{x}$\\ 
		PERMIT & New Private Housing Permits (SAAR) & 4 & & $\mathrm{x}$\\
		PERMITNE & New Private Housing Permits, Northeast (SAAR) & 4 & & $\mathrm{x}$\\
		PERMITMW & New Private Housing Permits, Midwest (SAAR) & 4 & & $\mathrm{x}$\\
		PERMITS & New Private Housing Permits, South (SAAR) & 4 & & $\mathrm{x}$\\
		PERMITW & New Private Housing Permits, West (SAAR) & 4 & & $\mathrm{x}$\\
		AMDMNOx & New Orders for Durable goods & 5 & & $\mathrm{x}$\\
		AMDMUOx & Unfilled Orders for Durable goods & 5 & & $\mathrm{x}$\\
		BUSINVx & Total Business Inventories & 5 & & $\mathrm{x}$\\
		ISRATIOx & Total Business: Inventories to Sales Ratio & 1 & & $\mathrm{x}$\\
		BOGMBASE & Monetary Base: Total & 5 & & $\mathrm{x}$\\
		TOTRESNS & Total Reserves of Depository Institutions & 5 & & $\mathrm{x}$\\
		BUSLOANS & Commercial and Industrial Loans, All Commercial Banks & 5 & & $\mathrm{x}$\\
		REALLN & Real Estate Loans & 5 & & $\mathrm{x}$\\
		NONREVSL & Total Nonrevolving Credit & 5 & & $\mathrm{x}$\\
		CONSPI & Credit to PI ratio & 1 & & $\mathrm{x}$\\
		CP3Mx & 3-Month AA Financial Commercial Paper Rate & 1 & & $\mathrm{x}$\\
		TB3MS & 3-Month Treasury Bill & 1 & & $\mathrm{x}$\\
		GS10 & 10-Year Treasury Rate & 1 & & $\mathrm{x}$\\
		COMPAPFFx & 3-Month Commercial Paper Minus FEDFUNDS & 1 & & $\mathrm{x}$\\
		TB3SMFFM & 3-Month Treasury C Minus FEDFUNDS & 1 & & $\mathrm{x}$\\
		TB6SMFFM & 6-Month Treasury C Minus FEDFUNDS & 1 & & $\mathrm{x}$\\
		T5YFFM & 5-Year Treasury C Minus FEDFUNDS & 1 & & $\mathrm{x}$\\
		AAAFFM & Moodys Aaa Corporate Bond Minus FEDFUNDS & 1 & & $\mathrm{x}$\\
		EXSZUSx & Switzerland-U.S. Foreign Exchange Rate & 5 & & $\mathrm{x}$\\
		EXJPUSx & Japan-U.S. Foreign Exchange Rate & 5 & & $\mathrm{x}$\\
		EXCAUSx & Canada-U.S. Foreign Exchange Rate & 5 & & $\mathrm{x}$\\
		WPSFD49502 & PPI: Final Demand & 5 & & $\mathrm{x}$\\
		WPSID61 & PPI: Processed goods & 5 & & $\mathrm{x}$\\
		WPSID62 & PPI: Unprocessed goods & 5 & & $\mathrm{x}$\\
		CPIAUCSL & CPI: All Items & 5 & & $\mathrm{x}$\\
		CPIAPPSL & CPI: Apparel & 5 & & $\mathrm{x}$\\
		CPITRNSL & CPI: Transportation & 5 & & $\mathrm{x}$\\
		CPIMEDSL & CPI: Medical Care & 5 & & $\mathrm{x}$\\
		CUSR0000SAC & CPI: Commodities & 5 & & $\mathrm{x}$\\
		CUSR0000SAD & CPI: Durables & 5 & & $\mathrm{x}$\\
		CUSR0000SAS & CPI: Services & 5 & & $\mathrm{x}$\\
		CPIULFSL & CPI: All Items Less Food & 5 & & $\mathrm{x}$\\
		CUSR0000SA0L2 & CPI: All Items Less Shelter & 5 & & $\mathrm{x}$\\
		CUSR0000SA0L5 & CPI: All Items Less Medical Care & 5 & & $\mathrm{x}$\\
		DDURRG3M086SBEA & PCE: Durable goods & 5 & & $\mathrm{x}$\\
		DNDGRG3M086SBEA & PCE: Nondurable goods & 5 & & $\mathrm{x}$\\
		DSERRG3M086SBEA & PCE: Services & 5 & & $\mathrm{x}$\\
		CES2000000008 & Avg Hourly Earnings: Construction & 5 & & $\mathrm{x}$\\
		CES3000000008 & Avg Hourly Earnings: Manufacturing & 5 & & $\mathrm{x}$\\
		DTCOLNVHFNM & Consumer Motor Vehicle Loans Outstanding & 5 & & $\mathrm{x}$\\
		DTCTHFNM & Total Consumer Loans and Leases & 5 & & $\mathrm{x}$\\
		\bottomrule
	\end{longtable}
	
	\noindent Notes: The dataset described in \citet{McCracken:Ng:2016} is available for download at \href{https://research.stlouisfed.org/econ/mccracken/fred-databases/}{https://research.stlouisfed.org/econ/}. Column Transf. code indicates the transformation of a series $x_{t}$, where: (1) no transformation, (2) $\Delta x_{t}$, (5) $\Delta \log \left(x_{t}\right)$, (6) $\Delta^{2} \log \left(x_{t}\right)$ with $\Delta^{i}$ indicating $i$ th differences. Columns Medium and X-large refer to the different model sizes discussed in Section 3.
}

%% file: tables/Table4_group_model_stats.tex
\begin{longtable}{llcccc}
\caption{Median, Std. deviation, Min and Max RMSFE ratio by FRED-MD group and model.}\label{tab:RMSFE_by_group_model}\\
\toprule
\textbf{Group} & \textbf{Model} & \textbf{Median} & \textbf{Std} & \textbf{Min} & \textbf{Max} \\
\midrule
\endfirsthead
\multicolumn{6}{l}{\textit{(continued)}} \\
\toprule
\textbf{Group} & \textbf{Model} & \textbf{Median} & \textbf{Std} & \textbf{Min} & \textbf{Max} \\
\midrule
\endhead
\bottomrule
\endfoot
\multicolumn{6}{l}{\textbf{$h=1$}} \\
\textit{Consumption, Orders, and Inventories} & BVAR & 0.998 & 0.020 & 0.980 & 1.042 \\
 & Factor model & 0.972 & 0.060 & 0.847 & 1.030 \\
 & NNAR & 0.983 & 0.144 & 0.856 & 1.303 \\
 & BART & 1.005 & 0.014 & 0.977 & 1.015 \\
 & Moirai-large & 1.014 & 0.078 & 0.858 & 1.047 \\
 & Chronos & 1.002 & 0.064 & 0.862 & 1.018 \\
 & TimesFM & 1.035 & 0.081 & 0.881 & 1.079 \\
\addlinespace
\textit{Housing} & BVAR & 1.007 & 0.008 & 0.990 & 1.016 \\
 & Factor model & 0.938 & 0.047 & 0.854 & 1.005 \\
 & NNAR & 0.940 & 0.077 & 0.863 & 1.129 \\
 & BART & 1.022 & 0.027 & 0.978 & 1.061 \\
 & Moirai-large & 0.959 & 0.088 & 0.849 & 1.140 \\
 & Chronos & 0.949 & 0.054 & 0.856 & 1.051 \\
 & TimesFM & 0.917 & 0.112 & 0.853 & 1.139 \\
\addlinespace
\textit{Interest and Exchange Rates} & BVAR & 0.967 & 0.054 & 0.849 & 1.043 \\
 & Factor model & 1.001 & 0.068 & 0.872 & 1.103 \\
 & NNAR & 1.166 & 0.317 & 1.027 & 1.910 \\
 & BART & 1.176 & 0.111 & 1.035 & 1.420 \\
 & Moirai-large & 1.033 & 0.016 & 1.001 & 1.065 \\
 & Chronos & 0.964 & 0.050 & 0.853 & 1.032 \\
 & TimesFM & 1.165 & 0.092 & 1.014 & 1.289 \\
\addlinespace
\textit{Labor Market} & BVAR & 0.954 & 0.066 & 0.827 & 1.140 \\
 & Factor model & 0.930 & 0.073 & 0.766 & 1.017 \\
 & NNAR & 0.905 & 0.132 & 0.733 & 1.300 \\
 & BART & 0.988 & 0.075 & 0.858 & 1.182 \\
 & Moirai-large & 0.910 & 0.135 & 0.703 & 1.242 \\
 & Chronos & 0.909 & 0.094 & 0.715 & 1.061 \\
 & TimesFM & 0.933 & 0.114 & 0.706 & 1.170 \\
\addlinespace
\textit{Money and Credit} & BVAR & 1.014 & 0.060 & 0.951 & 1.157 \\
 & Factor model & 0.995 & 0.041 & 0.927 & 1.082 \\
 & NNAR & 0.978 & 0.250 & 0.934 & 1.838 \\
 & BART & 1.089 & 0.059 & 0.975 & 1.181 \\
 & Moirai-large & 0.963 & 0.145 & 0.912 & 1.436 \\
 & Chronos & 0.960 & 0.047 & 0.889 & 1.017 \\
 & TimesFM & 0.970 & 0.099 & 0.934 & 1.297 \\
\addlinespace
\textit{Output and Income} & BVAR & 0.946 & 0.036 & 0.911 & 1.027 \\
 & Factor model & 0.953 & 0.031 & 0.914 & 1.029 \\
 & NNAR & 0.962 & 0.081 & 0.927 & 1.272 \\
 & BART & 1.001 & 0.045 & 0.913 & 1.066 \\
 & Moirai-large & 0.966 & 0.043 & 0.925 & 1.089 \\
 & Chronos & 0.967 & 0.027 & 0.918 & 1.016 \\
 & TimesFM & 1.011 & 0.130 & 0.921 & 1.482 \\
\addlinespace
\textit{Prices} & BVAR & 0.998 & 0.051 & 0.913 & 1.128 \\
 & Factor model & 1.002 & 0.068 & 0.838 & 1.057 \\
 & NNAR & 0.990 & 0.082 & 0.825 & 1.091 \\
 & BART & 0.997 & 0.038 & 0.936 & 1.106 \\
 & Moirai-large & 1.024 & 0.090 & 0.800 & 1.082 \\
 & Chronos & 1.016 & 0.082 & 0.806 & 1.086 \\
 & TimesFM & 1.010 & 0.079 & 0.787 & 1.050 \\
\addlinespace
\textit{Stock Market} & BVAR & 1.034 & 0.088 & 0.988 & 1.158 \\
 & Factor model & 1.002 & 0.031 & 0.949 & 1.005 \\
 & NNAR & 1.024 & 0.398 & 0.985 & 1.694 \\
 & BART & 1.015 & 0.037 & 0.956 & 1.024 \\
 & Moirai-large & 1.034 & 0.041 & 0.979 & 1.059 \\
 & Chronos & 1.020 & 0.026 & 0.974 & 1.020 \\
 & TimesFM & 1.020 & 0.068 & 0.988 & 1.119 \\
\addlinespace
\multicolumn{6}{l}{\textbf{$h=3$}} \\
\textit{Consumption, Orders, and Inventories} & BVAR & 0.988 & 0.041 & 0.911 & 1.043 \\
 & Factor model & 0.983 & 0.096 & 0.749 & 1.019 \\
 & NNAR & 0.988 & 0.112 & 0.783 & 1.150 \\
 & BART & 0.995 & 0.034 & 0.946 & 1.028 \\
 & Moirai-large & 0.995 & 0.077 & 0.796 & 1.000 \\
 & Chronos & 0.989 & 0.075 & 0.805 & 1.003 \\
 & TimesFM & 0.992 & 0.059 & 0.852 & 1.012 \\
\addlinespace
\textit{Housing} & BVAR & 0.961 & 0.038 & 0.901 & 1.005 \\
 & Factor model & 0.932 & 0.085 & 0.783 & 1.039 \\
 & NNAR & 0.947 & 0.107 & 0.785 & 1.146 \\
 & BART & 1.019 & 0.069 & 0.897 & 1.099 \\
 & Moirai-large & 0.941 & 0.121 & 0.770 & 1.158 \\
 & Chronos & 0.913 & 0.077 & 0.776 & 1.008 \\
 & TimesFM & 0.791 & 0.100 & 0.739 & 1.002 \\
\addlinespace
\textit{Interest and Exchange Rates} & BVAR & 0.964 & 0.057 & 0.866 & 1.089 \\
 & Factor model & 1.015 & 0.062 & 0.959 & 1.183 \\
 & NNAR & 1.135 & 0.083 & 1.018 & 1.280 \\
 & BART & 1.225 & 0.109 & 1.051 & 1.381 \\
 & Moirai-large & 1.016 & 0.062 & 0.907 & 1.111 \\
 & Chronos & 0.990 & 0.046 & 0.851 & 1.035 \\
 & TimesFM & 1.126 & 0.068 & 0.972 & 1.193 \\
\addlinespace
\textit{Labor Market} & BVAR & 0.905 & 0.057 & 0.791 & 0.998 \\
 & Factor model & 0.874 & 0.095 & 0.708 & 1.000 \\
 & NNAR & 0.899 & 0.115 & 0.719 & 1.108 \\
 & BART & 0.921 & 0.114 & 0.763 & 1.375 \\
 & Moirai-large & 0.897 & 0.106 & 0.732 & 1.069 \\
 & Chronos & 0.880 & 0.101 & 0.692 & 1.002 \\
 & TimesFM & 0.826 & 0.098 & 0.743 & 1.001 \\
\addlinespace
\textit{Money and Credit} & BVAR & 0.994 & 0.036 & 0.951 & 1.072 \\
 & Factor model & 0.979 & 0.052 & 0.873 & 1.055 \\
 & NNAR & 0.986 & 0.119 & 0.906 & 1.368 \\
 & BART & 1.082 & 0.063 & 0.915 & 1.114 \\
 & Moirai-large & 0.975 & 0.066 & 0.886 & 1.130 \\
 & Chronos & 0.969 & 0.054 & 0.881 & 1.033 \\
 & TimesFM & 0.981 & 0.069 & 0.899 & 1.142 \\
\addlinespace
\textit{Output and Income} & BVAR & 0.982 & 0.028 & 0.882 & 1.000 \\
 & Factor model & 0.979 & 0.025 & 0.911 & 1.008 \\
 & NNAR & 0.992 & 0.035 & 0.937 & 1.061 \\
 & BART & 1.006 & 0.040 & 0.910 & 1.079 \\
 & Moirai-large & 0.989 & 0.020 & 0.932 & 1.010 \\
 & Chronos & 0.981 & 0.023 & 0.913 & 1.002 \\
 & TimesFM & 0.985 & 0.090 & 0.920 & 1.318 \\
\addlinespace
\textit{Prices} & BVAR & 0.982 & 0.076 & 0.767 & 1.071 \\
 & Factor model & 1.007 & 0.122 & 0.682 & 1.070 \\
 & NNAR & 0.967 & 0.129 & 0.653 & 1.100 \\
 & BART & 1.036 & 0.097 & 0.825 & 1.162 \\
 & Moirai-large & 0.947 & 0.113 & 0.637 & 1.024 \\
 & Chronos & 0.945 & 0.115 & 0.632 & 1.015 \\
 & TimesFM & 1.015 & 0.137 & 0.635 & 1.151 \\
\addlinespace
\textit{Stock Market} & BVAR & 1.010 & 0.003 & 1.008 & 1.014 \\
 & Factor model & 1.011 & 0.001 & 1.011 & 1.013 \\
 & NNAR & 1.036 & 0.124 & 1.020 & 1.243 \\
 & BART & 1.046 & 0.077 & 1.002 & 1.152 \\
 & Moirai-large & 1.000 & 0.042 & 1.000 & 1.073 \\
 & Chronos & 1.008 & 0.044 & 0.983 & 1.068 \\
 & TimesFM & 0.995 & 0.020 & 0.993 & 1.029 \\
\addlinespace
\multicolumn{6}{l}{\textbf{$h=6$}} \\
\textit{Consumption, Orders, and Inventories} & BVAR & 0.997 & 0.027 & 0.945 & 1.035 \\
 & Factor model & 0.998 & 0.073 & 0.864 & 1.074 \\
 & NNAR & 1.001 & 0.069 & 0.883 & 1.085 \\
 & BART & 1.016 & 0.040 & 0.921 & 1.045 \\
 & Moirai-large & 1.002 & 0.049 & 0.893 & 1.009 \\
 & Chronos & 0.996 & 0.058 & 0.878 & 1.015 \\
 & TimesFM & 0.996 & 0.042 & 0.906 & 1.020 \\
\addlinespace
\textit{Housing} & BVAR & 0.892 & 0.083 & 0.760 & 1.002 \\
 & Factor model & 0.922 & 0.131 & 0.675 & 1.053 \\
 & NNAR & 0.923 & 0.159 & 0.673 & 1.186 \\
 & BART & 0.989 & 0.151 & 0.778 & 1.227 \\
 & Moirai-large & 0.910 & 0.157 & 0.666 & 1.144 \\
 & Chronos & 0.872 & 0.116 & 0.682 & 1.001 \\
 & TimesFM & 0.696 & 0.095 & 0.618 & 0.889 \\
\addlinespace
\textit{Interest and Exchange Rates} & BVAR & 0.994 & 0.063 & 0.836 & 1.106 \\
 & Factor model & 0.999 & 0.049 & 0.859 & 1.101 \\
 & NNAR & 1.095 & 0.055 & 1.003 & 1.229 \\
 & BART & 1.220 & 0.104 & 1.039 & 1.457 \\
 & Moirai-large & 1.008 & 0.082 & 0.827 & 1.159 \\
 & Chronos & 0.994 & 0.056 & 0.824 & 1.061 \\
 & TimesFM & 1.140 & 0.085 & 0.934 & 1.253 \\
\addlinespace
\textit{Labor Market} & BVAR & 0.984 & 0.068 & 0.734 & 1.016 \\
 & Factor model & 0.972 & 0.068 & 0.769 & 1.011 \\
 & NNAR & 0.972 & 0.074 & 0.751 & 1.060 \\
 & BART & 0.980 & 0.088 & 0.794 & 1.332 \\
 & Moirai-large & 0.966 & 0.072 & 0.757 & 1.081 \\
 & Chronos & 0.965 & 0.069 & 0.752 & 1.010 \\
 & TimesFM & 0.964 & 0.065 & 0.765 & 1.010 \\
\addlinespace
\textit{Money and Credit} & BVAR & 1.003 & 0.036 & 0.918 & 1.044 \\
 & Factor model & 0.999 & 0.063 & 0.856 & 1.105 \\
 & NNAR & 0.991 & 0.073 & 0.902 & 1.192 \\
 & BART & 1.050 & 0.107 & 0.933 & 1.274 \\
 & Moirai-large & 0.996 & 0.054 & 0.855 & 1.057 \\
 & Chronos & 0.983 & 0.060 & 0.862 & 1.041 \\
 & TimesFM & 0.998 & 0.062 & 0.880 & 1.135 \\
\addlinespace
\textit{Output and Income} & BVAR & 0.996 & 0.022 & 0.910 & 1.007 \\
 & Factor model & 0.998 & 0.008 & 0.972 & 1.004 \\
 & NNAR & 0.998 & 0.009 & 0.987 & 1.018 \\
 & BART & 1.008 & 0.021 & 0.990 & 1.061 \\
 & Moirai-large & 0.999 & 0.021 & 0.923 & 1.019 \\
 & Chronos & 0.989 & 0.019 & 0.922 & 1.006 \\
 & TimesFM & 0.999 & 0.038 & 0.970 & 1.138 \\
\addlinespace
\textit{Prices} & BVAR & 0.978 & 0.091 & 0.725 & 1.032 \\
 & Factor model & 1.003 & 0.141 & 0.602 & 1.044 \\
 & NNAR & 0.963 & 0.149 & 0.591 & 1.056 \\
 & BART & 1.021 & 0.119 & 0.757 & 1.146 \\
 & Moirai-large & 0.946 & 0.131 & 0.600 & 1.014 \\
 & Chronos & 0.934 & 0.139 & 0.573 & 1.016 \\
 & TimesFM & 0.960 & 0.214 & 0.593 & 1.629 \\
\addlinespace
\textit{Stock Market} & BVAR & 1.005 & 0.007 & 1.004 & 1.016 \\
 & Factor model & 1.015 & 0.006 & 1.007 & 1.019 \\
 & NNAR & 1.088 & 0.088 & 1.017 & 1.192 \\
 & BART & 1.096 & 0.149 & 1.049 & 1.327 \\
 & Moirai-large & 1.035 & 0.030 & 1.004 & 1.065 \\
 & Chronos & 1.004 & 0.062 & 0.980 & 1.097 \\
 & TimesFM & 1.026 & 0.038 & 0.994 & 1.069 \\
\addlinespace
\multicolumn{6}{l}{\textbf{$h=12$}} \\
\textit{Consumption, Orders, and Inventories} & BVAR & 0.999 & 0.050 & 0.924 & 1.080 \\
 & Factor model & 1.003 & 0.032 & 0.966 & 1.074 \\
 & NNAR & 1.001 & 0.033 & 1.000 & 1.089 \\
 & BART & 1.013 & 0.036 & 0.983 & 1.090 \\
 & Moirai-large & 0.999 & 0.015 & 0.973 & 1.018 \\
 & Chronos & 0.998 & 0.032 & 0.931 & 1.026 \\
 & TimesFM & 1.003 & 0.041 & 0.913 & 1.038 \\
\addlinespace
\textit{Housing} & BVAR & 0.862 & 0.114 & 0.671 & 1.015 \\
 & Factor model & 0.939 & 0.169 & 0.642 & 1.125 \\
 & NNAR & 0.936 & 0.192 & 0.625 & 1.218 \\
 & BART & 1.052 & 0.207 & 0.670 & 1.277 \\
 & Moirai-large & 0.871 & 0.162 & 0.630 & 1.104 \\
 & Chronos & 0.886 & 0.145 & 0.628 & 1.046 \\
 & TimesFM & 0.555 & 0.078 & 0.482 & 0.744 \\
\addlinespace
\textit{Interest and Exchange Rates} & BVAR & 1.009 & 0.070 & 0.849 & 1.162 \\
 & Factor model & 0.969 & 0.067 & 0.794 & 1.064 \\
 & NNAR & 1.124 & 0.107 & 1.005 & 1.418 \\
 & BART & 1.195 & 0.149 & 1.049 & 1.659 \\
 & Moirai-large & 1.032 & 0.119 & 0.771 & 1.309 \\
 & Chronos & 0.998 & 0.065 & 0.761 & 1.083 \\
 & TimesFM & 1.105 & 0.091 & 0.920 & 1.248 \\
\addlinespace
\textit{Labor Market} & BVAR & 0.996 & 0.076 & 0.735 & 1.062 \\
 & Factor model & 0.994 & 0.079 & 0.742 & 1.046 \\
 & NNAR & 0.994 & 0.085 & 0.769 & 1.221 \\
 & BART & 1.003 & 0.174 & 0.840 & 1.905 \\
 & Moirai-large & 0.995 & 0.099 & 0.763 & 1.324 \\
 & Chronos & 0.990 & 0.072 & 0.771 & 1.031 \\
 & TimesFM & 0.990 & 0.081 & 0.785 & 1.083 \\
\addlinespace
\textit{Money and Credit} & BVAR & 1.006 & 0.043 & 0.909 & 1.048 \\
 & Factor model & 1.009 & 0.048 & 0.909 & 1.086 \\
 & NNAR & 1.003 & 0.068 & 0.929 & 1.210 \\
 & BART & 1.071 & 0.070 & 0.988 & 1.191 \\
 & Moirai-large & 1.000 & 0.039 & 0.890 & 1.049 \\
 & Chronos & 0.988 & 0.064 & 0.856 & 1.053 \\
 & TimesFM & 1.013 & 0.051 & 0.920 & 1.113 \\
\addlinespace
\textit{Output and Income} & BVAR & 0.999 & 0.026 & 0.899 & 1.014 \\
 & Factor model & 1.002 & 0.006 & 0.983 & 1.007 \\
 & NNAR & 0.998 & 0.011 & 0.978 & 1.026 \\
 & BART & 1.020 & 0.036 & 1.004 & 1.146 \\
 & Moirai-large & 1.004 & 0.022 & 0.924 & 1.025 \\
 & Chronos & 0.998 & 0.020 & 0.922 & 1.006 \\
 & TimesFM & 1.005 & 0.017 & 0.974 & 1.048 \\
\addlinespace
\textit{Prices} & BVAR & 0.990 & 0.083 & 0.739 & 1.041 \\
 & Factor model & 0.990 & 0.119 & 0.633 & 1.029 \\
 & NNAR & 0.964 & 0.137 & 0.608 & 1.088 \\
 & BART & 0.992 & 0.094 & 0.769 & 1.118 \\
 & Moirai-large & 0.942 & 0.125 & 0.619 & 1.014 \\
 & Chronos & 0.942 & 0.136 & 0.581 & 1.022 \\
 & TimesFM & 0.972 & 0.174 & 0.662 & 1.440 \\
\addlinespace
\textit{Stock Market} & BVAR & 1.003 & 0.031 & 0.980 & 1.042 \\
 & Factor model & 1.019 & 0.013 & 1.004 & 1.030 \\
 & NNAR & 1.079 & 0.137 & 1.008 & 1.273 \\
 & BART & 1.105 & 0.182 & 1.051 & 1.391 \\
 & Moirai-large & 1.023 & 0.054 & 1.009 & 1.108 \\
 & Chronos & 1.001 & 0.039 & 0.984 & 1.058 \\
 & TimesFM & 1.050 & 0.049 & 0.990 & 1.086 \\
\end{longtable}